\documentclass[pra, notitlepage, superscriptaddress,  reprint, twocolumn,nofootinbib]{revtex4-1}

\usepackage{comment}
\usepackage{enumerate}
\usepackage{amssymb}
\usepackage{amsmath}
\usepackage{braket}
\usepackage{graphicx}
\usepackage[usenames,dvipsnames]{color}
\usepackage{tensor}
\usepackage{empheq}
\usepackage{capt-of}
\usepackage[normalem]{ulem} % either use this (simple) or
\usepackage{soul} % 
\usepackage{MnSymbol}
\usepackage{cancel}

\usepackage[colorlinks,bookmarks=false,citecolor=NavyBlue,linkcolor=OliveGreen,urlcolor=blue]{hyperref}
\newcommand{\be}{\begin{equation}}
\newcommand{\ee}{\end{equation}}
\newcommand{\ba}{\begin{aligned}}
\newcommand{\ea}{\end{aligned}}
\newcommand{\bw}{\begin{widetext}}
\newcommand{\ew}{\end{widetext}}

\renewcommand{\vec}[1]{\boldsymbol{#1}}
\newcommand{\bea}{\begin{eqnarray}}
\newcommand{\eea}{\end{eqnarray}}

\newcommand{\II}{\mathrm{i}}

\def\doi{http://dx.doi.org/}

\begin{document}
\title{Closure of the entanglement gap at quantum criticality: The case of the Quantum Spherical Model}
\author{Sascha Wald}
\email{swald@pks.mpg.de}
\affiliation{Max-Planck-Institut f\"ur Physik Komplexer Systeme,
N\"othnitzer Stra{\ss}e 38, D-01187, Dresden, Germany\looseness=-1}
\author{Ra\'ul Arias} 
\affiliation{SISSA and INFN, via Bonomea 265, 34136 Trieste, Italy}
\affiliation{Instituto de F\'isica La Plata - CONICET and Departamento de F\'isica, Universidad Nacional de La Plata C.C. 67, 1900, La Plata, Argentina}
\author{Vincenzo Alba}
\affiliation{Institute  for  Theoretical  Physics, Universiteit van Amsterdam,
Science Park 904, Postbus 94485, 1098 XH Amsterdam,  The  Netherlands}

\begin{abstract}

\noindent
The study of entanglement spectra is a powerful tool to detect or elucidate universal 
behaviour in quantum many-body systems. We investigate the scaling of the 
entanglement (or Schmidt) gap $\delta\xi$, i.e., the lowest laying gap of the 
entanglement spectrum, at a two-dimensional quantum critical point. 
We focus on the paradigmatic quantum spherical model, 
which exhibits a second-order transition, and is mappable to free bosons with an
additional external constraint. 
We analytically show that the Schmidt gap vanishes at the critical point, 
although only logarithmically. 
For a system on a torus and the half-system bipartition, the entanglement gap vanishes 
as $\pi^2/\ln(L)$, with $L$ the linear system size. 
The entanglement gap is nonzero in the paramagnetic phase and exhibits a faster decay  
in the ordered phase. 
The rescaled gap $\delta\xi\ln(L)$ exhibits  a crossing for different system sizes 
at the transition, although logarithmic corrections prevent a precise verification of
the finite-size scaling. 
Interestingly, the change of the entanglement gap across the phase diagram is 
reflected in the zero-mode eigenvector of the spin-spin correlator. 
At the transition quantum fluctuations give rise to a non-trivial structure of the 
eigenvector, whereas in the ordered phase it is 
flat. 
We also show that the vanishing of the entanglement gap at criticality can be qualitatively 
but not quantitatively captured by neglecting the structure of the zero-mode 
eigenvector. 

\end{abstract}

\maketitle

%################################################################
\section{Introduction}

\begin{figure}[t]
 \includegraphics[width=.38\textwidth]{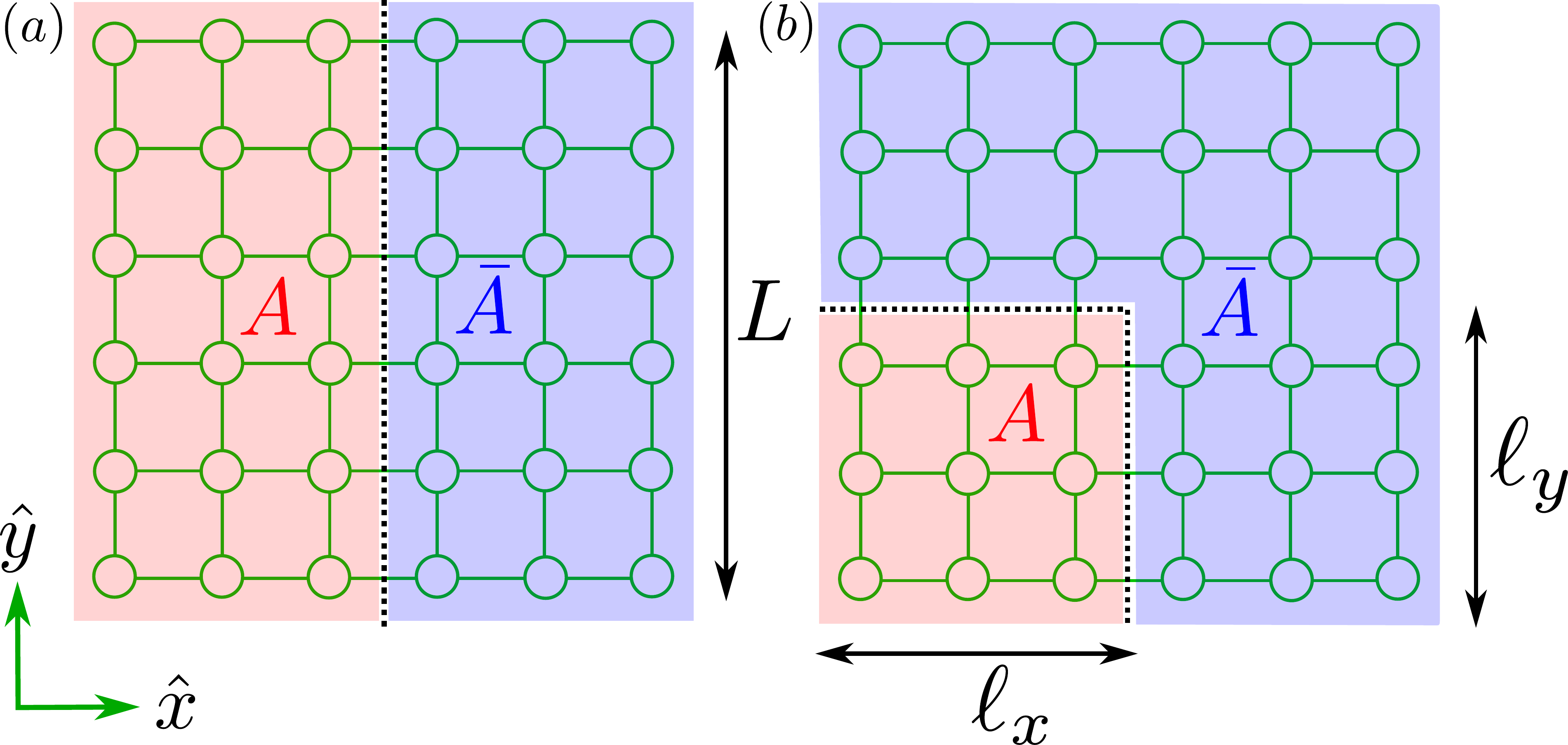}
 \caption{Bipartition of the two dimensional lattice as $A\cup\bar A$.
 Periodic boundary conditions in both directions are used. (a) 
 A bipartition with straight boundary between $A$ 
 and $\bar A$. $A$ contains $|A|=L\times\ell_x$ sites and spans the full lattice 
 along the $\hat y$ direction. (b) Bipartition with a corner. Now $|A|=\ell_x\ell_y$. 
 We also define the ratios $\omega_{x(y)}=\ell_{x(y)}/L$. We mostly consider the 
 the case with $\omega_y=1$. 
}
 \label{fig:partition1}
\end{figure}

\noindent
In the last two decades the study of quantum entanglement has revolutionised our 
understanding of quantum many-body systems~\cite{area,amico-2008,calabrese-2009,laflorencie-2016}. 
The main ingredient to address entanglement-related questions in a quantum system
$\mathcal{S}$ is the reduced 
density matrix $\rho_A$ of a subsystem $A \subset \mathcal{S}$. Given the ground-state $|\Psi\rangle$ 
of $\mathcal{S}$ and a spatial bipartition of $\mathcal{S}= A\cup\bar A$ (see e.g. 
Fig.~\ref{fig:partition1}), 
$\rho_A$ is defined as 
\begin{equation}
	\rho_A=\mathrm{Tr}_{_{\bar A}}|\Psi\rangle\langle\Psi|. 
\end{equation}
The entanglement spectrum (ES) $\{\xi_i = -\ln(\lambda_i)\,|\, \lambda_i \in \operatorname{spec}(\rho_A)\}$ has been the subject of intense investigation. 
Pioneering studies~\cite{peschel-1999,chung-2000,peschel-2004,viktor} were 
fueled by the rapid success of the density matrix renormalisation 
group~\cite{white-1992,uli-2011} to simulate one-dimensional 
quantum many-body systems. 
The interest in the ES was revived after it was discovered
that for fractional quantum Hall states  
the lower part of the ES contains universal information about the edge modes 
and the conformal field theory (CFT) describing them~\cite{li-2008}. This sparked intense 
theoretical activity to clarify the nature of the ES in fractional quantum 
Hall systems~\cite{thomale-2010,andreas-2010,haque-2007,thomale-2010a,hermanns-2011,chandran-2011,qi-2012,liu-2012,sterdyniak-2012,dubail-2012,dubail-2012a,Chan14}, topologically ordered 
phases of matter~\cite{pollmann-2010,turner-2011,bauer-2014}, frustrated and 
magnetically ordered systems~\cite{poilblanc-2010,cirac-2011,de-chiara-2012,alba-2011,metlitski-2011,alba-2012,Alba13,lepori-2013,james-2013,kolley-2013,Chan14,rademaker-2015,kolley-2015,frerot-2016}, CFT systems~\cite{calabrese-2008,lauchli-2013,Alba-2017,cardy-talk}, and  
systems with impurities~\cite{bayat-2014}. 

In this work we investigate the ES in {\it critical} two-dimensional quantum many-body systems. 
We focus on the lowest laying entanglement gap $\delta\xi$ defined as  
\begin{equation}
	\label{eq:gap-def}
\delta\xi=\xi_1-\xi_0,
\end{equation}
where $\xi_0$ and $\xi_1$ are the lowest and the first excited ES level, 
respectively. The behaviour of the entanglement gap at quantum critical points has not been 
thoroughly addressed, except for one-dimensional 
systems~\cite{truong-1987,peschel-1999,chung-2000,peschel-2004,alba-2011,andreas-2010,
de-chiara-2012,lepori-2013,di-giulio-2020}. Several exact results suggest 
that at one-dimensional quantum critical points $\delta\xi$ vanishes. 
For instance, in CFT systems $\delta\xi$ decays logarithmically
as $\propto 1/\ln(L)$ with the subsystem's length $L$~\cite{calabrese-2008}. Similar scaling 
is found in corner transfer matrix calculations~\cite{truong-1987} (see also 
\cite{viktor} for a review). Higher-dimensions are far less explored.
Interestingly, it has been argued that the closing of 
the entanglement gap does not necessarily signal critical behaviour~\cite{Chan14}. 
Similar conclusions have been reached by considering the ES of a bipartition in 
momentum space~\cite{lundgren-2014}. 
Still, the ES can be useful to distinguish different phases of matter. 
This is the case for systems that exhibit order by breaking of a 
continuous symmetry~\cite{metlitski-2011}. 
It has been suggested that deep in the ordered phase 
the lower part of the ES contains the fingerprints 
of symmetry breaking, being reminiscent of the so-called 
Anderson tower-of-states~\cite{lhullier,beekman-2020,Wietek-2017}. This has been verified 
by analytical calculations in the quantum rotor model~\cite{metlitski-2011}, 
numerical simulations in the two-dimensional Bose-Hubbard model 
in the superfluid phase~\cite{Alba13} (see also \cite{frerot-2016}), 
and also in two-dimensional Heisenberg models on the square~\cite{kolley-2013} and 
on the kagome lattice~\cite{kolley-2015}. 
A signature of the tower-of-states scenario is that 
the gaps in the lower part of the ES decay as a power-law 
with the subsystem volume, with multiplicative logarithmic 
corrections~\cite{metlitski-2011}. 
Higher ES levels are expected to exhibit a much slower 
decay~\cite{metlitski-2011,Alba13,rademaker-2015}. 
The behaviour of the entanglement gap upon approaching the 
critical point has not been investigated thoroughly. 

Here we address this issue in the quantum spherical 
model~\cite{Ober72,Henk84,Vojta96,Wald15,Bien12} (QSM). 
The QSM is a paradigmatic many-body system in which the effects
of strongly interacting degrees of freedom may be studied at a 
considerably low cost, as the model can be mapped to free 
bosons subject to an additional external constraint.
Despite its simplicity it exhibits several salient 
features of realistic quantum many-body systems. 
For instance, its classical version served as a testing ground for the 
theory of critical phenomena and finite size scaling~\cite{brezin-1982}. 
In two dimensions the QSM exhibits a standard paramagnetic (disordered)
phase and a ferromagnetic (ordered) one, which are separated by 
a second order quantum phase transition. The universality class of the 
transition is that of the three-dimensional classical $O(N)$ 
vector model~\cite{zinn-justin-1998} in the large $N$ limit~\cite{Stan68,Henk84,Vojta96}.
Surprisingly, entanglement properties of the QSM are rather unexplored, 
although there is recent interest~\cite{Lu19,Lu20,Wald20}. 
We should stress that although the results that we are going to derive for the ES 
cannot be considered general, they certainly represent an interesting case 
study, and can be useful to understand the generic behaviour of ES in quantum many-body 
systems. 

Here we consider a two-dimensional lattice of linear size $L$ with 
periodic boundary conditions in both directions. 
The typical bipartitions that we use are reported in Fig.~\ref{fig:partition1}. 
Figure~\ref{fig:partition1} (a) shows a bipartition with a straight boundary between 
$A$ and its complement, with $A$ spanning the full lattice 
along the $\hat y$ direction. This is not the case in 
Fig.~\ref{fig:partition1} (b), where the boundary has a corner. 
The effect of corners in the scaling of the entanglement entropies 
is nontrivial, and it has been studied intensely in the 
last decade~\cite{Casini:2008as,Casini:2006hu,PhysRevLett.110.135702,pitch,2014arXiv1401.3504K,Singh2012,Helmes2014,laflorencie-2016,Seminara-2017}. 
%
%%%%%%%%%%%%%%%%%%%%%%%%%%%%%%%%%%%%%%%%%%%%%%%%%%%%%%% 
%%%%%%%%%%%%%%%%%%%%%%%%%%%%%%%%%%%%%%%%%%%%%%%%%%%%%%% 
%
Since the QSM is mappable to free bosons, entanglement-related 
observables can be calculated from the 
two-point correlations functions~\cite{viktor}. 

Here we show that $\delta\xi$ (cf.~\eqref{eq:gap-def}) 
is nonzero in the paramagnetic phase, whereas it 
vanishes in the ordered phase, as expected~\cite{metlitski-2011}. 
This is compatible with the numerical results in 
\cite{Alba13} (see also 
\cite{kolley-2013,kolley-2015}). At the 
quantum critical point, in the case of straight boundary
the entanglement gap vanishes as $\pi^2/\ln(L)$. However, we show that  
logarithmic corrections are present, which make it difficult to 
robustly verify the finite-size scaling of $\delta\xi$. 
We also show that the behaviour of the entanglement gap is reflected 
in the zero-mode eigenvector of 
the spin-spin correlation matrix. As the transition is 
approached from the paramagnetic side, the eigenvector flattens, 
meaning that all its components become equal. This 
reflects the presence of a zero mode. Exactly at criticality, the 
eigenvector is not flat in the thermodynamic limit, due to  
the presence of strong fluctuations, whereas it  is flat 
in the ordered phase. Interestingly, we show that 
by neglecting the structure of the eigenvector at the 
critical point, i.e., by approximating the eigenvector with the flat 
vector, we obtain that $\delta\xi=
A/\sqrt{\ln(L)}$, which accounts for the vanishing of the 
entanglement gap, although it is not quantitatively accurate. 
We clarify how the behaviour as $A/\sqrt{\ln(L)}$ arises from some  
interesting multiplicative logarithmic corrections 
in the expectation values of the QSM correlators with the 
flat vector. Interestingly, the constant $A$ 
depends only on low-energy properties of the model and on the 
geometry of the bipartition. 

The manuscript is organised as follows. In section~\ref{sec:model} we introduce the 
QSM and its phase diagram. In section~\ref{sec:es-def} we define the quantities of 
interest. In section~\ref{sec:fs} we discuss the finite-size scaling in 
the QSM. Specifically, in subsection~\ref{sec:mu} we focus on the so-called 
gap equation, which ensures the external constraint in the QSM.
In subsections~\ref{sec:corr-S} and~\ref{sec:corr-p} 
we derive the finite-size 
scaling of the spin and momentum correlation functions, respectively. 
In section~\ref{sec:gap} we investigate the critical behaviour of $\delta\xi$. 
Our  prediction is discussed in section~\ref{sec:dim-red}, and it is compared 
against numerical results in section~\ref{sec:num-res}. 
We describe the behaviour of $\delta\xi$ across the phase diagram of the QSM 
in subsection~\ref{sec:overview}, whereas we address the vanishing of $\delta\xi$ and its 
finite-size scaling in subsections~\ref{sec:cr-gap} and~\ref{sec:fs-gap}, 
respectively. 
In section~\ref{sec:approx} we discuss how the entanglement gap is 
related to the zero-mode eigenvector of the correlator, which we 
introduce in subsection~\ref{sec:approx-1}. In subsection~\ref{sec:approx-2} 
we show that by assuming that the eigenvector is flat at 
criticality one can {\it qualitatively} explain the vanishing of the 
entanglement gap. We conclude in 
section~\ref{sec:concl}. In Appendix~\ref{app:app0} we report the 
derivation of the finite-size scaling of the correlation functions in the 
QSM. In Appendix~\ref{app:app1} we derive the expectation values of the 
correlators with the flat vector.

%################################################################
\section{Quantum Spherical Model}
\label{sec:model}

\noindent
The QSM~\cite{Henk84, Vojta96,Wald15} on a two dimensional cubic lattice of
linear size $L$ and volume $V=L^2$ is defined by the Hamiltonian
\begin{equation}
\label{ham}
H = \frac{g}{2}\sum_{\vec{n}} p_{\vec{}}^2 
- J\sum_{\langle \vec{n},\vec{m}\rangle} s_{\vec{n}}s_{\vec{m}} 
+ (\mu-2)\sum_{\vec{n}} s_{\vec{n}}^2. 
\end{equation}
Here, $\vec{n}=(n_x,n_y)\in [1,\ldots,L]^2$ denotes a generic lattice site, and 
$\langle \vec{n}, \vec{m}\rangle$ a lattice bond joining two nearest-neighbour sites. 
$J>0$ is the ferromagnetic exchange constant and we choose $J=1$ in the 
remainder of the paper.
The canonically conjugated variables $s_{\vec{n}}$  and $p_{\vec{n}}$ satisfy 
the standard bosonic commutation relations 
\begin{equation}
	[p_{\vec{n}},p_{\vec{m}}]=[s_{\vec{n}},s_{\vec{m}}]=0,
	\quad[s_{\vec{n}},p_{\vec{m}}]=\II\delta_{\vec{nm}}. 
\end{equation}
We refer to $p_{\vec{n}}$ as momentum variable, and to 
the parameter $g$ as {\it quantum coupling} as the model 
reduces to the famous classical 
spherical model~\cite{Berl52,Lew52} in the limit $g\to 0$. 
The Lagrange multiplier $\mu$ is called {\it spherical parameter} 
and fixes the spherical constraint, i.e.
\begin{equation}
\label{constr-mu}
\sum_{\vec{n}} \langle s_{\vec{n}}^2\rangle=V.
\end{equation}
This means that all allowed configurations of the QSM are located around the
sphere in configuration space that is defined by Eq.~(\ref{constr-mu}).
Critical properties of the QSM are 
determined through the self-consistent behaviour of $\mu$ \cite{Vojta96}. 
The two dimensional QSM does not exhibit a finite temperature phase 
transition \cite{Berl52,Lew52}, although it possesses a ground-state transition, i.e.,  
at $T=0$ \cite{Vojta96,Henk84,Wald15}.

We now briefly review how to diagonalise the Hamiltonian~\eqref{ham} 
and describe its critical behaviour.
First, we exploit the translational invariance of the model by 
performing a  Fourier transform as 
\begin{equation}
	p_{\vec{n}} = \frac{1}{\sqrt{V}}
	\sum_{\vec{k}} e^{-\II \vec{n} \vec{k}}\pi_{\vec{k}}\ ,
	\qquad s_{\vec{n}} = \frac{1}{\sqrt{V}}
	\sum_{\vec{k}} e^{\II \vec{n} \vec{k}}q_{\vec{k}}. 
\end{equation}
Here the sum over $\vec{k}= (k_x,k_y)$ runs in the first Brillouin zone 
$k_i= 2\pi/L \,j$, with $j\in[-L/2,L/2]$ integer. The Hamiltonian~\eqref{ham} 
in Fourier space reads
\begin{equation}
\label{ham-k}
 H = \sum_{\vec{k}} \frac{g}{2}\pi_{\vec{k}} \pi_{-{\vec{k}}}
 + \Lambda_{\vec{k}}^2 \, q_{\vec{k}}q_{-{\vec{k}}} 
\end{equation}
with the single-particle dispersion relation
\begin{equation}
\label{disp}
	\Lambda_{\vec{k}} = \sqrt{\mu + \omega_{\vec{k}}}\quad \textrm{with} \quad 
	\omega_k = 2-\cos k_x-\cos k_y
\end{equation}
In order to fully diagonalise~\eqref{ham-k} we introduce bosonic ladder 
operators $b_{\vec{k}}$ and $b_{\vec{k}}^\dagger$ obeying standard bosonic commutation 
relations viz. 
\begin{equation}
\label{eq:quantisation}
q_{\vec{k}} = \alpha_{\vec{k}} \frac{b_{\vec{k}}+b_{-\vec{k}}^\dagger}{\sqrt{2}} \ , 
\qquad \pi_{\vec{k}} = \frac{\II}{\alpha_{\vec{k}}}
\frac{b_{\vec{k}}^\dagger - b_{-\vec{k}}}{\sqrt{2}} 
\end{equation}
with the parameter $\alpha_{\vec{k}}^2 = \sqrt{g/2}\Lambda_{\vec{k}}^{-1}$.
In terms of these ladder operators, the Hamiltonian~\eqref{ham-k} is diagonal
and reads
\begin{equation}
\label{ham-diag}
H = \sum_{\vec{k}} E_{\vec{k}} (b_{\vec{k}}^\dagger b_{\vec{k}} + 1/2),
\quad \textrm{with} \quad E_{\vec{k}} = \sqrt{2g} \Lambda_{\vec{k}}  .
\end{equation}
Entanglement-related properties of Gaussian systems such as the QSM
stem from the two-point correlation functions $\langle s_{\vec{n}}
s_{\vec{m}}\rangle$ and $\langle p_{\vec{n}}p_{\vec{m}}\rangle$. 
In equilibrium at zero temperature $T=0$, the eigenmodes $\vec{k}$ 
of the system are occupied according to 
\begin{equation}
\label{eq:bose}
 \langle b_{\vec{k}} b_{\vec{k}'} \rangle=
 \langle b^\dagger_{\vec{k}} b^\dagger_{\vec{k}'} \rangle 
 =\langle b^\dagger_{\vec{k}'} b_{\vec{k}}  \rangle = 0  , \quad 
 \langle b_{\vec{k}'} b^\dagger_{\vec{k}}  \rangle = \delta_{\vec{k}\vec{k}'}.
\end{equation}
From Eq.~\eqref{eq:bose}, we can thus immediately derive the two-point 
correlation functions~\cite{Wald15}
\begin{align}
\label{eq:snsm}
\mathbb{S}_{\vec{nm}}=\langle s_{\vec{n}} s_{\vec{m}} \rangle &= \frac{1}{2V}\sum_{\vec{k}} 
e^{\II (\vec{n}-\vec{m})\cdot \vec{k}}  \alpha_{\vec{k}}^2, \\
\label{eq:pnpm}
\mathbb{P}_{\vec{nm}} =\langle p_{\vec{n}} p_{\vec{m}} \rangle &= \frac{1}{2V}\sum_{\vec{k}}
e^{-\II (\vec{n}-\vec{m})\cdot \vec{k}}  \alpha_{\vec{k}}^{-2},\\
\label{eq:snpm}
\mathbb{K}_{\vec{nm}}=\langle s_{\vec{n}} p_{\vec{m}}\rangle &= \frac{\II}{2}\delta_{\vec{nm}}.
\end{align}
Importantly, from~\eqref{eq:snsm} and~\eqref{eq:pnpm} one obtains the relation
\begin{equation}
\label{eq:id}
\mathbb{P}_{\vec{n}\vec{m}}=\frac{1}{g}\int d\mu\, \mathbb{S}_{\vec{n}\vec{m}}, 
\end{equation}
which allows to relate the critical behaviour of the spin correlator to that 
of the momentum correlator. 
From~\eqref{eq:snsm}, one can rewrite the spherical constraint~\eqref{constr-mu} as
\begin{equation}
\label{mu-fs}
\sum_{\vec{n}}\mathbb{S}_{\vec{n}\vec{n}} = V \quad \Rightarrow \quad
\frac{2}{g} = \frac{1}{V}\sum_{\vec{k}}\frac{1}{E_{\vec{k}}}.
% =
% \frac{2}{g}\mathbb{S}_{\vec{n}\vec{n}}. 
\end{equation}
This equation is also called gap equation~\cite{Amit84}
% in the context of the large-$N$ model
and implies that only the average number of bosons is fixed. 
From the finite-size expressions~\eqref{eq:snsm}~\eqref{eq:pnpm} and~\eqref{mu-fs}, 
the thermodynamic limit $L\to\infty$  is obtained in the usual way by replacing 
\begin{align}
	\frac{2\pi k_j}{L}&\to k'_j, \quad
\frac{1}{L^2}\sum_{k_x,k_y}\to\prod_{j=x,y}\int_{-\pi}^\pi \frac{dk'_j}{2\pi}. 
\end{align}
A crucial observation is that the correlator~\eqref{eq:snsm} 
and the spherical parameter~\eqref{mu-fs} exhibit a singularity for 
$\vec{k}=0$, due to the zero mode. We anticipate that this will play an important
role in the behaviour of the entanglement gap. This contribution of the zero mode 
to the entanglement entropy was previously investigated
focusing on the harmonic chain~\cite{Botero04}. 

We now summarise the zero-temperature critical behaviour of the QSM. In two 
dimensions the model exhibits a second order phase transition at a critical 
value $g_c$. For $g<g_c$ the ground-state of~\eqref{ham} exhibits magnetic 
order. At $g>g_c$ the ground-state is paramagnetic. The behaviour of the QSM 
is determined by the scaling of the spherical 
parameter $\mu$. In the thermodynamic limit, in the paramagnetic phase one has 
that $\mu$ is finite and nonzero. On the other hand, one has 
$\mu=0$  at the critical point, and in the ordered phase. The value of 
$g_c$ can be determined analytically. In the thermodynamic limit the spherical 
constraint~\eqref{mu-fs} is rewritten as 
\begin{equation}
 \sqrt{\frac{2}{g}} 
= \frac{2}{\pi^2}
\frac{ K^2\left(\frac{1}{2}-\frac{\mu+2}{4} \sqrt{2-\frac{2 \sqrt{\mu  (\mu +4)}}{\mu +2}}\right)}
{\Big[(1+\frac{\mu}{2} ) \left(\mu +\sqrt{\mu  (\mu +4)}+2\right)-1\Big]^\frac{1}{4}} 
\end{equation}
with the complete elliptic integral~\cite{Abra65}
\begin{equation} 
	K(x)=\int_0^\frac{\pi}{2}\frac{d\theta}{\sqrt{1-x^2\sin^2(\theta)}}. 
	\label{eq:K}
\end{equation}
The critical coupling $g_c$ follows by imposing the condition $\mu = 0$. 
This yields 
\begin{align}
  g_c= \frac{\pi^4}{2} K^{-4}\left(1/2-1/\sqrt{2}\right) \simeq 9.67826. 
  \label{gcrit}
\end{align}
The different phases of the model correspond to different finite-size scaling behaviours 
of $\mu$. In the paramagnetic phase one has $\mu={\mathcal O}(1)$ in the limit $L\to\infty$. 
At the critical point one can show that $\mu={\mathcal O}(1/L^2)$, 
whereas in the ordered phase $\mu={\mathcal O}(1/L^4)$ (see section~\ref{sec:fs}). These behaviours are 
numerically illustrated in Fig.~\ref{fig:mu}. The universality class of the ground 
state transition~\cite{Vojta96} is that of the large-$N$ vector model 
in three dimensions, as expected from general renormalisation group arguments. 
Critical properties of the large-$N$ vector model have been characterised 
analytically~\cite{brezin-1982} and finite-size corrections have also been 
investigated~\cite{singh-1987,chamati-1997,caracciolo-1998,chamati-2000,caracciolo-2001}.

%################################################################
\section{Entanglement spectra and entanglement gaps}
\label{sec:es-def}

\noindent
Here we are interested in the ground-state entanglement spectrum of the QSM,
focussing on the two bipartitions depicted in Fig.~\ref{fig:partition1}. The 
lattice, with periodic boundary conditions, is divided into two regions 
$A$ and $\bar{A}$. Region $A$ is of size $|A|=\ell_x\times\ell_y$ and 
we define the corresponding aspect ratios $\omega_x=\ell_x/L$ and $\omega_y=\ell_y/L$, with 
$0\le \omega_{x,y}\le1$. 
In Fig.~\ref{fig:partition1} (a) the subsystem $A$ spans the full lattice along 
the $\hat y$ direction implying that the boundary between the two subsystems $A$ and $\bar{A}$
is straight. This case corresponds to $\omega_y=1$. 
In Fig.~\ref{fig:partition1} (b), the boundary presents a 
corner and is thus not straight. The presence of corners has striking 
consequences for entanglement entropies, giving rise to 
sub-leading universal logarithmic 
corrections~\cite{Casini:2008as,Casini:2009sr,Casini:2006hu,PhysRevLett.110.135702,pitch,2014arXiv1401.3504K}. The effects of corners in the scaling of the ES 
have not been investigated yet. 

For the case of a straight boundary with periodic boundary conditions 
the momentum $k_y$ is a good quantum 
number for the correlation matrices~\eqref{eq:snsm} and~\eqref{eq:pnpm}, and for 
the ES. This will be exploited in section~\ref{sec:gap} to 
reduce the computation of the ES of the QSM to that 
of an effective one-dimensional model. This dimensional reduction 
has been employed
to study symmetry-resolved entanglement entropies~\cite{sara2D}. 
This rather simple observation will also allow to obtain analytically the scaling 
of the entanglement gap at the critical point, by exploiting 
corner transfer matrix
results~\cite{truong-1987,peschel-2004,chung-2000,peschel-1999}.

We now review the calculation of entanglement-related quantities in the QSM. 
Since the QSM is essentially mappable to a free bosonic model (see section~\ref{sec:model}), 
its entanglement properties are derived from the two-point 
correlation functions~\eqref{eq:snsm} and~\eqref{eq:pnpm} (see Ref.~\cite{viktor} for a 
review). 
The crucial ingredient is the correlation matrix $\mathbb{C}$ restricted to the subsystem $A$, viz.
\begin{equation}
\label{eq:ca}
\mathbb{C}_A=\mathbb{S}_A \cdot \mathbb{P}_A, 
\end{equation}
with $\mathbb{S}_A$ and $\mathbb{P}_A$ being the correlation matrices 
defined in~\eqref{eq:snsm} and~\eqref{eq:pnpm}, restricted to the subsystem 
$A$. Since in 
the remainder we mostly consider the restricted correlation matrices 
$\mathbb{S}_A$ and $\mathbb{P}_A$,  we will often omit the 
subscript $A$ to lighten the notation. 

For free bosons the reduced density matrix of subsystem $A$ 
is a quadratic operator and is written as~\cite{viktor}  
\begin{equation}
	\label{eq:rdm-fb}
	\rho_A=Z^{-1}e^{-{\mathcal H}_A},\quad{\mathcal H}_A=\sum_k\epsilon_k b^\dagger_k 
	b_k. 
\end{equation}
Here ${\mathcal H}_A$ is the so-called entanglement Hamiltonian, $\epsilon_k$ are 
the single-particle ES levels, $b_k$ are free-bosonic 
operators and $Z$ ensures the normalisation of the 
reduced density matrix $\mathrm{Tr}\rho_A=1$.  
The spectrum  $\{e_k\}_{k=1,\dots,|A|}$
of the correlation matrix $\mathbb{C}_A$ is simply related to 
that of ${\mathcal H}_A$ viz.
\begin{equation}
\label{eq:spectra}
\sqrt{e_k} =\frac{1}{2}\coth\left(\frac{\epsilon_k}{2}\right).
\end{equation}
The normalisation factor $Z$ is obtained as 
\begin{equation}
	Z=\prod_{j=1}^{|A|}\Big(\sqrt{e}_j+\frac{1}{2}\Big). 
\end{equation}
The ES, i.e., the spectrum of the entanglement 
Hamiltonian ${\mathcal H}_A$, is obtained by filling 
the single-particle levels $\epsilon_k$ in all the possible 
ways. To construct the ES, 
it is convenient to introduce the bosonic occupation numbers 
$\alpha_k=0,1,\dots$ in the levels $\epsilon_k$. The generic ES level 
$\xi(\{\alpha_k\})$ is written as 
\begin{equation}
\label{eq:es}
\xi(\{\alpha_k\})=\ln Z+\sum_{j=1}^{|A|}\alpha_j\epsilon_j. 
\end{equation}
The eigenvalues $e_k$ satisfy the constraint  $e_k>1/4$, implying that 
$\epsilon_k>0$. Clearly, the lowest ES level $\xi_0$ 
corresponds to the vacuum state with $\alpha_k=0$ for all $k$. 
Let us order the $\epsilon_k$ as  
$\epsilon_1\le\epsilon_2\le\dots\le\epsilon_{|A|}$. 
The first excited ES level is obtained by populating the smallest 
single particle level $\epsilon_1$. Thus, the lowest 
entanglement gap $\delta\xi$ (Schmidt gap) is defined as 
\begin{equation}
\label{eq:dxi-def}
\delta\xi\equiv \xi_1-\xi_0=\epsilon_1.
\end{equation}
Here we focus on $\delta\xi$, although one can define 
higher gaps~\cite{DiGiulio19}.

%################################################################
\section{Finite-size critical correlators in the QSM}
\label{sec:fs}

\noindent
As explained in section~\ref{sec:es-def}, entanglement-related observables, 
and also the entanglement gap, in the QSM are entirely encoded in the two-point correlation 
functions~\eqref{eq:snsm} and~\eqref{eq:pnpm}. 
In the following sections we derive the finite-size behaviour of these two-point correlation 
functions.
In section~\ref{sec:mu} we discuss the gap equation~\eqref{mu-fs}. 
In sections~\ref{sec:corr-S} and~\ref{sec:corr-p} we the focus on the 
spin and momentum correlators respectively. For the classical spherical model similar results 
were obtained~\cite{singh-1987,singh-1989}.

%
%%%%%%%%%%%%%%%%%%%%%%%%%%%%%%%%%%%%%%%%%%%%%%%%%%%%%%% 
\begin{figure}[t]
\includegraphics[width=.43\textwidth]{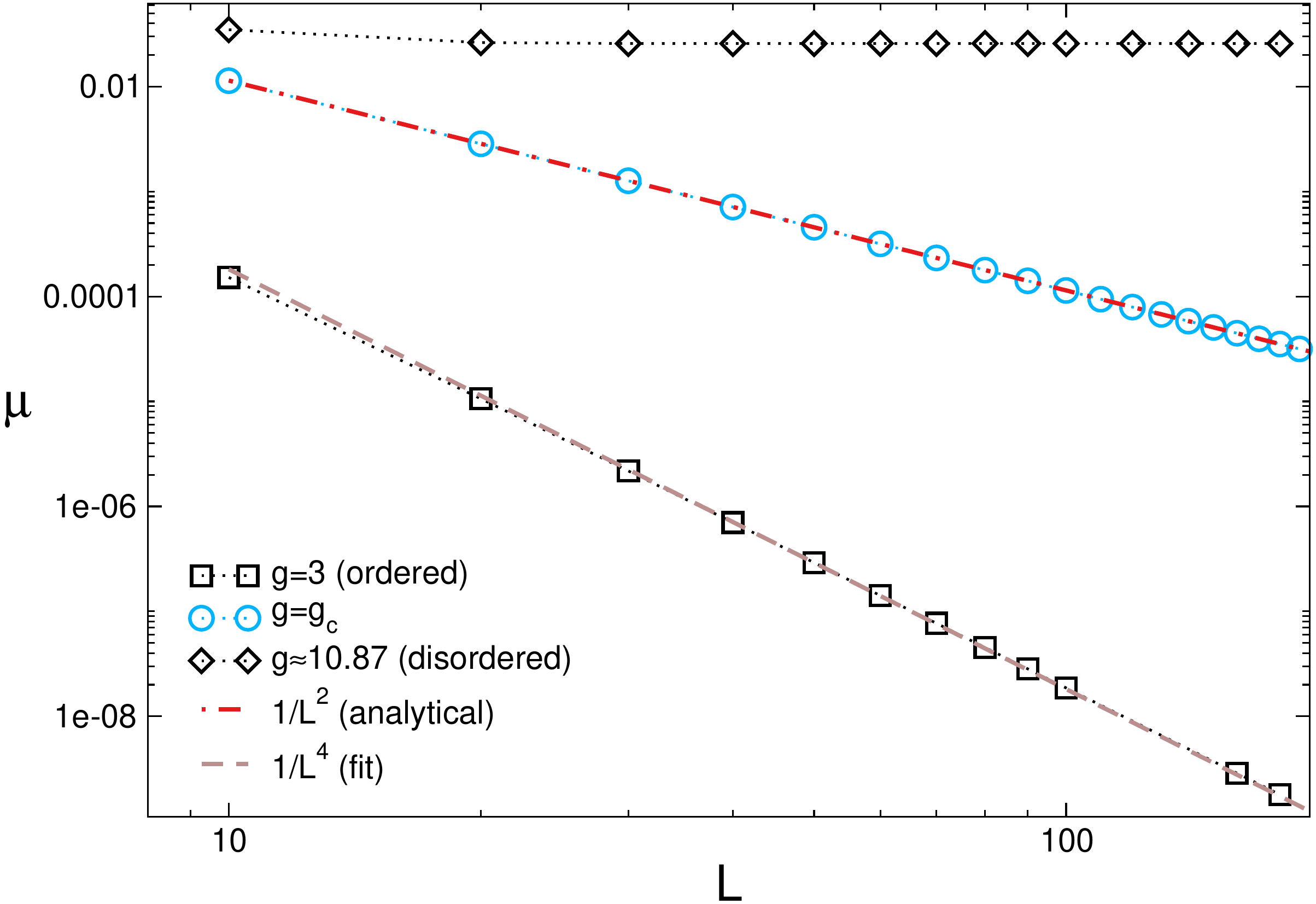}
\caption{
 Spherical parameter $\mu$ as a function of 
 linear size $L$ at the quantum critical point 
 at $g_c$ (circles), in the ordered phase 
 (squares), and in the paramagnetic phase (diamonds). 
 Note the different scaling with $L$ in the different phases 
 and at the critical point. The dashed-dotted line 
 is the analytic behaviour $\gamma^2_2/(2L^2)$. The dashed 
 line is a fit. 
}
 \label{fig:mu}
\end{figure}
%%%%%%%%%%%%%%%%%%%%%%%%%%%%%%%%%%%%%%%%%%%%%%%%%%%%%%% 
%

%################################################################
\subsection{Spherical parameter}
\label{sec:mu}

\noindent
Here we derive the finite-size scaling of the spherical parameter 
$\mu$ at the quantum phase transition. The 
result is not new~\cite{brezin-1982} but it is 
a useful initiation for the discussion of the correlators. 
To treat the sum over $\vec{k}$ in~\eqref{eq:snsm} we
observe that the following identity holds  
\begin{align}
\begin{split}
	\frac{1}{L^2}\sum_{\vec{k}}
	\frac{1}{\sqrt{\mu+\omega_{\vec{k}}}}=2
\int_0^\infty \frac{dt}{\sqrt{\pi}}&e^{-(\mu+2)t^2}
\big[I_0(t^2)\\
&\hspace{.75cm} +\sideset{}{'}\sum_{l=-\infty}^{\infty}I_{l L}(t^2)\big]^2,
\label{eq:1}
\end{split}
\end{align}
where the prime in the sum indicates that the $l=0$ contribution is removed, 
and $I_\nu$ are modified Bessel functions of the first kind~\cite{Abra65}. 
To derive~\eqref{eq:1}, we introduce an 
auxiliary integration~\cite{Amit84} over 
$t$ to represent the term $(\mu+\omega_{\vec{k}})^{-1/2}$,  
then we employ Poisson's summation formula. 
Further details are reported in Appendix~\ref{app:app0}. 
The first term in the brackets in~\eqref{eq:1} does not depend explicitly on $L$, 
and gives the thermodynamic contribution. However, there is an implicit 
dependence on $L$ through $\mu$. The second term is the genuine finite-size contribution. 
We are interested in the leading finite-size behaviour for large $L$. 
In this limit the integral in~\eqref{eq:1} can be treated by using a 
saddle point approximation. 

In order to use~\eqref{mu-fs}, we decompose the diagonal correlator 
$\mathbb{S}_{\vec{n}\vec{n}}$ as 
\begin{equation}
\label{eq:decomp}
	\mathbb{S}_{\vec{n}\vec{n}}=\mathbb{S}_{\vec{n}\vec{n}}^{(th)}+
	\mathbb{S}_{\vec{n}\vec{n}}^{(L)},
\end{equation}
with the thermodynamic contribution 
\begin{equation}
	\label{eq:s-th-0}
\mathbb{S}_{\vec{n}\vec{n}}^{(th)}=\frac{1}{8\pi^2}\int d\vec{k}\,\alpha^2_{\vec{k}}
\end{equation}
corresponding to the term $I_0(t^2)^2$ in~\eqref{eq:1}. 
The remaining terms in~\eqref{eq:1} are collected in  
$\mathbb{S}_{\vec{n}\vec{n}}^{(L)}$.\footnote{A similar
decomposition as~\eqref{eq:decomp} holds for the generic 
spin-spin correlator $\mathbb{S}_{\vec{n}\vec{m}}$ (see section~\ref{sec:corr-S}).} 
After expanding the square in~\eqref{eq:1}, we observe that 
$\mathbb{S}_{\vec{n}\vec{n}}^{(L)}$ is written as 
\begin{equation}
	\label{eq:int1}
	\mathbb{S}_{\vec{n}\vec{n}}^{(L)}=\frac{\sqrt{g_c}}
	{\sqrt{2\pi}}\int_0^\infty\!\!\!\! dt 
	e^{-(\mu+2)t^2}\sum\limits_{l,l'=-\infty}^\infty I_{lL}(t^2)I_{l'L}(t^2). 
\end{equation}
In order to extract the large $L$ behaviour of~\eqref{eq:int1} we 
employ a standard saddle point approximation. 
The calculation is straightforward and details are
reported in Appendix~\ref{app:app0}. 

A striking simplification occurs at the critical point and in the ordered 
phase, where $\mu\to0$. One can verify numerically 
that at the thermodynamical critical point $\mu\propto 1/L^2$. 
This is expected because $\mu\propto m^2=1/\xi_\mathrm{corr}^2$, with $m$ the mass 
of the theory and $\xi_\mathrm{corr}$ the correlation length, and at the critical point  
$\xi_\mathrm{corr}\propto L$. In the limit $\mu\to 0$, one obtains the  
surprisingly elegant result (see Appendix~\ref{app:app0})
\begin{equation}
	\label{eq:s1x}
	\mathbb{S}_{\vec{n}\vec{n}}^{(L)}= -\frac{\sqrt{g_c}}
	{\pi L}\Big[\ln \left(1-e^{-\sqrt{2\mu } L}\right)
	-\sum_{l,l'=1}^{\infty}
	\frac{e^{-L \sqrt{2\mu  \left(l^2+l'^2\right)}}}
{\sqrt{l^2+l'^2}}\Big].
\end{equation}
Interestingly, in~\eqref{eq:s1x} the first term is of one-dimensional nature, and 
it is obtained by isolating the terms with either $l=0$ or $l'=0$ in the sum 
in~\eqref{eq:int1}. In the second term in~\eqref{eq:s1x} the 
scaling as $\mu\propto 1/L^2$ gives rise to a non-trivial 
behaviour of the correlator as it cancels the factor $L$ in the 
exponential. It also implies that terms with large $l,l'$ are 
exponentially suppressed, and the sums converge quickly. 
Double sums as in~\eqref{eq:s1x} appear often in lattice 
calculations, and have been investigated in the past~\cite{chamati-1997, singh-1987, singh-1989}.
In some cases they can be expressed in 
terms of generalised Riemann zeta functions~\cite{contino-2002}.

Using Eqs.~\eqref{eq:s-th-0} and~\eqref{eq:s1x} in the gap equation~\eqref{mu-fs}
at criticality yields
\begin{multline}
\label{eq:sp}
1=\frac{\sqrt{g_c}}{8\sqrt{2}\pi^2}\int_{-\pi}^\pi
\frac{d\vec{k}}{\sqrt{\mu+\omega_{\vec{k}}}}
	+\frac{\sqrt{g_c}}{\pi L}\sum_{l,l'=1}^{\infty}
	\frac{e^{-L \sqrt{2\mu  \left(l^2+l'^2\right)}}}
	{\sqrt{l^2+l'^2}}\\
	-\frac{\sqrt{g_c}}{\pi L}\ln \left(1-e^{-\sqrt{2\mu } L}\right).
\end{multline}
The integral in~\eqref{eq:sp} has to be considered carefully due to
a $\propto 1/L$ contribution in the $\mu\to0$ 
limit which can be extracted as~\cite{Wald20} 
\begin{equation}
\label{eq:interm}
\int\frac{d\vec{k}}{\sqrt{\mu+\omega_{\vec{k}}}}=
\int\frac{d\vec{k}}{\sqrt{\omega_{\vec{k}}}} 
-4\pi\sqrt{\mu}+\dots, 
\end{equation}
where the dots denote subleading terms in $1/L$. 
The second term in~\eqref{eq:interm} is the singular term that determines the critical 
behaviour of three-dimensional QSM at the thermal phase transition~\cite{Wald20}. This 
is not surprising because  the universality class of the quantum phase transition 
in two dimensions is the same~\cite{Henk84,Vojta96}. Based on the expected finite-size scaling 
$\mu\propto 1/L^2$ it is convenient to define 
\begin{equation}
\label{eq:mu-red}
\mu=\frac{\gamma_2^2}{2L^2}, 
\end{equation}
where the constant $\gamma_2$ is to be determined and the factor $2$ is for 
later convenience. 
We substitute the Ansatz~\eqref{eq:mu-red} in the gap equation~\eqref{eq:sp}, and 
use the spherical constraint in the thermodynamic 
limit~\eqref{mu-fs} at criticality, where $\mu=0$. This yields
\begin{equation}
\label{eq:mu-gc}
\frac{\gamma_2}{4}-
	\sum_{l,l'=1}^{\infty}
	\frac{e^{-\gamma_2\sqrt{\left(l^2+l'^2\right)}}}
	{\sqrt{l^2+l'^2}}   
	+\ln \left(1-e^{-\gamma_2}\right)=0,
\end{equation}
where the first term is~\eqref{eq:interm} and 
the other two are obtained from~\eqref{eq:int1}. 
Eq.~\eqref{eq:mu-gc} can be solved numerically to obtain 
the universal constant $\gamma_2\simeq1.51196$. 
Note that Eq.~\eqref{eq:mu-gc} has also been found 
in the context of the large $N$ limit of the three dimensional $N-$vector 
model~\cite{brezin-1982,caracciolo-1998}. 
The behaviour of $\mu$ in the different regions of the phase 
diagram of the QSM and the accuracy of~\eqref{eq:mu-red} are verified in Fig.~\ref{fig:mu}
where we show the numerical solution of Eq.~\eqref{mu-fs}. 
In the paramagnetic region for $g>g_c$ one has $\mu={\mathcal O}(1)$. At the critical point 
and in the ferromagnetic phase $\mu\to0$ in the limit $L\to\infty$. The dashed-dotted line 
is the analytic result~\eqref{eq:mu-red} with $\gamma_2$ obtained from~\eqref{eq:mu-gc}. 
Below the critical point we expect $\mu\propto 1/L^4$~\cite{brezin-1982},
which is confirmed by the fit (dashed line).

%################################################################
\subsection{Spin-spin correlation function $\mathbb{S}_{\vec{n}\vec{m}}$}
\label{sec:corr-S}

\noindent
We now discuss the finite-size scaling of the spin-spin correlation 
function~\eqref{eq:snsm} at the quantum critical point. We only 
discuss the final result, reporting the details of the derivation 
in Appendix~\ref{app:app0}. First, one can again decompose the correlator as 
\begin{equation}
\label{eq:s-decomp}
{\mathbb{S}}_{\vec{n}\vec{m}}=\mathbb{S}_{\vec{n}\vec{m}}^{(th)}+\mathbb{S}_{\vec{n}\vec{m}}^{(L)},
\end{equation}
with the thermodynamic contribution
\begin{equation}
	\label{eq:snm-th}
	\mathbb{S}_{\vec{n}\vec{m}}^{(th)}=
	\frac{\sqrt{g_c}}{2\sqrt{2}(2\pi)^2}\int d\vec{k}\frac{e^{i\vec{k}(\vec{n}-\vec{m})}}
	{\sqrt{\mu+\omega_{\vec{k}}}}.
\end{equation}
As in Eq.~\eqref{eq:s-th-0} there is an implicit dependence on $L$ via $\mu$. 
The finite-size part has the surprisingly simple form  
\begin{align}
	\label{eq:s1}
	\mathbb{S}^{(L)}_{\vec{n}\vec{m}}=
	\frac{\sqrt{g_c}}{4\pi}\sideset{}{'}\sum_{l,l'=-\infty}^\infty\frac{e^{-\sqrt{2\mu}
	F_{ll'}(\vec{n},\vec{m})}}
	{F_{ll'}(\vec{n},\vec{m})}. 
\end{align}
Here we defined 
\begin{equation}
F_{ll'}({\vec{n},\vec{m}})=\sqrt{(lL+n_x-m_x)^2+(l'L+n_y-m_y)^2}.
\end{equation}
The prime in the sum means that the term $(l,l')=(0,0)$ has been removed. 
Again, Eq.~\eqref{eq:s1} holds in the limit $L\to\infty$ and $\mu\to0$. 
The general expression, which is valid also in the paramagnetic phase, 
is reported in Appendix~\ref{app:app0}. From Eq.~\eqref{eq:s1}, it is clear that the 
correlators $\mathbb{S}_{\vec{n}\vec{m}}$ depend only on $n_x-m_x$ and $n_y-m_y$, 
as expected due to translation invariance. Moreover, one has that 
$\mathbb{S}_{\vec{n}\vec{m}}$ is periodic along the two directions, 
i.e., it is invariant under $n_y-m_y\to n_y-m_y\pm L$ and $n_x-m_x\to n_x-m_x\pm L$. 
This is enforced by the infinite sums over $l,l'$. 
For a bipartition with straight boundary between the two 
subsystems (Fig.~\ref{fig:partition1} (a)) the invariance under $n_y-m_y\to n_y-m_y\pm L$ 
remains true also for the correlator restricted to $A$. 
Finally, $\mathbb{S}_{\vec{n}\vec{m}}^{(L)}$ exhibits an interesting 
singularity structure. For $\omega_y=1$
the denominator in Eq.~\eqref{eq:s1} is singular, whereas it is 
regular for $\omega_y<1$. Specifically, 
the terms with $l=0$ and $l'=\pm1$ in~\eqref{eq:s1} exhibit a singularity in the 
limit $n_x-m_x\to0$ and $n_y-m_y\to \pm L$. 
On the other hand, terms with $|l'|>1$ or $|l|>1$ in~\eqref{eq:s1} are not singular. 
The same singularity appears if $\omega_x=1$ and $\omega_y<1$. 
We anticipate that these singularities will 
give rise to multiplicative logarithmic corrections in the expectation 
value of the correlators that we will show in section~\ref{sec:approx}. 

%################################################################
\subsection{Momentum correlation function $\mathbb{P}_{\vec{n}\vec{m}}$}
\label{sec:corr-p}

\noindent
The same finite-size analysis as in section~\ref{sec:corr-S} 
can be carried out for the momentum correlator $\mathbb{P}_{\vec{n}\vec{m}}$ 
(cf.~\eqref{eq:pnpm}). Following the decomposition 
\begin{equation}
\label{eq:p-decomp}
\mathbb{P}_{\vec{n}\vec{m}}=\mathbb{P}_{\vec{n}\vec{m}}^{(th)}+\mathbb{P}_{\vec{n}\vec{m}}^{(L)},
\end{equation}
with
\begin{equation}
\label{eq:p-th}
	\mathbb{P}_{\vec{n}\vec{m}}^{(th)}=
	\frac{1}{4\sqrt{2 g_c}\pi^2}\int_{-\pi}^\pi d\vec{k} e^{i\vec{k}(\vec{n}-\vec{m})}
	\sqrt{\mu+\omega_{\vec{k}}} ,
\end{equation}
the finite-size part $\mathbb{P}_{\vec{n}\vec{m}}^{(L)}$ has the same 
structure as~\eqref{eq:s1},  and it reads 
\begin{multline}
	\label{eq:q2}
	\mathbb{P}^{(L)}_{\vec{n}\vec{m}}=\\
	-\frac{1}{4\pi\sqrt{g_c}}\,\,\sideset{}{'}\sum_{l,l'=-\infty}^\infty
	\frac{e^{-\sqrt{2\mu}F_{ll'}(\vec{n},\vec{m})}}{F^2_{ll'}(\vec{n},\vec{m})}
	\Big[\frac{1}{F_{ll'}(\vec{n},\vec{m})}
	+\sqrt{2\mu}\Big].
\end{multline}
This expression is obtained from the spin-spin correlator, cf. Eq~\eqref{eq:s1}, by using~\eqref{eq:id}.
As for~\eqref{eq:s1}, the finite-size term~\eqref{eq:q2} is singular 
if  subsystem $A$ spans the full lattice in one of the two directions, 
i.e., if $\omega_x=1$ or $\omega_y=1$.
For $\omega_y=1$ the singularity occurs  for $l=0$ and 
$l'=\pm1$ in the limit $n_x-m_x\to0$ and 
$n_y-m_y\to\pm L$. Note that the first term in Eq.~\eqref{eq:q2} exhibits a stronger 
singularity than the second one. 

%################################################################
\section{Critical behaviour of the entanglement gap}
\label{sec:gap}

\noindent
We now discuss the critical behaviour of the entanglement 
gap $\delta\xi$. In subsection~\ref{sec:dim-red}, by using 
a dimensional reduction, we provide an 
exact result for the case of a smooth boundary between the 
subsystems. In subsection~\ref{sec:num-res} we discuss 
numerical results. We first discuss the behaviour of the 
entanglement gap across the phase diagram of the QSM 
in subsection~\ref{sec:overview}. In subsection~\ref{sec:cr-gap} we 
show that at the critical point the entanglement gap 
vanishes logarithmically with the system size. Finally, 
in subsection~\ref{sec:fs-gap} we investigate the finite-size 
scaling $\delta\xi$ near criticality.

%################################################################
\subsection{Exact result via dimensional reduction}
\label{sec:dim-red}

\noindent
Let us focus on the bipartition with $\omega_y=1$ (see Fig.~\ref{fig:partition1} a).  
Periodic boundary conditions along the $\hat y$ direction imply
that the momentum $k_y$ is a good quantum number 
for the correlation matrix $\mathbb{C}_A$ (cf.~\eqref{eq:ca}) restricted 
to subsystem $A$. 
Moreover, translation invariance implies that by performing a Fourier 
transform along the $\hat y$ direction the Hamiltonian~\eqref{ham}  
can be written as the sum of $L$ {\it decoupled} quadratic one-dimensional 
systems~\cite{viktor}. This dimensional reduction is effective for any free system, 
and has been recently employed to 
study the so-called symmetry-resolved entanglement entropies \cite{sara2D}. 
The fact that $k_y$ is a good quantum number implies that the 
correlation matrix $\mathbb{C}_A$ has a block 
structure with each block corresponding to a different $k_y$ viz.
\begin{equation}
	\mathbb{C}_A=\bigoplus_{k_y}\mathbb{C}_A^{(k_y)}, \quad k_y=\frac{2\pi}{L}j,\,j=0,1,\dots L-1. 
\end{equation}
It is straightforward to diagonalise a given block with 
fixed $k_y$ by imposing that the 
eigenvectors of $\mathbb{C}_A$ are also eigenvectors 
of the momentum along $\hat y$ with the given eigenvalue $k_y$. 
Since we are interested only in the largest eigenvalue $e_1$ of $\mathbb{C}_A$ 
a further simplification occurs. As the critical behaviour  
is associated with the formation of a uniform magnetization, it 
is natural to expect that $e_1$ is in the sector 
with $k_y=0$. This can be readily checked numerically. Thus, in the 
following we restrict the calculation to $k_y=0$. By imposing that the 
eigenvectors of $\mathbb{C}_A$ are ``flat'' along $\hat y$, i.e., they 
do not depend on $y$, the problem is reduced to the diagonalisation of 
the reduced correlation matrix  
\begin{equation}
	\label{eq:ca-mom}
	\mathbb{C}_A^{(k_y=0)}=\mathbb{S}_A^{(k_y=0)}\cdot\mathbb{P}_A^{(k_y=0)}, 
\end{equation}
where we defined the reduced spin and momentum correlators as  
\begin{align}
\label{eq:snsm-1}
& \mathbb{S}_A^{(k_y=0)}(n_x-m_x)= \frac{1}{2L}\sum_{k_x} 
e^{\II (n_x-m_x)k_x}  \alpha_{k_x}^2, \\
\label{eq:pnpm-1}
& \mathbb{P}^{(k_y=0)}_A(n_x-m_x)= \frac{1}{2L}\sum_{k_x}
e^{-\II (n_x-m_x)k_x}  \alpha_{k_x}^{-2}. 
\end{align}
Eqs.~\eqref{eq:snsm-1} and~\eqref{eq:pnpm-1} 
depend only on the coordinates $n_x-m_x$ along the $\hat x$ direction, 
and subsystem $A$ is the interval of length $\ell_x$.
Here $\alpha_{k_x}$ corresponds to $\alpha_{\vec{k}}$ in Eq.~\eqref{eq:quantisation} with  
$k_y=0$. 
The correlators~\eqref{eq:snsm-1} and~\eqref{eq:pnpm-1} and hence~\eqref{eq:ca-mom}  
are formally the same as those of the so-called massive harmonic 
chain with frequency $\Omega=\sqrt{2\mu}$~\cite{viktor}.
The full ES of the massive harmonic chain for the bipartition 
in two semi-infinite chains 
has been calculated by using the corner transfer matrix 
approach~\cite{viktor}. The reduced density matrix 
$\rho_A$, up to a trivial renormalisation, is written as 
\begin{equation}
\label{eq:ctm}
\rho_A\sim e^{-{\mathcal H}_{\rm ctm}}, 
\end{equation}
with the corner transfer matrix Hamiltonian
\begin{equation}
\label{eq:ctm-h}
{\mathcal H}_{\rm ctm}=\sum_{j=0}^\infty \epsilon (2j+1)\beta_j^\dagger \beta_j, 
\,\,\epsilon=\frac{\pi K(\sqrt{1-\kappa^2})}{K(\kappa)},
\end{equation}
where $\beta_j$ are bosonic ladder operators. Here $K(x)$ is the complete elliptic 
integral of the first kind (see Eq.~(\ref{eq:K})).
The parameter $\kappa$ is given in terms of $\Omega$ 
as~\cite{sara2D} 
\begin{equation}
\label{eq:kapp}
\kappa=\frac{1}{2}(2+\Omega^2-\Omega\sqrt{4+\Omega^2}). 
\end{equation}
Eq.~\eqref{eq:ctm} holds if $A$ is the half-infinite line. 
In this limit, as it is clear from Eq.~\eqref{eq:ctm-h}, 
the single-particle ES levels are equally spaced~\cite{viktor} with spacing 
$\epsilon$. 
To determine the finite-size scaling of the entanglement gap $\delta\xi$ 
we use the fact that for $L\to\infty$ at criticality $\mu\propto 1/L^2$ 
(see Eq.~\eqref{eq:mu-red}).  
By substituting~\eqref{eq:mu-red} in the corner transfer matrix results~\eqref{eq:ctm-h}
and~\eqref{eq:kapp}, we obtain that in the large $L$ limit $\delta\xi$ decays logarithmically 
with $L$ as 
\begin{equation}
\label{eq:gap-decay}
\delta\xi=\frac{\pi^2}{\ln(\frac{8L}{\gamma_2})}+{\mathcal O}(\ln^{-3}(L)), 
\end{equation}
Note the dependence on the universal constant $\gamma_2$. 
To derive~\eqref{eq:gap-decay}, one can also observe that close to the 
critical point, on the paramagnetic side, Eq.~\eqref{eq:ctm-h} 
gives $\delta\xi=\pi^2/\ln(\xi_{\mathrm{corr}})$. 
Eq.~\eqref{eq:gap-decay} then follows from standard scaling arguments. 
A similar decay of the entanglement gap as in~\eqref{eq:gap-decay} 
is obtained for critical one-dimensional 
systems~\cite{viktor}, both fermionic and bosonic ones. 
An important remark is that the corner transfer matrix calculation is valid 
for the bipartition in two semi-infinite systems, which implies that there is 
only one boundary between the two subsystems, in contrast with the bipartitions 
Fig.~\ref{fig:partition1}, which contain two boundaries because we are using 
periodic boundary conditions. Despite that, as it will be clear in 
section~\ref{sec:num-res}, Eq.~\eqref{eq:gap-decay}  gives the leading behaviour 
for large $L$ of $\delta\xi$. 
We anticipate that a logarithmic subleading term as ${\mathcal O}(\ln^{-2}(L))$, 
which is missing in Eq.~\eqref{eq:gap-decay}, is present. 
From Eqs.~\eqref{eq:rdm-fb} and~\eqref{eq:gap-decay}
one obtains that the eigenvalue $e_1$ of $\mathbb{C}_A$ is given as 
\begin{equation}
\label{eq:e1}
e_1=\frac{1}{6}+\frac{1}{\pi^4}\ln^2\Big(\frac{8L}{\gamma_2}\Big)+{\mathcal O}(\ln^{-2}(L)).  
\end{equation}
Importantly, 
the missing ${\mathcal O}(\ln^{-2}(L))$ term in~\eqref{eq:gap-decay} will give 
a ${\mathcal O}(\ln(L))$ contribution in~\eqref{eq:e1}. 

%################################################################
\subsection{Numerical results}
\label{sec:num-res}

\noindent
In this section we discuss numerical results confirming the validity 
of the logarithmic scaling of the entanglement gap at criticality.
We provide numerical evidence that the prefactor of the logarithmic 
decay obeys the standard finite-size scaling behaviour. For instance, 
it exhibits a crossing for different system sizes at the critical point. However, 
logarithmic corrections are present, and a precise finite-size scaling 
analysis is very challenging. 

%################################################################
\subsubsection{Overview}
\label{sec:overview}

\begin{figure}[t]
 \includegraphics[height=5.5cm]{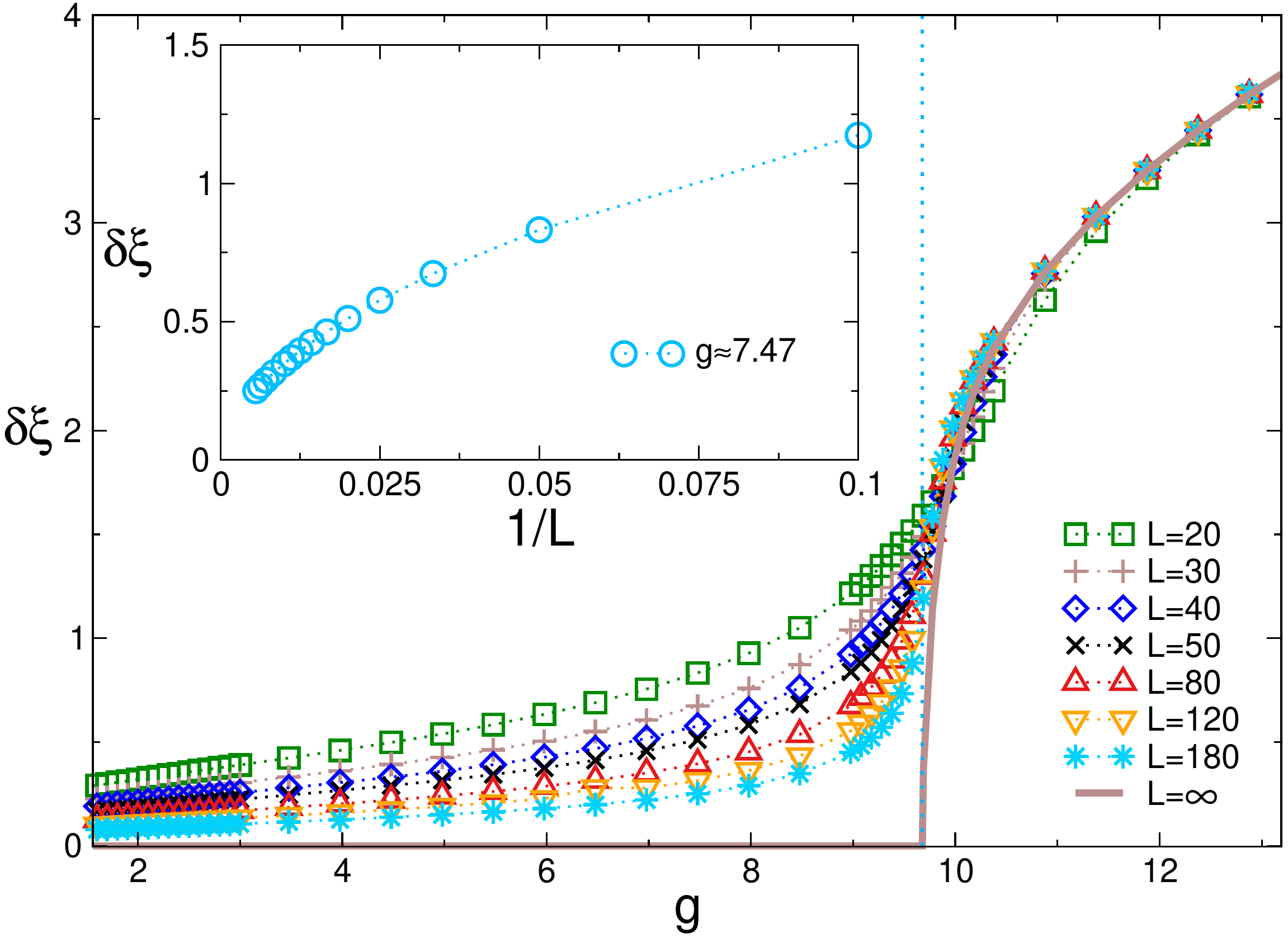}
 \caption{Entanglement gap $\delta\xi$ as a function of $g$ and 
 linear size $L$: Overview across the phase diagram. 
 The results are for the bipartition in Fig.~\ref{fig:partition1} 
 (a) with $\ell_x=L/2$. The vertical line marks the critical point 
 at $g_c$. 
 The continuous line is the result in the thermodynamic limit. 
 Inset: Scaling of the entanglement gap in the ordered phase 
 at $g<g_c$. 
}
\label{fig:gap_ov}
\end{figure}

\noindent
Before discussing the scaling of $\delta\xi$ at the critical point, it is useful 
to focus on its behaviour across the phase diagram of the QSM, see Fig.~\ref{fig:gap_ov}. 
%
%%%%%%%%%%%%%%%%%%%%%%%%%%%%%%%%%%%%%%%%%%%%%%%%%%%%%%% 
%%%%%%%%%%%%%%%%%%%%%%%%%%%%%%%%%%%%%%%%%%%%%%%%%%%%%%% 
%
The figure shows $\delta\xi$ as a function of $g$
for several system sizes $L$. The entanglement spectrum 
is calculated for the bipartition with straight boundary, i.e., $\omega_y=1$ and $\omega_x=1/2$
(see Fig.~\ref{fig:partition1} (a)).
In Fig.~\ref{fig:gap_ov} the solid line is $\delta\xi$ as  obtained by using the 
value of the spherical constraint $\mu$ in the thermodynamic limit $L\to\infty$
(cf.~\eqref{eq:sp}). This yields $\mu={\mathcal O}(1)$ in the paramagnetic 
phase and $\mu = 0$ in the ferromagnetic phase and at criticality ($g\le g_c$). The thermodynamic entanglement gap is 
obtained by substituting the thermodynamic value of $\mu$ in the finite-size 
expressions for the correlators (cf.~\eqref{eq:snsm} and~\eqref{eq:pnpm}) and 
taking the limit $L\to\infty$ after. This procedure gives the correct thermodynamic 
behaviour of $\delta\xi$, at least  away from the critical point. 
Although we use the finite-size expressions for the correlators, 
we observe that $\delta\xi$ converges quickly to its thermodynamic 
value. This is expected because the behaviour of the QSM is 
determined by the scaling of $\mu$. 
In the ordered phase and at the critical point the spin correlator~\eqref{eq:snsm} 
diverges due to the zero mode. Thus, we regularise the zero-mode by fixing $\mu=10^{-6}$. 
As it is clear from Fig.~\ref{fig:gap_ov}, this analysis, although it is not 
rigorous, suggests that 
$\delta\xi=0$ in the ordered phase, whereas $\mu$ is finite and nonzero in the 
paramagnetic phase. 

Let us now discuss the finite-size behaviour of $\delta\xi$. 
In the paramagnetic phase, i.e. $g>g_c$, the approach to the thermodynamic limit is exponential, 
which is expected because the model is massive. For $g<g_c$, i.e., in the 
ferromagnetic phase, the data suggest a vanishing gap. The 
scaling of the entanglement gap in magnetically ordered phases has 
been investigated extensively~\cite{metlitski-2011,Alba13,kolley-2013,kolley-2015,frerot-2016}. 
For instance, in Ref.~\cite{metlitski-2011} it was predicted that 
in the presence of continuous symmetry breaking in generic dimension $d$, 
$\delta\xi$ should decay as 
\begin{equation}
	\label{eq:grover}
	\delta\xi\propto (L^{d-1}\ln(L))^{-1}. 
\end{equation}
In $d=1$ one recovers the logarithmic decay as $1/\ln(L)$, reflecting the 
absence of symmetry breaking. In $d>1$ Eq.~\eqref{eq:grover} yields a  
``fast'' power-law decay with a multiplicative logarithmic correction. 
An important remark is that Eq.~\eqref{eq:grover} applies to the gaps in the 
lower part of the entanglement spectrum, i.e., the part which is related 
to the Anderson tower of states. Gaps in the higher part of the entanglement spectrum 
are expected to vanish logarithmically~\cite{metlitski-2011}. 

%################################################################
\subsubsection{Vanishing of the entanglement gap at the quantum critical point}
\label{sec:cr-gap}

\noindent
%
%%%%%%%%%%%%%%%%%%%%%%%%%%%%%%%%%%%%%%%%%%%%%%%%%%%%%%% 
\begin{figure}[t]
\includegraphics[height=5.5cm]{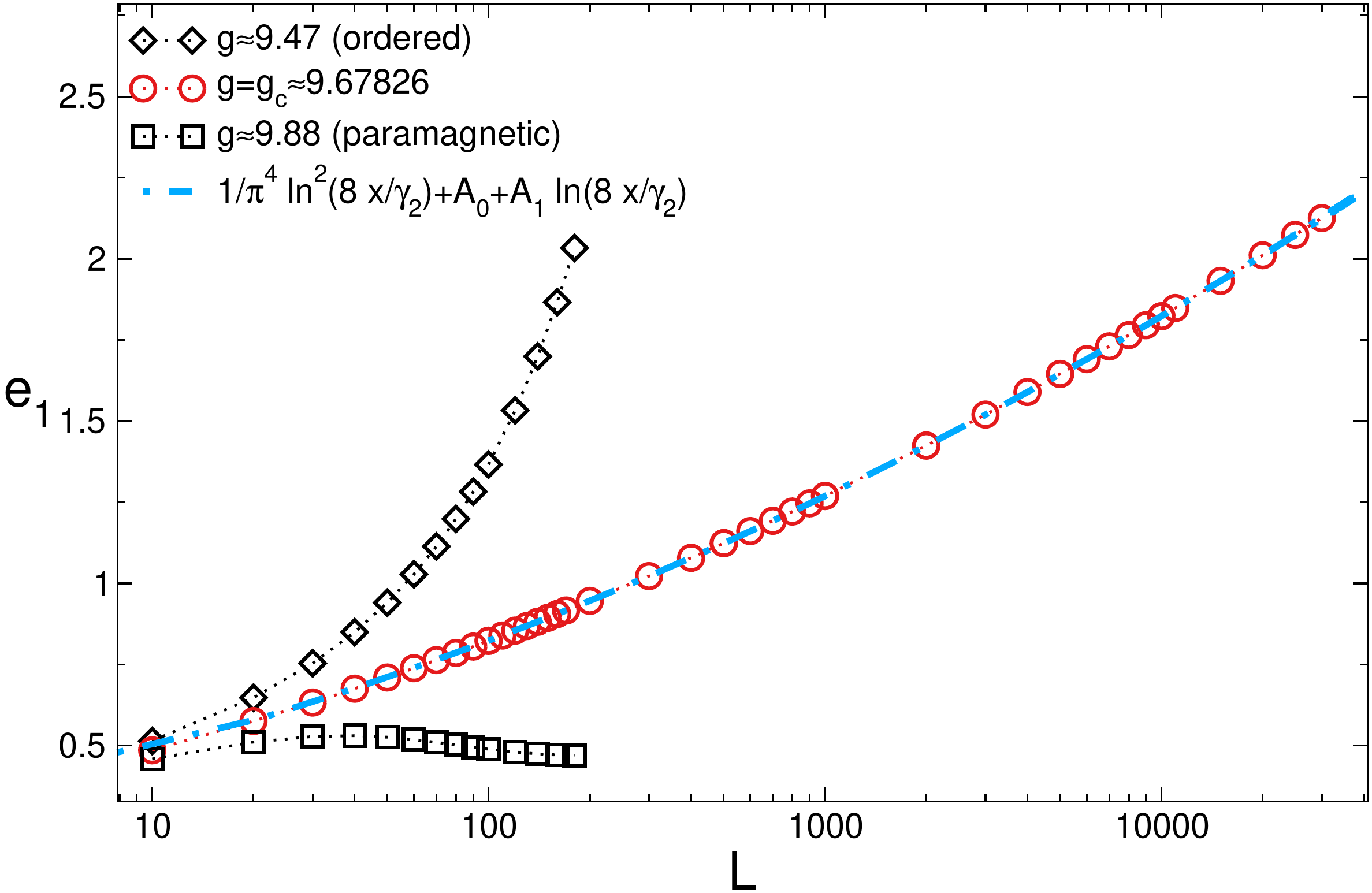}
\caption{Largest eigenvalue $e_1$ of the correlation matrix. 
 Data are for the bipartition in Fig.~\ref{fig:partition1} (a) 
 with $\omega_x=1/2$ and $\omega_y=1$. $e_1$  is 
 plotted versus linear size $L$. In the ordered phase (diamonds) 
 we observe a fast increase with $L$, whereas in the paramagnetic phase 
 $e_1={\mathcal O}(1)$. Note the logarithmic divergence 
 as $e_1\propto \ln^2(L)$ at the critical point at $g_c$. 
 The dashed dotted line is a fit to $e_1=1/\pi^4\ln^2(8L/\gamma_2)+A_0+
 A_1\ln(8L/\gamma_2)$, with $A_0,A_1$ fitting parameters. 
}
\label{fig:gap_log}
\end{figure}
%%%%%%%%%%%%%%%%%%%%%%%%%%%%%%%%%%%%%%%%%%%%%%%%%%%%%%% 
%
%
%%%%%%%%%%%%%%%%%%%%%%%%%%%%%%%%%%%%%%%%%%%%%%%%%%%%%%% 
\begin{figure}[t]
\includegraphics[height=5.5cm]{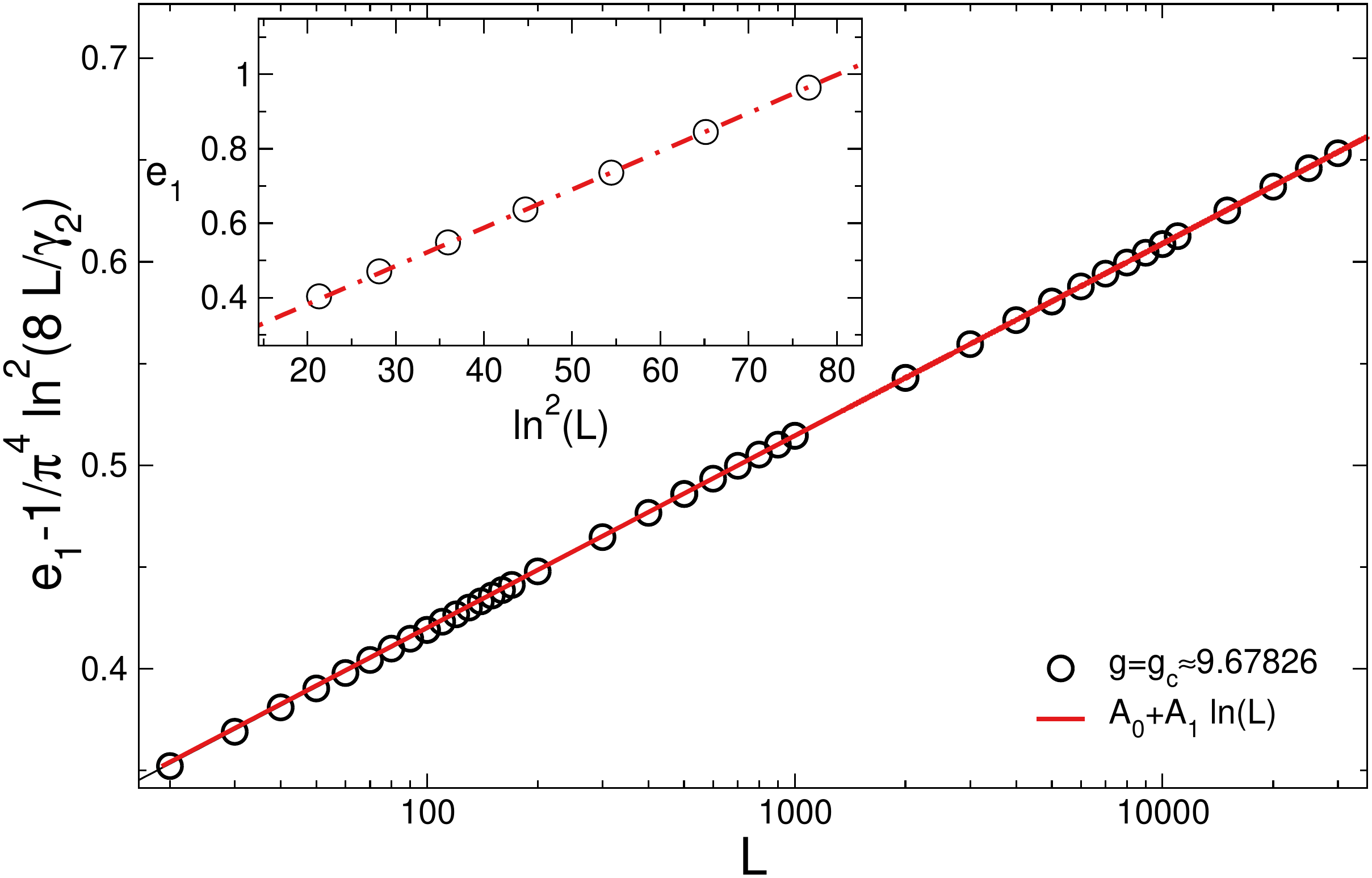}
\caption{Largest eigenvalue $e_1$ of the correlation matrix: Subleading 
 logarithmic correction. Plot of $e_1-1/\pi^4\ln^2(8L/\gamma_2)$ versus $\ln(L)$. 
 The data are the same as in Fig.~\ref{fig:gap_log}. The line is a 
 fit to $A_0+A_1 \ln(8L/\gamma_2)$, with $A_0,A_1$ fitting parameters. The fit 
 gives $A_1\simeq0.041$. The inset shows $e_1$ obtained by using 
 $\mu=\gamma_2^2/(2L)$ and fixing $\gamma_2=8$. $e_1$ is plotted 
 versus $\ln^2(L)$. The line is a fit to $A_0'+1/\pi^4\ln^2(L)$. 
}
\label{fig:gap_sub}
\end{figure}
%%%%%%%%%%%%%%%%%%%%%%%%%%%%%%%%%%%%%%%%%%%%%%%%%%%%%%% 
%
We now focus on the scaling of the entanglement gap at the quantum critical 
point $g_c\simeq9.67826$. First, instead of $\delta\xi$ we, equivalently, consider 
the scaling of the largest eigenvalue $e_1$ of 
$\mathbb{C}_A$. 
We show our numerical results for $e_1$ in Fig.~\ref{fig:gap_log} 
as a function of $L$ (note the logarithmic scale on 
the $x$-axis). To highlight the different scaling as compared to other regions of the 
phase diagram, we report also data in the paramagnetic phase (square symbols) and in
the ferromagnetic phase (diamonds). Within the ordered phase
$e_1$ increases faster than logarihmically. 
In the paramagnetic region 
$e_1$ exhibits a mild increase for small $L$, saturating at $L\to\infty$. 
This is a consequence of the finite correlation length in the paramagnetic phase. 
A dramatically different behaviour is visible at criticality (circles), 
for which we report data up to $L\sim 40000$.\footnote{Note that since $\omega_y=1$, 
we can use dimensional reduction to attain large system sizes (see section~\ref{sec:dim-red}).}
Interestingly, for moderately large $L$ the behaviour of $\delta\xi$  
is compatible with a logarithmic increase, although 
Eq.~\eqref{eq:e1} suggests a $\ln^2(L)$ scaling. 
This should be attributed to the presence of a sub-leading logarithmic term 
$\ln(L)$ (cf.~\eqref{eq:e1}). 
A fit to $A_2\ln^2(8L/\gamma_2)+A_0+A_1\ln(8L/\gamma_2)$ 
(dashed-dotted line) gives $A_2\simeq 0.01$, which is 
in good agreement with the prediction $1/\pi^4$. One also obtains  
$A_1\simeq 0.04$ and $A_0\simeq 0.16$. Note that $A_0\simeq 1/6$, as 
predicted by~\eqref{eq:e1}. 

To further corroborate our results, in Fig.~\ref{fig:gap_sub} we 
show $e_1-1/\pi^4\ln(8L/\gamma_2)$ versus $L$ using a logarithmic scale on the 
$x$-axis. The data are the same as in Fig.~\ref{fig:gap_log}. 
The continuous line is a fit to 
\begin{equation}
\label{eq:fit}
e_1-\frac{1}{\pi^4}\ln\Big(\frac{8L}{\gamma_2}\Big)=A_0+A_1\ln\Big(\frac{8L}{\gamma_2}\Big)
\end{equation}
with $A_0$ and $A_1$ fitting constants. The logarithmic behaviour is perfect. Note that 
this logarithmic term is not predicted by~\eqref{eq:e1}. Its origin could 
be attributed to the fact that the corner transfer matrix result is obtained 
for the semi-infinite system, i.e., the biparititon with one boundary. 
It is interesting to investigate the dependence on $\gamma_2$ 
of the constant $A_1$ in~\eqref{eq:fit}. In the inset in Fig.~\ref{fig:gap_sub} we show $e_1$ 
obtained by fixing  $\mu=\gamma_2^2/(2L^2)$ with $\gamma_2=8$ 
in~\eqref{eq:snsm} and~\eqref{eq:pnpm}. In the inset $e_1$ is 
plotted versus $\ln^2(L)$. The dashed-dotted line is a fit to 
$1/\pi^4\ln^2(L)+A_0'$. The perfect linear behaviour suggests that 
the subleading logarithmic term is absent or 
its prefactor is small. A fit to $1/\pi^4\ln^2(8L/\gamma_2)+A_0'+A_1'\ln(L)$ 
gives $A_1'\approx 0.0007$. It would be interesting to investigate 
this behaviour more systematically. One possible scenario is that 
the prefactor of the logarithmic term is of the form 
$A_1'=\ln(\gamma_2/8)$.

%################################################################
\subsubsection{Finite-size scaling analysis}
\label{sec:fs-gap}
%
%%%%%%%%%%%%%%%%%%%%%%%%%%%%%%%%%%%%%%%%%%%%%%%%%%%%%%% 
\begin{figure}[t]
\includegraphics[width=.42\textwidth]{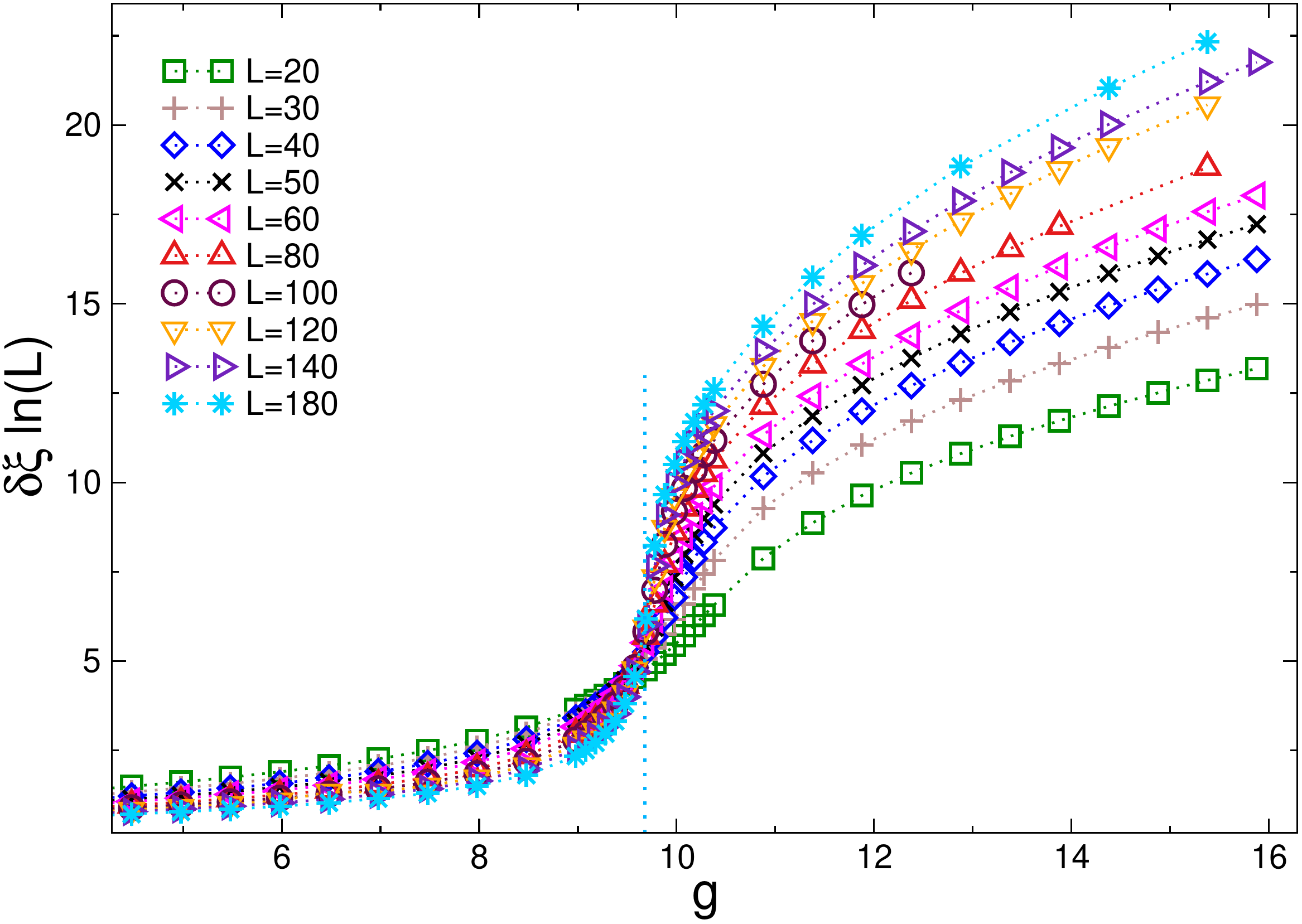}
\caption{Finite-size scaling of the rescaled entanglement gap 
 $\delta\xi \ln(L)$ plotted as function of $g$. 
 Here $L$ is the system size. The vertical line marks the 
 critical point. 
}
\label{fig:gap-1}
\end{figure}
%%%%%%%%%%%%%%%%%%%%%%%%%%%%%%%%%%%%%%%%%%%%%%%%%%%%%%% 
%
\noindent
Having established the logarithmic vanishing of $\delta\xi$ 
at the critical point, it is natural to investigate its behaviour in 
the vicinity of the quantum phase transition. A natural idea is that 
$\delta\xi$ obeys standard finite-size scaling~\cite{vicari}  
\begin{equation}
\label{eq:fs-an}
\delta\xi\ln(L)=f((g-g_c)L^{1/\nu})+\dots, 
\end{equation}
where the dots stand for scaling corrections, $f(x)$ is a scaling 
function and $\nu$ is the exponent that governs the 
divergence of the correlation length  at the 
critical point. For the QSM one has $\nu=1$~\cite{Vojta96} . 
The scaling function $f(x)$ is determined by the universality 
class of the QSM, and, in principle, 
can be calculated. 
Under the assumption that the $f(x)$ is analytic, one can 
expand~\eqref{eq:fs-an} near $g_c$ to obtain 
\begin{equation}
	\label{eq:cross}
\delta\xi\ln(L)=f(0)+(g-g_c)L^{1/\nu}+\dots,  
\end{equation}
From the analysis in section~\ref{sec:dim-red} one should expect 
$f(0)=\pi^2\simeq 9.8$. Eq.~\eqref{eq:cross} implies that the data 
for $\delta\xi$ for different system sizes should exhibit a crossing  
at $g_c$. This crossing method for the entanglement gap has been used to detect 
a quantum phase transition in a system of coupled one-dimensional 
models~\cite{james-2013}. However, since $\delta\xi$ has logarithmic corrections, 
one should expect strong limitations, 
as we are going to show. The scaling Ansatz~\eqref{eq:fs-an} implies that 
by plotting the rescaled gap $\delta\xi\ln(L)$ as a function of the 
scaling variable $(g-g_c)L^{1/\nu}$ one should observe a data 
collapse for different system sizes, provided that scaling corrections 
can be neglected. 

Our finite-size data for $\delta\xi$ as a function of $g$ 
for several system sizes $L$ are shown in Fig.~\ref{fig:gap-1} focussing 
on the vicinity $g\approx g_c$. We only show data for moderately large 
system sizes $L\lesssim 200$. Clearly, the data exhibit a crossing 
at $g\approx 9.6$, which is close to the critical point $g_c\simeq 9.67826$. 
This is quite remarkable because logarithmic corrections are 
present. In fact, we observe that even including larger system sizes, 
it is challenging to obtain a more precise estimate of $g_c$. 
%
%%%%%%%%%%%%%%%%%%%%%%%%%%%%%%%%%%%%%%%%%%%%%%%%%%%%%%% 
\begin{figure}[t]
\includegraphics[width=.42\textwidth]{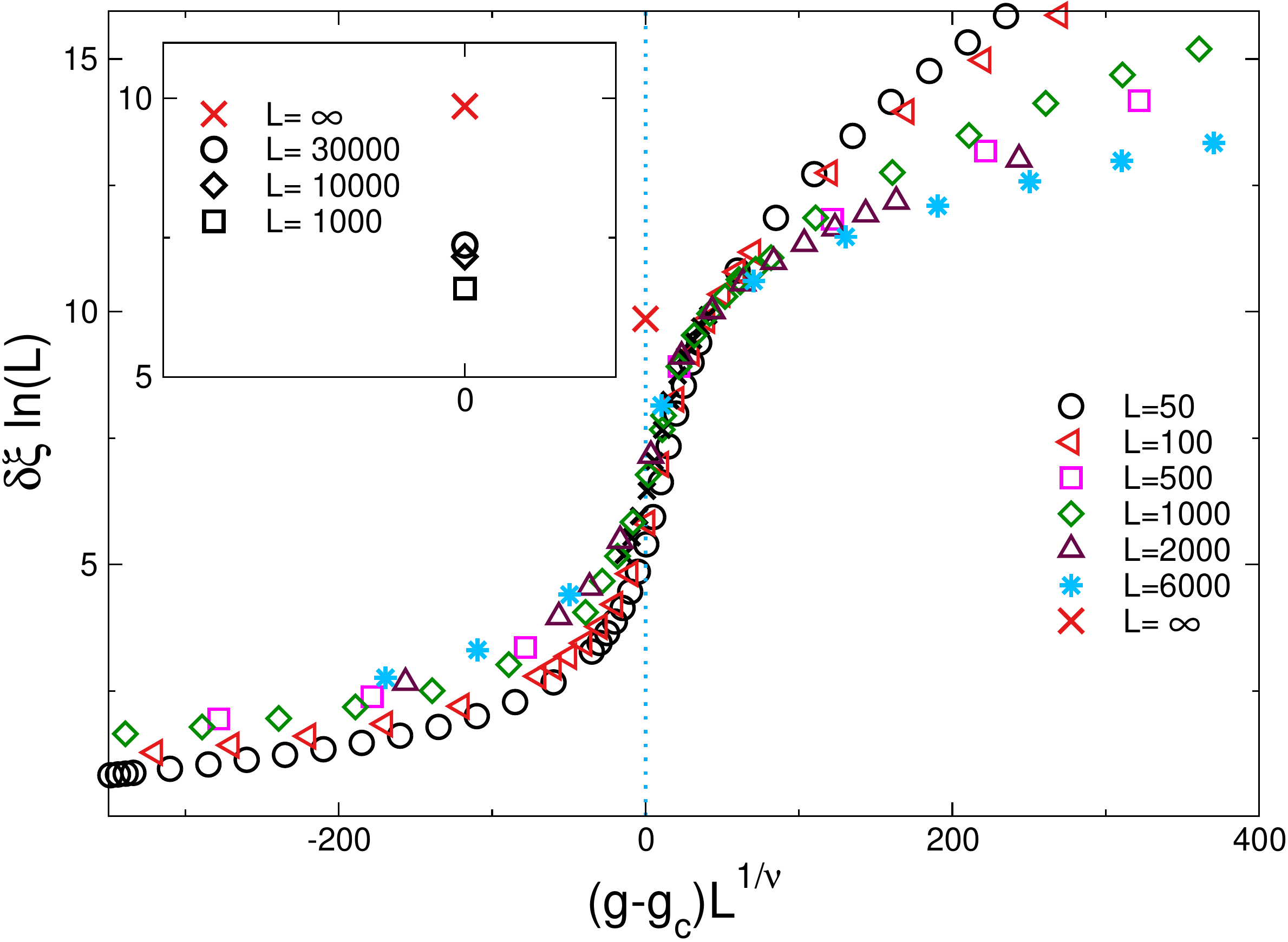}
\caption{Scaling behaviour of the rescaled entanglement gap. 
 $\delta\xi\ln(L)$ plotted against $(g-g_c)L^{1/\nu}$. 
 Here $g_c\simeq9.67826$,  and $\nu=1$  is the correlation 
 length critical exponent. 
}
\label{fig:gap-2}
\end{figure}
%%%%%%%%%%%%%%%%%%%%%%%%%%%%%%%%%%%%%%%%%%%%%%%%%%%%%%% 
%
In Fig.~\ref{fig:gap-2} we perform a data collapse analysis plotting 
the rescaled entanglement gap $\delta\xi\ln(L)$ versus the scaling variable 
$(g-g_c)L^{1/\nu}$. Since we expect that the scaling behaviour is  determined 
by the QSM universality class, we fix $\nu=1$. 
Due to the logarithmic scaling corrections, the data collapse 
is poor. 
From section~\ref{sec:dim-red} one should expect $f(0)=\pi^2$. On the other 
hand, the data up to $L\lesssim 10^4$ suggest $f(0)\approx 7$, which is 
quite far from the expected value $f(0)\simeq 9.8$. As it is shown in the inset, 
a very slow drift towards the asymptotic value is visible, compatible with 
the presence of logarithmic corrections. 
In conclusion, our analysis suggests that the 
scaling of the entanglement gap  can be used to estimate the position 
of the quantum critical point, although extracting the critical 
exponent $\nu$ and the scaling function requires 
knowledge of the logarithmic corrections.

%################################################################
\section{Entanglement gap and the zero-mode eigenvector}
\label{sec:approx}

\noindent
In this section we discuss how the vanishing of the entanglement gap is 
reflected in the eigenstate of the correlation matrix that corresponds to 
the zero mode. Moreover, we show that assuming a flat structure of the 
zero-mode eigenvector at criticality allows one to capture 
qualitatively the logarithmic vanishing of the entanglement gap. 
Within this approximation the vanishing of $\delta\xi$ 
is related to some interesting multiplicative logarithmic corrections 
in the correlators. Finally, the result suggests that the presence of 
corners in the bipartition affects the vanishing of the gap. 

%################################################################
\subsection{The zero-mode eigenvector}
\label{sec:approx-1}

%%%%%%%%%%%%%%%%%%%%%%%%%%%%%%%%%%%%%%%%%%%%%%%%%%%%%%% 
\begin{figure}[t]
\includegraphics[width=.42\textwidth]{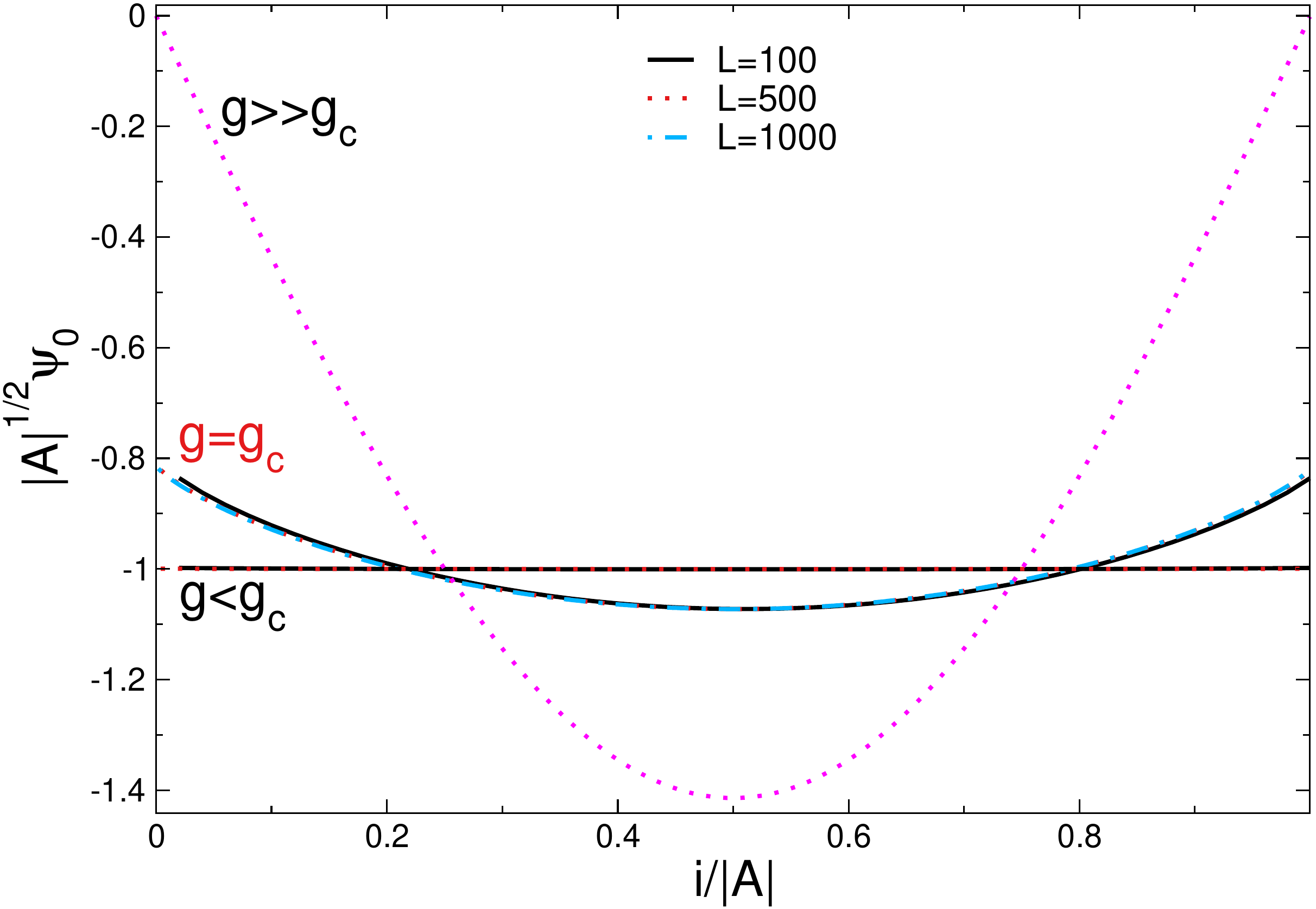}
\caption{Eigenvector corresponding to the largest eigenvalues 
 of the correlation matrix $\mathbb{S}_{\vec{n}\vec{m}}$ (zero-mode 
 eigenvector). Data are for the straight bipartition with $\omega_y=1$ 
 and $\omega_x=1/2$. 
 Eigenvector's components are rescaled by 
 $|A|^{1/2}$. On the $x$-axis $i$ is a label.  
 In the ordered phase for $g<g_c$ 
 the eigenvector becomes flat in the thermodynamic limit, 
 in contrast with the behaviour at the critical point 
 at $g_c$, and in the paramagnetic phase. 
}
\label{fig:eig}
\end{figure}
%%%%%%%%%%%%%%%%%%%%%%%%%%%%%%%%%%%%%%%%%%%%%%%%%%%%%%% 

\noindent
Let us consider the eigenvector $|\psi_0\rangle$ corresponding to the 
largest eigenvalue of the spin-spin correlator $\mathbb{S}_{A}$. 
This eigenvector is closely related 
to that of $\mathbb{C}_A$ corresponding to $e_1$, which gives 
the smallest single-particle ES level. 
Its behaviour is summarised in Fig.~\ref{fig:eig}, showing the
components of the eigenvector for different system sizes and in different regions 
of the phase diagram. 
We consider the bipartition with straight boundary $\omega_y=1$ and $\omega_x=1/2$ 
(see Fig.~\ref{fig:partition1} a). 
Upon increasing $L$ all the components decay to zero. Thus, it is convenient to 
rescale by $|A|^{1/2}=\sqrt{\ell_x\ell_y}$ 
(see Fig.~\ref{fig:partition1}). 
We define the flat vector $|\vec{1}\rangle$ in region $A$ as 
\begin{equation}
\label{eq:flat}
|\vec{1}\rangle=\frac{1}{\sqrt{|A|}}(1,1,\dots,1)^T.
\end{equation}
It is clear from Fig.~\ref{fig:eig} in the thermodynamic limit in the ordered phase one has that 
$|\psi_0\rangle\to|\vec{1}\rangle$, up to an irrelevant global phase. 

The structure of $|\psi_0\rangle$ for $g>g_c$ can be understood as follows. 
Deep in the paramagnetic phase the correlation length is small. 
In the limit $g\to\infty$ spin-spin correlators become ultra-local, 
viz.
\begin{equation}
	\label{eq:S-app}
	\mathbb{S}_{\vec{n}\vec{m}}=\delta_{\vec{n}\vec{m}}+\varepsilon
	(\delta_{|n_x-m_x|,1}+\delta_{|n_y-m_y|,1}), 
\end{equation}
with $\varepsilon$ vanishing for $g\to\infty$. 
In the case $\omega_y=1$, it is straightforward to determine the eigenvector 
of~\eqref{eq:S-app} corresponding to the largest eigenvalue 
in the sector with $k_y=0$. 
Due to $\omega_y=1$, the eigenvector is ``flat'' along $\hat y$, and has a 
non-trivial dependence only on the $x$ coordinate. The  components of the eigenvector 
are given as 
\begin{equation}
\label{eq:eig-lg}
\psi_{n_x,n_y}=\frac{1}{|A|^{1/2}}\sin\Big(\frac{\pi n_x}{\ell_x}\Big). 
\end{equation}
The dotted line in Fig.~\ref{fig:eig} shows the eigenvector $|\psi_0\rangle$ for 
$g=10$ and the data are in perfect agreement with~\eqref{eq:eig-lg}. 

Upon approaching the quantum critical point, the zero-mode eigenvector 
flattens, reflecting that the system develops 
ferromagnetic order. 
To understand that, let us consider the spin correlator~\eqref{eq:snsm} 
in the thermodynamic limit. Upon increasing $L$, as $\mu\to0$, 
the correlator develops a singularity for $\vec{k}=0$ which encodes the 
critical behaviour of the QSM. In the limit of large $L$ 
one can isolate the contribution of the 
zero mode as~\cite{Botero04} 
\begin{equation}
	\label{eq:flat-d}
	\mathbb{S}_{\vec{n}\vec{m}}=\mathbb{S}_{\vec{n}\vec{m}}^{(th)}+
	\frac{c}{\sqrt{\mu}}+\dots,
\end{equation}
where $c$ is a constant. 
Here the first term is obtained by setting $\mu=0$ and by replacing the sum 
in~\eqref{eq:snsm} with an integral and the second term is the contribution of the 
zero mode $\vec{k}=0$. 
The second contribution in~\eqref{eq:flat-d} does not 
depend on $\vec{n}$ and $\vec{m}$, and is divergent in the limit 
$\mu\to0$. In this limit one has that the flat vector becomes an exact eigenvector 
of $\mathbb{S}_{\vec{n}\vec{m}}$ with an eigenvalue that is proportional to $L$. 
However, the decomposition in~\eqref{eq:flat-d} is not justified because the limit $\mu\to0$ and 
the limit $L\to\infty$ cannot be taken independently, because $\mu\propto 1/L^2$. 
Figure~\ref{fig:eig} shows that at the critical point the rescaled 
components of $|\psi_0\rangle$ collapse on the same curve. The 
structure of the eigenvector is not flat. 
On the other hand, in the ordered phase, where $\mu\propto 1/L^4$ 
(see Fig.~\ref{fig:mu}) upon increasing $L$ the eigenvector becomes flat. 
This suggests that the decomposition~\eqref{eq:flat-d} 
holds if $\mu$ decays sufficiently fast for large $L$. 

%################################################################
\subsection{An interesting logarithmic correction}
\label{sec:approx-2}

\noindent
In this section we investigate the scaling of the entanglement gap 
assuming that the eigenvector $|\psi_0\rangle$ is flat also at 
the critical point, and that the decomposition in Eq.~\eqref{eq:S-app} holds. 
A similar analysis for the massive harmonic chain was presented in Ref.~\cite{Botero04}. 
%
%%%%%%%%%%%%%%%%%%%%%%%%%%%%%%%%%%%%%%%%%%%%%%%%%%%%%%% 
\begin{figure}[t]
\includegraphics[width=.42\textwidth]{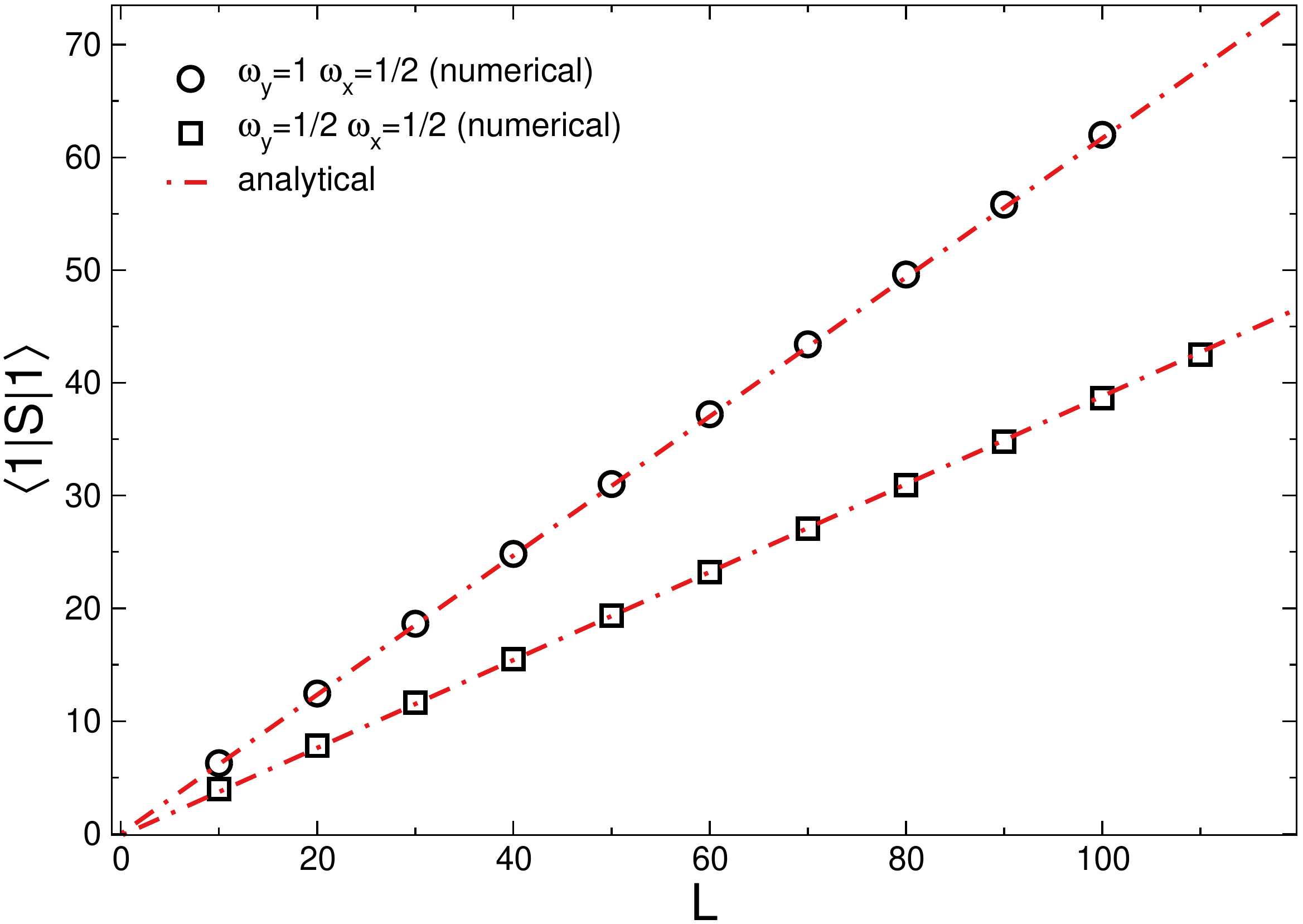}
\caption{Expectation value $\langle1|\mathbb{S}|1\rangle$ of the 
 correlation matrix $\mathbb{S}$ (cf.~\eqref{eq:snsm}) over the 
 flat vector $|1\rangle$. Symbols are numerically exact 
 data for the bipartition with several values of 
 $\omega_x$ and $\omega_y$ (see Fig.~\ref{fig:partition1}). 
 The dashed dotted line is the analytic result $s_0 L$. 
 Note that $s_0$ is obtained by summing~\eqref{eq:s-th} 
 and~\eqref{eq:sl}. 
}
\label{fig:s0}
\end{figure}
%%%%%%%%%%%%%%%%%%%%%%%%%%%%%%%%%%%%%%%%%%%%%%%%%%%%%%% 
% 
Here we assume that $\mathbb{S}_{\vec{n}\vec{m}}$ 
can be decomposed as   
\begin{equation}
	\label{eq:wrong-d}
	{\mathbb{S}}_{\vec{n}\vec{m}}=s_0 L |\vec{1}\rangle\langle\vec{1}|+\mathbb{S}'_{\vec{n}\vec{m}}
\end{equation}
and we assume that $\mathbb{S}'_{\vec{n}\vec{m}}$ is negligible. 
%
%
%%%%%%%%%%%%%%%%%%%%%%%%%%%%%%%%%%%%%%%%%%%%%%%%%%%%%%% 
\begin{figure}[t]
\includegraphics[width=.42\textwidth]{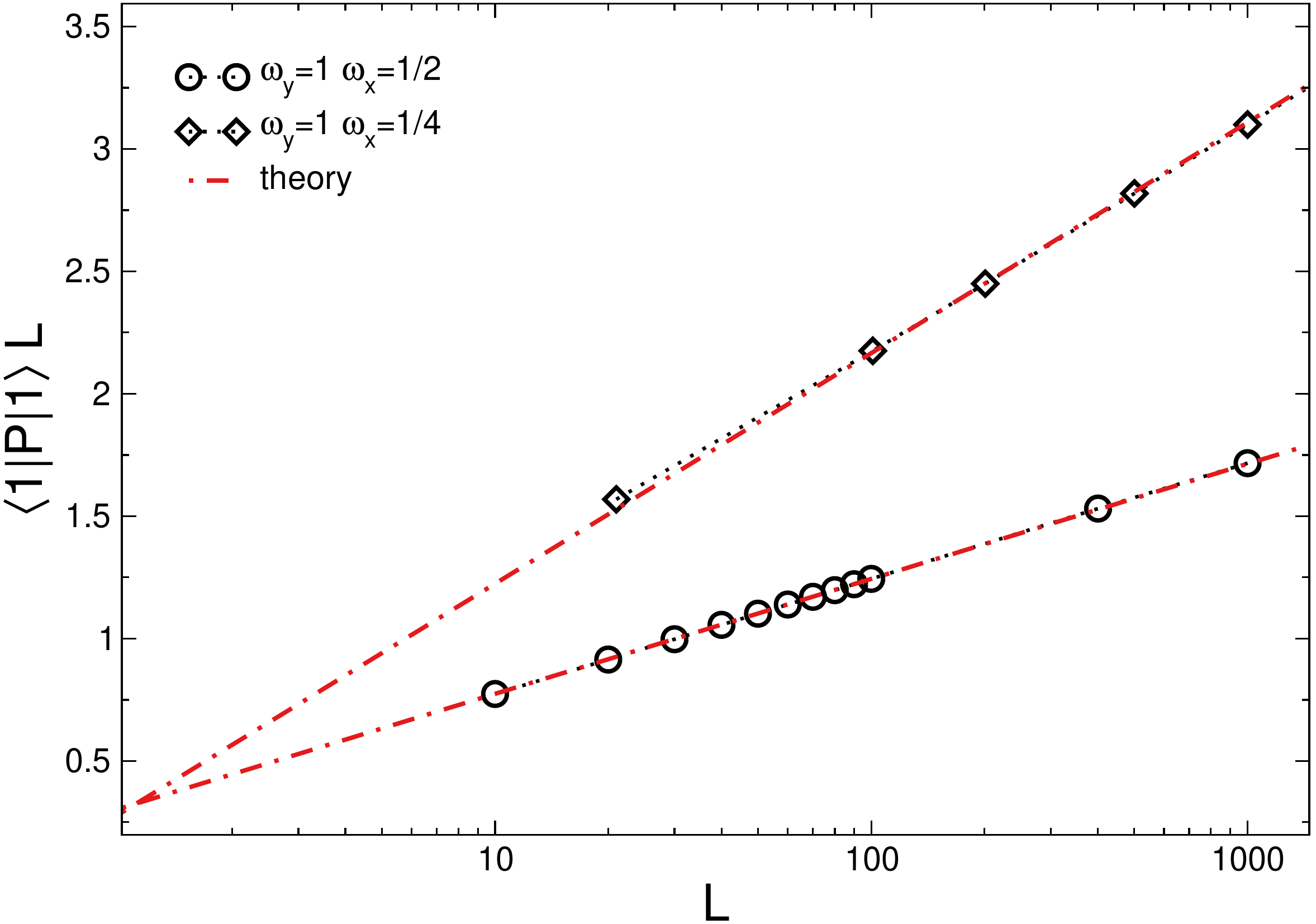}
\caption{ Rescaled expectation value $\langle\vec{1}|\mathbb{P}|\vec{1}\rangle L$ 
 over the flat vector $|\vec{1}\rangle$. The data are for the straight 
 bipartition with $\omega_y=1$ and $\omega_x=1/2,1/4$. Note the logarithmic 
 scale on the $x$ axis. The dashed-dotted line is the analytical result. 
}
\label{fig:p0}
\end{figure}
%%%%%%%%%%%%%%%%%%%%%%%%%%%%%%%%%%%%%%%%%%%%%%%%%%%%%%% 
%
The product $\mathbb{P}\cdot\mathbb{S}$ is thus decomposed as 
\begin{equation}
\label{eq:wrong-d1}
	\mathbb{P}\cdot\mathbb{S}=s_0L\mathbb{P}|\vec{1}\rangle\langle\vec{1}|
	+\mathbb{P}\cdot\mathbb{S}', 
\end{equation}
where we suppress the indices $\vec{n},\vec{m}$ to lighten the notation. 
Consistently with~\eqref{eq:wrong-d}, we are going to neglect the second term in~\eqref{eq:wrong-d1}. 
The matrix $\mathbb{P}\cdot\mathbb{S}$ is not hermitian, whereas 
$\mathbb{P}$ and $\mathbb{S}$ are hermitian. This means that one has to introduce 
right and left eigenvectors. We define two vectors $u_R$ and $u_L$ 
as 
\begin{align}
	& u_R=\mathbb{P}|\vec{1}\rangle\\
	& u_L=|\vec{1}\rangle. 
\end{align}
It is now straightforward to check that 
$u_R$ and $u_L$ are the right and left eigenvectors of 
$\mathbb{P}\cdot\mathbb{S}$, respectively. 
The eigenvalue is given as 
\begin{equation}
\label{eq:e1-exp}
	e_1=\langle\vec{1}|\mathbb{S}|\vec{1}\rangle\langle\vec{1}|\mathbb{P}|\vec{1}\rangle. 
\end{equation}
Eq.~\eqref{eq:e1-exp} implies that the problem of calculating the eigenvalue 
$e_1$ of $\mathbb{C}_A$ (cf.~\eqref{eq:ca}) is reduced to the simpler problem 
of calculating the flat-vector expectation values 
in~\eqref{eq:e1-exp}. 
In the following we are going to calculate 
\begin{align}
\label{eq:ss}
& \langle\vec{1}|\mathbb{S}|\vec{1}\rangle=\frac{1}{|A|}\sum\limits_{\vec{n},\vec{m}\in A}\mathbb{S}_{\vec{n}\vec{m}},\\
\label{eq:pp-1}
& \langle\vec{1}|\mathbb{P}|\vec{1}\rangle=\frac{1}{|A|}\sum\limits_{\vec{n},\vec{m}\in A}\mathbb{P}_{\vec{n}\vec{m}}. 
\end{align}
Note that~\eqref{eq:ss} has the same form as the spin susceptibility. 
To obtain~\eqref{eq:ss} and~\eqref{eq:pp-1}, we 
use the expansion of the spin and momentum correlators discussed in 
section~\ref{sec:corr-S} and section~\ref{sec:corr-p}. 
Importantly, both the thermodynamic and the finite-size contributions 
in~\eqref{eq:s-decomp} and~\eqref{eq:p-decomp} have to be taken into account. 

We start discussing the expectation value $\langle\vec{1}|\mathbb{S}|
\vec{1}\rangle$ and first consider the contribution of the 
thermodynamic part of the correlator in~\eqref{eq:snm-th}. 
From~\eqref{eq:snm-th} we can perform the sums over $\vec{n}$ and $\vec{m}$, 
and after using the explicit form of the spherical parameter~\eqref{eq:mu-red},
taking the limit $L\to\infty$, we obtain  
for a bipartition with generic $\omega_x$ and $\omega_y$ 
\begin{align}
\begin{split}
\label{eq:s-th}
\langle\vec{1}|\mathbb{S}^{(th)}|\vec{1}\rangle&=\\
&\hspace{-.75cm}\frac{2\sqrt{g_c}L}{\pi^2\omega_x\omega_y}\iint_{-\infty}^\infty dk_x dk_y
\frac{\sin^2(\frac{k_x}{2}\omega_x)\sin^2(\frac{k_y}{2}\omega_y)}
{k_x^2 k_y^2(\gamma_2^2+k_x^2+k_y^2)^\frac{1}{2}}.
\end{split}
\end{align}
Note that this expectation value grows linearly with $L$. 
The constant $\gamma_2$ is defined in~\eqref{eq:mu-gc}. 
The integral in~\eqref{eq:s-th} depends on the 
universal low-energy behaviour of the QSM, i.e., at $k_x,k_y\to0$, 
although it is not fully universal.  
We now show that the finite-size term~\eqref{eq:s1} yields a linear 
contribution in $L$ in~\eqref{eq:ss}. Indeed, it is straightforward 
to take the limit $L\to\infty$ in~\eqref{eq:s1} to obtain  
\begin{align}
 \begin{split}
  \label{eq:sl}
\langle\vec{1}|\mathbb{S}^{(L)}|\vec{1}\rangle&=
\frac{\sqrt{g_c}L}{4\pi\omega_x\omega_y}\,\,\sideset{}{'}\sum_{l,l'=-\infty}^\infty
\iint_{0}^{\omega_x}dxdx'\\
&\hspace{-.25cm}\iint_{0}^{\omega_y}dydy' 
\frac{e^{-\gamma_2\sqrt{(l+x-x')^2+(l'+y-y')^2}}}{\sqrt{(l+x-x')^2+(l'+y-y')^2}}. 
 \end{split}
\end{align}
Note that the integral in~\eqref{eq:sl} is finite, 
although the denominator in~\eqref{eq:sl} is singular for 
$l=0$ and $l'=\pm1$ (see section~\ref{sec:corr-S}). 
It is straightforward to integrate the contributions~\eqref{eq:s-th} 
and~\eqref{eq:sl} numerically. 
We conclude that the expectation value~\eqref{eq:ss} grows  linearly  
with $L$ in the limit $L\to\infty$. 
The accuracy of~\eqref{eq:s-th} and~\eqref{eq:sl} is 
numerically verified in Fig.~\ref{fig:s0}. 
The symbols are exact numerical data for~\eqref{eq:ss}, whereas the 
dashed-dotted lines are the theoretical predictions obtained by 
summing~\eqref{eq:s-th} and~\eqref{eq:sl}. 

We now show that, surprisingly, the expectation value~\eqref{eq:pp-1} 
decays as $\ln(L)/L$, i.e., it exhibits  a multiplicative logarithmic correction. 
The derivation is quite cumbersome, although it requires standard methods 
such as Poisson's summation formula and the Euler-Maclaurin formula. 
The details are reported in Appendix~\ref{app:app1}. 
Here we solely discuss the final result. Similar to~\eqref{eq:ss} one can treat 
separately the thermodynamic contribution 
of~\eqref{eq:pp-1} (cf.~\eqref{eq:p-th}) and the finite-size one 
(cf.~\eqref{eq:q2}). 
For simplicity we consider the bipartition with $\omega_x=1/p$ and 
$\omega_y=1/q$, with $p,q\in\mathbb{N}$. Clearly, for 
$\omega_y<1$ the boundary between the two subsystems is not 
straight, i.e., it has a corner (see Fig.~\ref{fig:partition1} b). 
One obtains  
\begin{multline}
\label{eq:p-th-1}
\langle\vec{1}|\mathbb{P}^{(th)}|\vec{1}\rangle=
\sum_{p'=0}^{p-1}\sum_{q'=0}^{q-1}\int_0^{1/p}d k_x\int_0^{1/q}d k_y\\
\sin^2(\pi(k_x+p'/p))\sin^2(\pi(k_y+q'/q))\eta_{p',q'}(k_x,k_y). 
\end{multline}
The function $\eta_{p',q'}(k_x,k_y)$ reads as 
\begin{multline}
\label{eq:eta-f}
	\eta_{p',q'}(k_x,k_y)=
	\frac{4}{\pi^3\sqrt{g_c}}\Big[
\frac{q}{(k_x+p'/p)^2}
	+\frac{p}{(k_y+q'/q)^2}\\
+p\psi'(1+k_y+q'/q)
+\frac{q}{1+k_x+p'/p}
+\frac{q}{2(1+k_x+p'/p)^2}\\
\frac{q}{6(1+k_x+p'/p)^3}
+\dots\Big]\frac{\ln(L)}{L}. 
\end{multline}
The dots in the square brackets denote terms of higher 
powers of $1/(k_x+p'/p)$ that can be derived systematically 
by using the Euler-Maclaurin formula (see Appendix~\ref{app:app1}). 
The function $\psi'(x)$ is the first derivative of the digamma 
function $\psi(x)$ with respect to its argument~\cite{Abra65}. 
As anticipated above, the behaviour as $\ln(L)/L$ is clearly visible 
in~\eqref{eq:eta-f}.  
As for~\eqref{eq:s-th} and~\eqref{eq:sl}, it is clear that  
$\eta_{p',q'}$ is determined by the low-energy part of the 
dispersion of the QSM. 

Let us now consider the finite-size contribution~\eqref{eq:q2}. Interestingly, as it is clear 
from~\eqref{eq:q2}, the finite-size correlator is smooth for $\omega_y<1$ 
and $\omega_x<1$, whereas it exhibits a singularity if either $\omega_y=1$ or 
$\omega_x=1$, i.e., if the boundary between $A$ and its complement is straight. 
Similar to~\eqref{eq:p-th-1}, the singular contribution is
\begin{equation}
	\label{eq:p-fs}
	\langle\vec{1}|\mathbb{P}^{(L)}|\vec{1}\rangle=-\frac{1}{\sqrt{g_c}\pi}
	\frac{\ln(L)}{L}. 
\end{equation}
Interestingly, the minus sign in~\eqref{eq:p-fs} suggests that the presence of corners 
increases the prefactor of the logarithmic correction. 
Finally, after combining Eqs.~\eqref{eq:s-th},~\eqref{eq:sl} and~\eqref{eq:p-th-1},~\eqref{eq:p-fs} 
with~\eqref{eq:e1-exp}, one obtains that $e_1\propto\ln(L)$. 
The prefactor of the logarithmic growth depends on the low-energy properties of 
the QSM. As anticipated, by approximating the zero-mode eigenvector with the 
flat vector one obtains that $\delta\xi$ decays logarithmically upon increasing $L$. However, 
from~\eqref{eq:rdm-fb} one 
obtains that $\delta\xi\propto 1/\sqrt{\ln(L)}$, instead of the correct 
behaviour as $1/\ln(L)$ established in section~\ref{sec:dim-red}. 

\section{Conclusions}
\label{sec:concl}

\noindent
We investigated the entanglement gap $\delta\xi$ in the two-dimensional 
critical QSM. Our main result is that in the QSM there is a 
relationship between critical behaviour and 
vanishing of the entanglement gap. 

There are several intriguing directions for future research. 
First, it would be interesting to study the behaviour of the 
entanglement gap below the transition, i.e., in the ordered phase. 
Furthermore, an interesting question is how the scenario outlined in this work 
survives beyond the large $N$ limit. This, however, is a very demanding task 
because entanglement-related observables cannot be calculated 
efficiently at finite $N$. Still, the flat-vector approximation 
discussed in section~\ref{sec:approx} could be generalized, at 
least perturbatively in $1/N$.  
It would be interesting to check whether the 
logarithmic correction that is responsible of the vanishing of 
the entanglement gap persists at finite $N$. 
Another natural direction is to understand if the vanishing of the entanglement gap at the 
critical point is an artifact of the large $N$ limit. The question 
is whether at finite $N$ a spurious transition appears, as observed 
in Ref.~\cite{Chan14}. 

It would be also interesting to study the negativity 
spectrum~\cite{ruggiero-2016,mbeng-2017,shapourian-2019,xhek-2020_A} at the 
quantum phase transition, and in particular the effect of the 
zero mode. A very interesting direction is to understand how the 
fluctuations of the number of particles between the two subsystems is 
reflected in the entanglement spectrum and the entanglement gap. 
Very recently, the symmetry resolved entanglement entropies emerged as 
ideal tools to do that~\cite{alba-2011,lauchli-2013,Laflorencie2014,Xavier2018,Murciano2019,Goldstein2018,Goldstein2018B,Feld2019,Calabrese2020,Bonsignori2019,Fraenkel2020,crc-20,sara2D,clss-19,cms-13,d-16,d-16-1,
matsuura,SREE,xhek-2020_B}. However, an important remark is that in 
the QSM the number of bosons is not conserved, and the symmetry-resolved 
entanglement entropies are not well defined. The particle number conservation is 
only enforced on average via the gap equation~\eqref{eq:gap-def}. Still, 
it should be possible to generalize the QSM to investigate this issue, e.g. by
studying spin-anisotropy in the QSM~\cite{Wald15}. 
It would be also important to understand how our results can be 
generalized to long-range spherical models. 
Finally, it would be interesting to consider higher-dimensional 
fermionic models. An interesting question is whether the area-law violation~\cite{Wolf-2004,Gioev-2006,Farkas-2007,Li-2006,Swingle-2010,Calabrese-2012,Ding-2012}  
affects the scaling of the entanglement gap. 

\begin{acknowledgments}
\noindent
V.A. would like to thank Paola Ruggiero for drawing to our attention 
Ref.~\onlinecite{sara2D} and for discussions. 
We also thank Pasquale Calabrese for useful comments on the manuscript. 
V.A. acknowledges support from 
the European Research Council under ERC Advanced grant 743032 DYNAMINT. 
\end{acknowledgments}

\appendix

%################################################################
\section{Critical behaviour of the spin correlator}
\label{app:app0}

\noindent
In this appendix we derive the large $L$ behaviour of the correlation function 
$\mathbb{S}_{\vec{n}\vec{m}}$ in the QSM. 
Specifically, we provide exact expressions for the leading and the first 
subleading terms in powers of $1/L$. 
The correlator to evaluate is defined as (cf. Eq.~\eqref{eq:snsm}) 
\begin{equation}
\label{eq:s-def-app}
\mathbb{S}_{\vec{n}\vec{m}}=
\frac{\sqrt{g}}{2\sqrt{2}V}\sum_{\vec{k}}
\frac{e^{i\vec{k}(\vec{n}-\vec{m})}}{\sqrt{\mu+\omega_{\vec{k}}}}. 
\end{equation}
The correlation depends only on the distance $\vec{d}=\vec{n}-\vec{m}$, 
reflecting translation invariance. 
Eq.~\eqref{eq:s-def-app} can be rewritten as 
\begin{equation}
\label{eq:p-app}
	\mathbb{S}_{\vec{n}\vec{m}}=\sqrt{\frac{g}{2\pi}}
	\int_0^\infty dt e^{-(\mu+2) t^2}
	\prod_{j=x,y}\frac{1}{L}\sum_{k_j} e^{-\cos(k_j)t^2+i k_j d_j}
\end{equation}
We now apply Poisson's summation formula which,
for a periodic function $G(q)=G(q+2\pi)$, is stated as 
\begin{equation}
\label{eq:poisson}
\frac{1}{L}\sum_{n=-(L-1)/2}^{(L-1)/2}G\Big(\frac{2\pi n}{L}\Big)=\sum_{l=-\infty}^\infty
\int_{-\pi}^\pi\frac{dq}{2\pi}G(q)e^{i q l L}. 
\end{equation}
The application of~\eqref{eq:poisson} to~\eqref{eq:p-app} yields  
\begin{equation}
\label{eq:app-2}
\mathbb{S}_{\vec{n}\vec{m}}
=\frac{\sqrt{g}}{\sqrt{2\pi}}
\int_0^\infty dt e^{-(\mu+2) t^2}
\prod_{j=x,y}\Big(\sum_{l_j=-\infty}^\infty I_{l_j L+d_j}(t^2)\Big).
\end{equation}
Here $I_n(t)$ is the modified Bessel function of the first kind~\cite{Abra65}. 
It is convenient to isolate the terms with $l_x=l_y=0$ in~\eqref{eq:app-2}, viz. 
\begin{multline}
\label{eq:split}
\prod_{j=x,y}\sum_{l_j=-\infty}^\infty I_{l_j L+d_j}(t^2)
=\\
\prod_{j=x,y}\Big(I_{d_j}(t^2)+\sideset{}{'}
\sum_{l_j=-\infty}^\infty I_{l_j L+d_j}(t^2)\Big).
\end{multline}
The first term on the right-hand side gives the thermodynamic contribution to the 
correlator $\mathbb{S}_{\vec{n}\vec{m}}$, i.e., in the limit $L\to\infty$, 
whereas the other terms are finite-size corrections. The prime in the sum 
is to stress that the terms with $l_x=l_y=0$ is removed. 
We now derive the large $L$ behaviour of~\eqref{eq:split}. 
Upon expanding~\eqref{eq:split}, it is clear that we have 
to derive the asymptotic behaviour of integrals
of the type 
\begin{equation}
	\label{eq:eq-a-2}
	K_{l,l'}(x,x')=\frac{\sqrt{g}}{\sqrt{2\pi}}\int_0^\infty e^{-(\mu+2)t^2}
	I_{l L+x}(t^2)I_{l' L+x'}(t^2).
\end{equation}
Without loss of generality we can restrict ourselves to the case with 
$l,l'>0$. The generalization to arbitrary $l,l'$ is straightforward by 
using the symmetry of the Bessel function $I_{-n}=I_{n}$. 
It is convenient to change variables in~\eqref{eq:eq-a-2} to 
$z^2=t^2/(l L+x)$, viz. 
\begin{multline}
	K_{l,l'}(x,x')=\frac{\sqrt{g}\sqrt{L l+x}}{\sqrt{2\pi}}\int_0^\infty dz 
e^{-(\mu+2)(l L+x)z^2}\\
I_{l L+x}(z^2(l L+x))I_{l' L+x'}(r(l' L+x') z^2), 
\end{multline}
where we introduced the ratio $r$ as 
\begin{equation}
r=\frac{lL+x}{l'L+x'}. 
\end{equation}
We can now perform a saddle point analysis for large $L l+x$. 
For large $L$, the integral $K_{l,l'}$ is determined by the 
saddle point
\begin{equation}
\label{eq:t2star}
t^*=
\Big(\frac{(\mu +2)(1+r^2)+2 \sqrt{r^4+(\mu  (\mu +4)+2) r^2+1}}{\mu  (\mu +2) (\mu +4) r^2}\Big)^{\frac{1}{4}}. 
\end{equation}
Finally, a standard calculation yields  
\begin{multline}
\label{eq:P2}
K_{l,l'}=\frac{\sqrt{g}\sqrt{l L+x}}{(2(l' L+x'))^{3/2}\sqrt{r}\pi}\\\times
	\left. e^{-(l'L+x')(r(2+\mu)t^2-r\eta(t^2)-\eta(t^2 r))}\frac{g'(t)}{\sqrt{f(t)}}\right|_{t\to t^*}. 
\end{multline}
Here we defined 
\begin{align}
\label{eq:gp}
& g'(t)=\frac{1}{(t^4+1)^{1/4}(r^2 t^4+1)^{1/4}}\\
\label{eq:ft}
& f(t)=-\frac{r^2 t^4-1}{t^2 \sqrt{r^2 t^4+1}}+(\mu +2) r-\frac{r \left(t^4-1\right)}{t^2
\sqrt{t^4+1}}, 
\end{align}
and the function $\eta(t)$ as 
\begin{equation}
	\eta(t)=(1+t^2)^\frac{1}{2}+\ln\Big(\frac{t}{1+(1+t^2)^\frac{1}{2}}\Big). 
\end{equation}
The main ingredient to derive~\eqref{eq:P2} is the asymptotic behaviour of the 
Bessel function $I_z(z)$ for $z\to\infty$~\cite{Abra65} together with the standard 
saddle point analysis~\cite{Cop65}.

Since we are interested in the critical behaviour of the correlators, 
it is useful to consider the limit $\mu\to0$, because $\mu$ vanishes at 
criticality. Specifically, we consider the limit $L\to\infty$ with $\mu\propto1/L^2$. 
In this limit we obtain the expression 
\begin{equation}
	\label{eq:app-4}
	K_{l,l'}(x,x')=\frac{\sqrt{g_c}}{4\pi}\frac{e^{-\sqrt{2\mu}\sqrt{(l L+x)^2+(l' L+x')^2}}}
	{\sqrt{(l L+x)^2+(l'L+x')^2}}, 
\end{equation}
where we fixed $g=g_c$. 
Finally, we now obtain that in the large $L$ limit  and for 
$\mu\to0$ the correlator $\mathbb{S}_{\vec{n}\vec{m}}$ is given as 
\begin{multline}
\label{eq:app-1-fin}
\mathbb{S}_{\vec{n}\vec{m}}=\frac{\sqrt{g_c}}{\sqrt{2\pi}}\int_0^\infty dt e^{-(\mu+2)t^2}I_{n_x-m_x}(t^2)I_{n_y-m_y}(t^2)\\
+\sideset{}{'}\sum_{l,l'=-\infty}^\infty K_{l,l'}(n_x-m_x,n_y-m_y), 
\end{multline}
where $K_{l,l'}$ is defined in~\eqref{eq:app-4} and the prime in the 
sum is to stress that the term with $l=l'=0$ has been removed. 
In~\eqref{eq:app-1-fin} one can recognize the two contributions 
in~\eqref{eq:snm-th} and~\eqref{eq:s1}. 
Note that 
the finite-size term (second term in~\eqref{eq:app-1-fin}) is ${\mathcal O}(1/L)$, 
whereas the thermodynamic one (first term in~\eqref{eq:app-1-fin}) is 
${\mathcal O}(1)$. In~\eqref{eq:app-1-fin} we neglect higher order corrections in 
powers of $1/L$. The large $L$ expansion for the momentum correlator 
$\mathbb{P}_{\vec{n}\vec{m}}$ can be obtained from~\eqref{eq:app-1-fin} by 
using~\eqref{eq:id}. 

%################################################################
\section{Derivation of the flat-vector expectation value  
	$\langle\vec{1}|\mathbb{P}|\vec{1}\rangle$}
\label{app:app1}

\noindent
In this appendix we derive the large $L$ behaviour of the 
expectation value of the momentum correlator with the flat vector $\langle\vec{1}|\mathbb{P}
|\vec{1}\rangle$. We consider the leading, i.e, the thermodynamic, 
as well as the first subleading contribution. The main goal 
is to show that the expectation value exhibits multiplicative logarithmic 
corrections. Two types of contributions are present. One 
originating from the thermodynamic limit of the correlator, 
whereas the second one is due to the first subleading term. The latter 
is present only for a straight boundary between the two subsystems,
and it vanishes if the bipartition has corners.

%################################################################
\subsection{Thermodynamic contribution}
\label{sec:p-th-c}

\noindent
Here derive the thermodynamic contribution, which is given 
as $\langle\vec{1}|\mathbb{P}^{(th)}|\vec{1}\rangle$, cf.~\eqref{eq:p-decomp}. 
Here $|\vec{1}\rangle$ is the flat vector restricted to region $A$, i.e, 
\begin{equation}
	|\vec{1}\rangle=\frac{1}{\sqrt{|A|}}(1,1,\dots,1),\quad |A|=\ell_x\ell_y. 
\end{equation}
The thermodynamic part of the momentum correlator reads 
\begin{equation}
\label{eq:app-p-th}
	\mathbb{P}_{\vec{n}\vec{m}}^{(th)}=
	\frac{1}{4\sqrt{2g}\pi^2}\int_{-\pi}^\pi d\vec{k} e^{i\vec{k}(\vec{n}-\vec{m})}
	\sqrt{\mu+\omega_{\vec{k}}}. 
\end{equation}
After performing the sum over $\vec{n}$ and $\vec{m}$ in~\eqref{eq:app-p-th}, and after 
changing variables to $k_x'=L\omega_x k_x/\pi$  and $k'_y=L\omega_y k_y/\pi$,  
we obtain  
\begin{widetext}
\begin{equation}
\label{eq:app-p3}
\langle \vec{1}|\mathbb{P}^{(th)}|\vec{1}\rangle=
\frac{2\sqrt{2}}{\sqrt{g}L^4\omega_x^2\omega_y^2}
\int_{0}^{L\omega_x/2} dk_x\int_{0}^{L\omega_y/2} dk_y
\frac{\sin^2(\pi k_x)\sin^2(\pi k_y)}{\sin^2\big(\frac{\pi}{L\omega_x}k_x\big)
\sin^2\big(\frac{\pi}{L\omega_y}k_y\big)}
\times\Big[\mu+2-\cos\big(\frac{2\pi}{L\omega_x}k_x\big)-\cos\big(\frac{2\pi}{L\omega_y}k_y\big)\Big]^\frac{1}{2}.
\end{equation}
\end{widetext}
In order to extract the large $L$ behaviour of~\eqref{eq:app-p3} it is useful to 
split the integration domains $[0,L\omega_x/2]$ and $[0,L\omega_y/2]$ and to write  
\begin{widetext}
\vspace{-.25cm}
\begin{align}
\begin{split}
	&\langle\vec{1}|\mathbb{P}^{(th)}|\vec{1}\rangle=\\
	&\frac{2\sqrt{2}}{\sqrt{g}L^4\omega_x^2\omega_y^2}
	\sum_{l_x,l_y=0}^{L/2-1}
	\int_{l_x\omega_x}^{(l_x+1)\omega_x}dk_x \int_{l_y\omega_y}^{(l_y+1)\omega_y}dk_y
	\frac{\sin^2(\pi k_x)\sin^2(\pi k_y)}{\sin^2\big(\frac{\pi}{L\omega_x}k_x\big)\sin^2\big(\frac{\pi}{L\omega_y}
k_y\big)}\Big[\mu+2-\cos\big(\frac{2\pi}{L\omega_x}k_x\big)-
\cos\big(\frac{2\pi}{L\omega_y}k_y\big)\Big]^\frac{1}{2}.
\end{split}
\end{align}
\vspace{-.25cm}
\end{widetext}
We now restrict ourselves to the case with $\omega_x=1/p$ and $\omega_y=1/q$, with 
$p,q$ positive integers. After a simple shift of the integration 
variables as $k_x\to k_x-l_x\omega_x$ and $k_y\to k_y-l_y\omega_y$, one obtains  
\begin{widetext}
\vspace{-.25cm}
\begin{multline}
\label{eq:app-p4}
	\langle\vec{1}|\mathbb{P}^{(th)}|\vec{1}\rangle=
	\frac{2\sqrt{2}p^2q^2}{\sqrt{g}L^4}
	\sum_{p'=0}^{p-1}\sum_{q'=0}^{q-1}\sum_{l_x=0}^{L/(2p)-1}
	\sum_{l_y=0}^{L/(2q)-1}\int_{0}^{1/p}dk_x \int_0^{1/q}dk_y
	\frac{\sin^2(\pi(k_x+l_x+p'/p))
	\sin^2(\pi (k_y+l_y+q'/q))}
	{\sin^2\big(\frac{p\pi}{L}(k_x+l_x+p'/p)\big)
	\sin^2\big(\frac{q\pi}{L}(k_y+l_y+q'/q)\big)}\\
\times\Big[\mu+2-\cos\big(\frac{2 p\pi}{L}(k_x+l_x+p'/p)\big)-
\cos\big(\frac{2q\pi}{L}(k_y+l_y+q'/q)\big)\Big]^\frac{1}{2}.
\end{multline}
\vspace{-.25cm}
\end{widetext}
We now focus on the behaviour at the quantum phase transition. We set $g=g_c$, 
$\mu=\gamma^2_2/(2L^2)$, and we 
expand~\eqref{eq:app-p4} in the limit $L\to\infty$, using the periodicity of the sine function.
This yields
\begin{widetext}
\vspace{-.25cm}
\begin{align}
\begin{split}
\label{eq:app-p5}
	\langle\vec{1}|\mathbb{P}^{(th)}|\vec{1}\rangle=
	\frac{4}{\sqrt{g_c}\pi^3L}
	\sum_{p'=0}^{p-1}\sum_{q'=0}^{q-1}\sum_{l_x=0}^{L/(2p)-1}
	\sum_{l_y=0}^{L/(2q)-1}\int_{0}^{1/p}dk_x \int_0^{1/q}dk_y
	\frac{\sin^2(\pi(k_x+p'/p))
	\sin^2(\pi (k_y+q'/q))}
	{(k_x+l_x+p'/p)^2(k_y+l_y+q'/q)^2}\\
\times
\Big[\frac{\gamma_2^2}{4\pi^2}+ p^2(k_x+l_x+p'/p)^2+
q^2(k_y+l_y+q'/q)^2\Big]^\frac{1}{2}.
 \end{split}
\end{align}
\vspace{-.25cm}
\end{widetext}
Importantly, as 
a result of the large $L$ limit, Eq.~\eqref{eq:app-p5} 
depends only on the low-energy part of the dispersion of the QSM, 
although it contains non-universal information. 
To proceed we determine the large $L$ behaviour of the sum over $l_x,l_y$ 
in~\eqref{eq:app-p5}, i.e., of the function $\eta_{p',q'}(k_x,k_y)$ defined 
as 
\begin{widetext}
\vspace{-.25cm}
\begin{align}
 \begin{split}
  \label{eq:app-sum}
\eta_{p',q'}(k_x,k_y)=
\frac{4}{\sqrt{g_c}\pi^3L}
\sum_{l_x=0}^{L/(2p)-1}
	\sum_{l_y=0}^{L/(2q)-1}
\frac{\sqrt{\frac{\gamma_2^2}{4\pi^2}+ p^2(k_x+l_x+p'/p)^2+
q^2(k_y+l_y+q'/q)^2}}{(k_x+l_x+p'/p)^2(k_y+l_y+q'/q)^2}.
 \end{split}
\end{align}
\vspace{-.25cm}
\end{widetext}
The asymptotic behaviour of $\eta_{p,q}$ in the limit $L\to\infty$ can be 
obtained by using the Euler-Mclaurin formula. Given a function $f(x)$ 
this is stated as 
\begin{multline}
\label{eq:eml}
\sum_{x=x_1}^{x_2} f(x)=\int_{x_1}^{x_2} f(x) dx \\+\frac{f(x_1)+f(x_2)}{2}+\frac{1}{6}\frac{f'(x_2)-f'(x_1)}{2!}+\dots
\end{multline}
Here the dots denote terms with higher derivatives of $f(x)$ 
calculated at the integration boundaries $x_1$ and $x_2$, that can be derived to arbitrary order. 
To proceed, we first isolate the term with either $l_x=0$ or $l_y=0$ in~\eqref{eq:app-sum}. 
The remaining sum after fixing $l_x=0$ or $l_y=0$ can be treated with~\eqref{eq:eml}. 
We define this contribution to the large $L$  behaviour of $\eta_{p',q'}$ 
as $\eta_0$, which is given as  
\begin{equation}
\label{eq:eta0}
\eta_0=\frac{4}{\sqrt{g_c}\pi^3}\Big[\frac{q}{(k_x+p'/p)^2}
	+\frac{p}{(k_y+q'/q)^2}
	\Big]\frac{\ln(L)}{L}. 
\end{equation}
In the derivation of~\eqref{eq:eta0}  we neglected the boundary 
terms in~\eqref{eq:eml} because they are subleading. 
We are now left with the sums over $l_x\in[1,L/(2p)]$ and $l_y\in[1,L/(2q)]$ in~\eqref{eq:app-sum}. 
These can be evaluated again by using~\eqref{eq:eml}. We first apply~\eqref{eq:eml} to the 
sum over $l_x$ and obtain two contributions. The first one is obtained after 
evaluating the integral in~\eqref{eq:eml} at $x_2=L/(2p)$. After 
expanding the result for $L\to\infty$, we find the contribution 
$\eta_1$ given as 
\begin{equation}
	\label{eq:app-ln}
	\eta_{1}=\sum_{l_y=1}^{L/(2q)}\frac{4p}{\sqrt{g_c}\pi^3(k_y+l_y+q'/q)^2}\frac{\ln(L)}{L}. 
\end{equation}
Note the term $\ln(L)/L$ in~\eqref{eq:app-ln}. 
The sum over $l_y$ in~\eqref{eq:app-ln} can be performed 
exactly to obtain in the large $L$ limit 
\begin{equation}
\label{eq:eta1}
\eta_{1}=\frac{4}{\sqrt{g_c}\pi^3}p\psi'(1+k_y+q'/q)\frac{\ln(L)}{L}. 
\end{equation}
Here $\psi'(z)$ is the first derivative of the digamma function $\psi(z)$ 
with respect to its argument~\cite{Abra65}. 
The second contribution is obtained by evaluating the integral in~\eqref{eq:eml} 
at $x_1=1$. The remaining sum over $l_y$ cannot be evaluated analytically. 
However, one can again treat the sum over $l_y$ with the Euler-Mclaurin formula~\eqref{eq:eml}. After 
neglecting the boundary terms in~\eqref{eq:eml}, which are subleading for large $L$,  
and after evaluating the integral in~\eqref{eq:eml} at $x_2=L/(2q)$, we obtain 
the contribution $\eta_2$ as 
\begin{equation}
\label{eq:eta2}
\eta_2=\frac{4}{\sqrt{g_c}\pi^3}\frac{q}{1+k_x+p'/p}\frac{\ln(L)}{L}. 
\end{equation}
To obtain the full contribution of the sum over $l_x$ in~\eqref{eq:app-sum} 
we now have to consider the effect of the boundary terms in~\eqref{eq:eml}. 
Before doing that we check the accuracy of~\eqref{eq:eta1} and~\eqref{eq:eta2}
by defining
\begin{multline}
\label{eq:J}
{\mathcal J}=\frac{4}{\sqrt{g_c}\pi^3L}
\int_1^{L/(2p)}d l_x\\
\sum_{l_y=1}^{L/(2q)-1}
\frac{\sqrt{\frac{\gamma_2^2}{4\pi^2}+ p^2(k_x+l_x+p'/p)^2+
q^2(k_y+l_y+q'/q)^2}}{(k_x+l_x+p'/p)^2(k_y+l_y+q'/q)^2}.
\end{multline}
%
%%%%%%%%%%%%%%%%%%%%%%%%%%%%%%%%%%
\begin{figure}[t]
\centering
\includegraphics[width=1\linewidth]{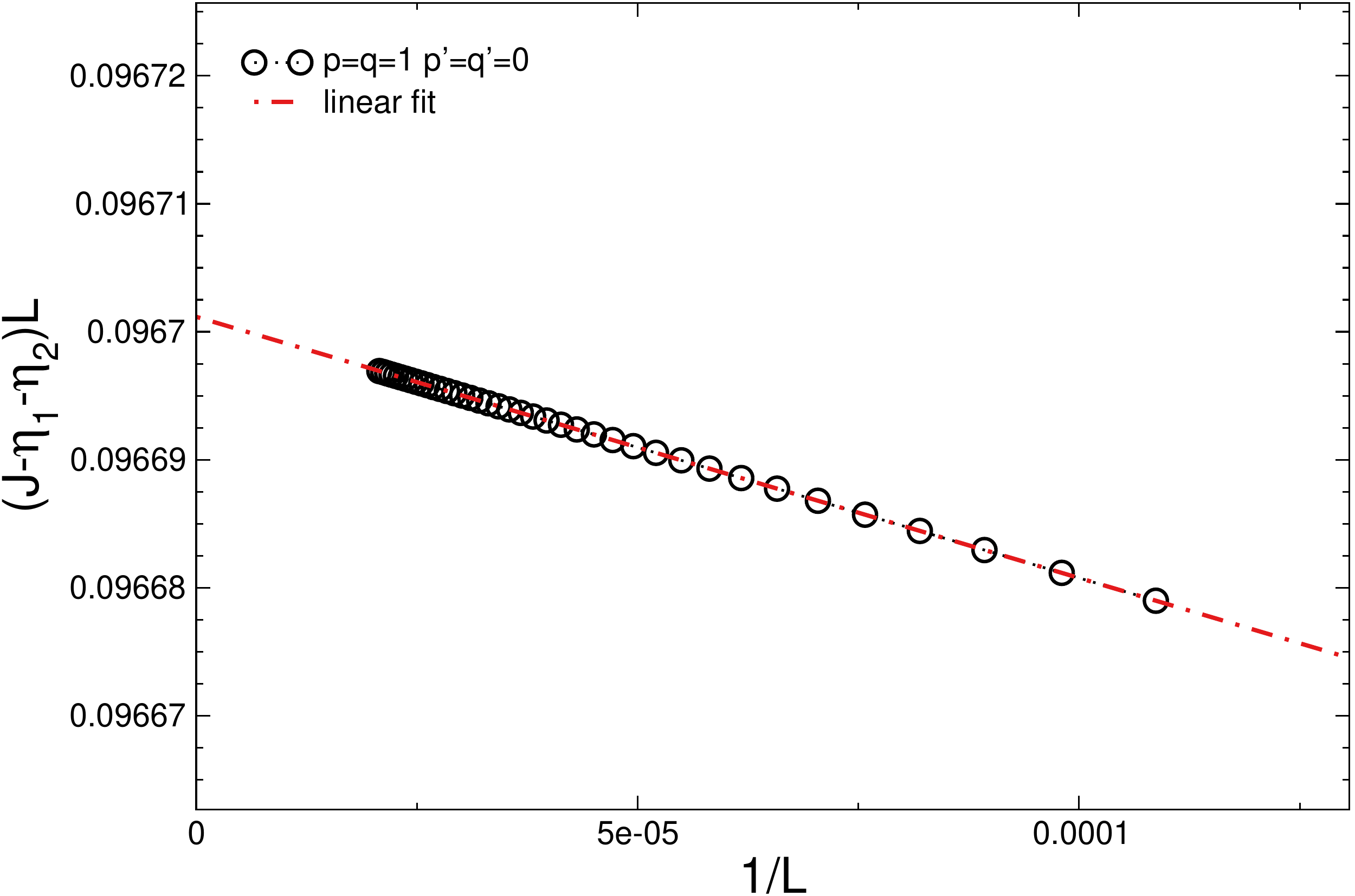}
\caption{ Check of the asymptotic behaviour in the large $L$ limit 
 of ${\mathcal J}$ (cf.~\eqref{eq:J}). The circles are numerical data for 
 $(J-\eta_1-\eta_2)L$, with $\eta_1$ and $\eta_2$ as defined in~\eqref{eq:eta1} 
 and~\eqref{eq:eta2}. The dashed-dotted line is a linear fit. Data 
 are for $p=q=1$ and $p'=q'=0$ (cf.~\eqref{eq:app-sum}). Note that 
 ${\mathcal J}-\eta_1-\eta_2\propto 1/L$ for $L\to\infty$. 
}
\label{fig:sum}
\end{figure}
%%%%%%%%%%%%%%%%%%%%%%%%%%%%%%%%%%
%
${\mathcal J}$ is obtained by neglecting the terms with either 
$l_x=0$ or $l_y=0$ in~\eqref{eq:app-sum}, which were treated in~\eqref{eq:eta0}, and 
by approximating the sum over $l_x$ in~\eqref{eq:eml} with an integral (see first term in~\eqref{eq:eml}), 
treating the sum over $l_y$ exactly. 
In Fig.~\ref{fig:sum} we show ${(\mathcal J}-\eta_1-\eta_2)L$ 
versus $1/L$. For large $L$ the data show a linear behaviour 
attaining a finite value in the limit $L\to\infty$. This shows 
that the leading order term $\propto\ln(L)/L$ of ${\mathcal J}$ 
is fully captured by $\eta_1+\eta_2$, the remaining contribution being 
$\propto1/L$, which we neglect. 

Having discussed the contribution which derives from approximating the 
sum over $l_x$ in~\eqref{eq:app-sum} with the integral in~\eqref{eq:eml}, we 
finally focus on the effect of the boundary terms in~\eqref{eq:eml}. 
Let us consider the first boundary term (first term in the second 
row in~\eqref{eq:eml}). The contribution as $\ln(L)/L$ is obtained by 
fixing $l_x=1$, other contributions are subleading. After performing the 
sum over $l_y$, one obtains the first boundary contribution $\eta_{b1}$ as 
\begin{equation}
\label{eq:etab1}
\eta_{b1}=\frac{2}{\sqrt{g_c}\pi^3}\frac{q}{(1+k_x+p'/p)^2}\frac{\ln(L)}{L}. 
\end{equation}
In a similar way the second boundary term (last term in~\eqref{eq:eml}) 
gives 
\begin{equation}
\label{eq:etab2}
\eta_{b2}=\frac{2}{3\sqrt{g_c}\pi^3}\frac{q}{(1+k_x+p'/p)^3}\frac{\ln(L)}{L}. 
\end{equation}
Note that boundary terms in~\eqref{eq:eml} are expected to be small. Specifically, 
the $k$-th term is suppressed as $1/(k+1)!$. 
The final result for $\eta(k_x,k_y,p,p',q,q')$ is obtained by putting 
together Eqs.~\eqref{eq:eta0},~\eqref{eq:eta1},~\eqref{eq:eta2},~\eqref{eq:etab1},~\eqref{eq:etab2} to 
obtain 
\begin{equation}
	\label{eq:group}
	\eta_{p',q'}(k_x,k_y)=\eta_0+\eta_1+\eta_2+\eta_{b1}+\eta_{b2}. 
\end{equation}
%
%
%%%%%%%%%%%%%%%%%%%%%%%%%%%%%%%%%%
\begin{figure}[t]
\centering
\includegraphics[width=0.47\textwidth]{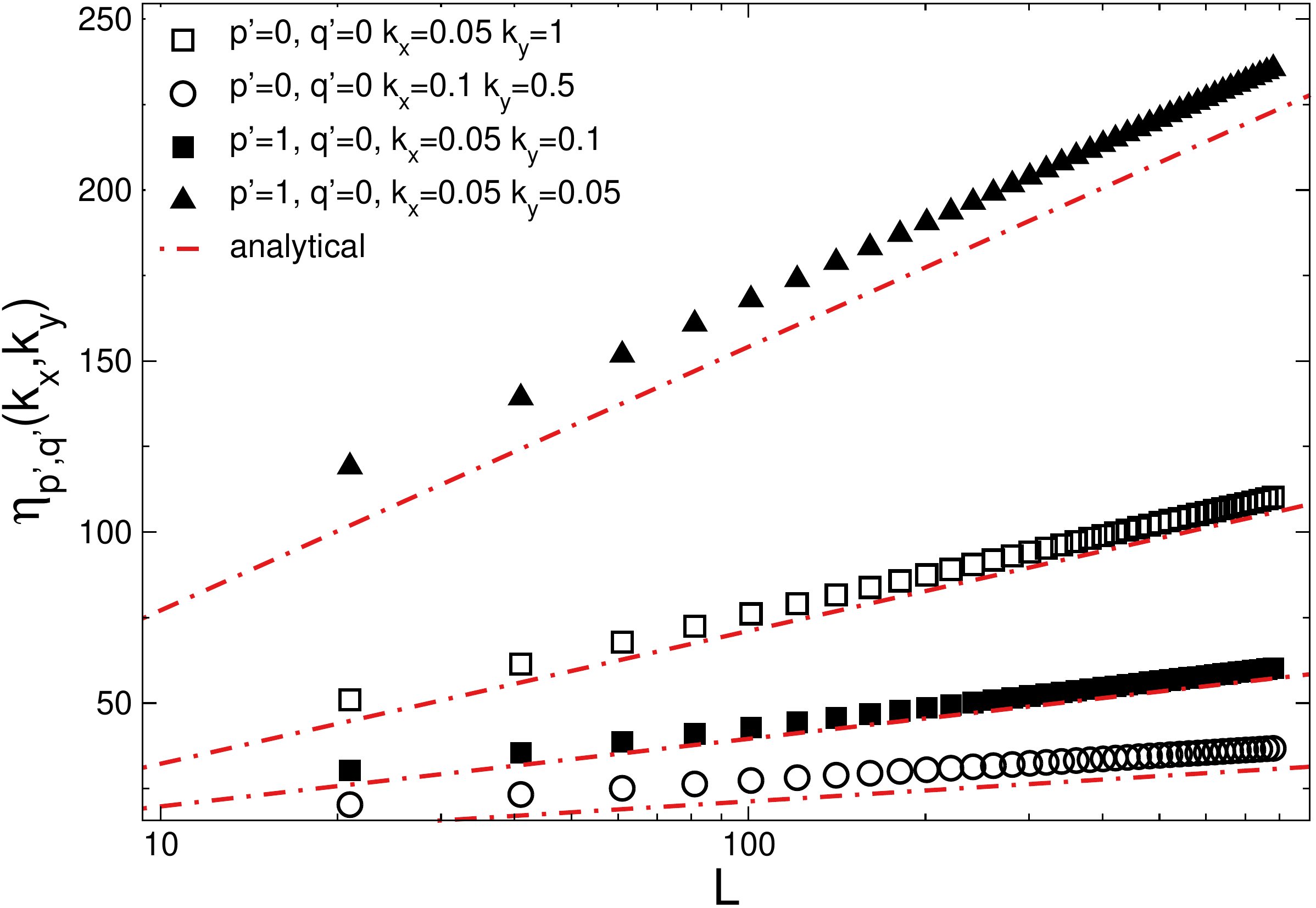}
\caption{ Large $L$ behaviour of the function $\eta_{p',q'}(k_x,k_y)$ 
 defined in~\eqref{eq:app-sum}. Here we consider a bipartition with 
 $\omega_x=1/p$ and $\omega_y=1/q$ (see Fig.~\ref{fig:partition1}). 
 We fix $q=1$ considering $q'=0$ and $p'=0,1$ (empty and full symbols, 
 respectively). Symbols are numerical results. Dashed-dotted lines 
 are the asymptotic behaviours in~\eqref{eq:group}. 
}
\label{fig:sum_p}
\end{figure}
%%%%%%%%%%%%%%%%%%%%%%%%%%%%%%%%%%
%

\noindent
In Fig.~\ref{fig:sum_p} we check the accuracy of~\eqref{eq:group}, 
showing the function $\eta_{p',q'}$ for fixed values of $q=1$, which corresponds 
to a straight partition between the two subsystems, 
and $p=1/2$. For all values of $p',q'$ and $k_x,k_y$ that we consider $\eta_{p',q'}$ 
is well described by~\eqref{eq:group}.

%################################################################
\subsection{Finite-size contribution}
\label{app:app2}

\noindent
In this section we derive the leading behaviour in the large $L$ limit of 
$\langle\vec{1}|\mathbb{P}^{(L)}|\vec{1}\rangle$. 
Interestingly, we show that in the presence of a straight boundary between 
the two subsystems the expectation value behaves as 
$\langle\vec{1}|\mathbb{P}^{(L)}|\vec{1}\rangle\propto \ln(L)/L$. 
On the other hand, in the presence of corners, the multiplicative 
logarithmic correction is absent. 
The finite-size correlator to calculate reads as
\begin{multline}
	\label{eq:q2-app}
	\mathbb{P}^{(L)}_{\vec{n}\vec{m}}=-\frac{1}{4\sqrt{g_c}
	\pi}\sideset{}{'}\sum_{l,l'=-\infty}^\infty
	e^{-\sqrt{2\mu}\sqrt{(l L+n_x-m_x)^2+(l'L+n_y-m_y)^2}}\\\times
	\Big[\frac{1}{[(l L+n_x-m_x)^2+(l' L+n_y-m_y)^2]^{3/2}}\\
	+\frac{\sqrt{2\mu}}{(l L+n_x-m_x)^2+(l' L+n_y-m_y)^2}\Big].
\end{multline}
Crucially, if $\omega_x<1$ and $\omega_y<1$, the denominators in~\eqref{eq:q2-app} 
are never singular. This implies that the logarithmic correction is not 
present, which can be straightforwardly checked numerically. 
Let us now consider the situation with $\omega_x<1$ and $\omega_y=1$. 
The other case with $\omega_x=1$ and $\omega_y<1$ can be treated similarly. 
A singularity appears in the limit $L\to\infty$ for $l=0$ and $l'=\pm1$. 
We numerically observe that only the first term in~\eqref{eq:q2-app} gives rise 
to a singular behaviour. Thus, we neglect the second term and fix $l=0$,  obtaining 
\begin{widetext}
\begin{equation}
\label{eq:pp}
	\langle\vec{1}|\mathbb{P}^{(L)}|\vec{1}\rangle\simeq
	-\frac{1}{4\sqrt{g_c}\pi L^2\omega_x}\,\sideset{}{'}
	\sum_{l'=-\infty}^\infty\sum_{n_x,m_x=0}^{L\omega_x}\sum_{n_y,m_y=0}^{L-1}
	\frac{
	e^{-\sqrt{2\mu}\sqrt{(n_x-m_x)^2+(l' L+n_y-m_y)^2}}}{((n_x-m_x)^2+(l' L+n_y-m_y)^2)^{3/2}}.
\end{equation}
\end{widetext}
Only the differences $n_x-m_x$ and $n_y-m_y$ appear in~\eqref{eq:pp}. Thus, it is convenient to 
change variables to $x=n_x-m_x$ and $y=n_y-m_y$, to obtain 
\begin{widetext}
 \begin{equation}
\label{eq:qtilde}
\langle\vec{1}|\mathbb{P}^{(L)}|\vec{1}\rangle\simeq
-\frac{1}{4\sqrt{g_c}\pi L^2\omega_x}\sideset{}{'}\sum_{l'=-\infty}^\infty\sum_{x=-L\omega_x}^{L\omega_x}\sum_{y=-(L-1)}^{L-1}\\
\Big(L\omega_x+1-|x|\Big)(L-|y|)\frac{e^{-\sqrt{2\mu}\sqrt{x^2+(l' L+y)^2}}}{(x^2+(l' L+y)^2)^{3/2}}. 
\end{equation}
\end{widetext}
Again, the singular behaviour occurs for $x\approx 0$ and 
$y\approx -l L$, with $l'=\pm1$.  In this limit we can neglect the 
exponential in~\eqref{eq:qtilde} because it is regular. Thus, we obtain 
\begin{widetext}
 \begin{equation}
\label{eq:qtilde}
\langle\vec{1}|\mathbb{P}^{(L)}|\vec{1}\rangle\simeq
-\frac{1}{4\sqrt{g_c}
\pi L^2\omega_x}\,\sideset{}{'}\sum_{l'=-\infty}^\infty\sum_{x=-L\omega_x}^{L\omega_x}\sum_{y=-(L-1)}^{L-1}
\frac{(L\omega_x+1-|x|)(L-|y|)}{(x^2+(l' L+y)^2)^{3/2}}. 
\end{equation}
\end{widetext}
To proceed, we consider the case $l=1$ and it is clear that the contribution 
from $l=-1$ is the same. We can restrict the sum over $x$ in~\eqref{eq:qtilde} to 
$x>0$ because of the symmetry $x\to-x$. We also restrict to $y<0$ because the singularity 
in~\eqref{eq:qtilde} occurs for $y<0$. We now have 
\begin{equation}
\label{eq:qtilde-1}
\langle\vec{1}|\mathbb{P}^{(L)}|\vec{1}\rangle\simeq
\frac{1}{2\sqrt{g_c}\pi L^2\omega_x}\sum_{x=0}^{L\omega_x}\sum_{y=0}^{L-1}
\frac{(L\omega_x+1-x)(y-L)}{(x^2+(L-y)^2)^{3/2}}. 
\end{equation}
Now the strategy is to treat the sum~\eqref{eq:qtilde-1} by using the Euler-Mclaurin 
formula~\eqref{eq:eml}. For instance, one can first apply~\eqref{eq:eml} to 
the sum over $x$ and obtain that  the leading term in the large $L$ limit 
is obtained by evaluating the integral in~\eqref{eq:eml} 
at $\omega_x L$. One can also verify that the boundary terms in~\eqref{eq:eml} 
can be neglected. 
A straightforward calculation gives the final result 
\begin{equation}
\label{eq:p-fs-fin}
\langle\vec{1}|\mathbb{P}^{(L)}|\vec{1}\rangle\simeq-\frac{1}{\sqrt{g_c}\pi}\frac{\ln(L)}{L}, 
\end{equation}
where the contribution of $l=-1$ in~\eqref{eq:qtilde} has been taken into account. 
%
%%%%%%%%%%%%%%%%%%%%%%%%%%%%%%%%%%
\begin{figure}[t]
\centering
\includegraphics[width=0.45\textwidth]{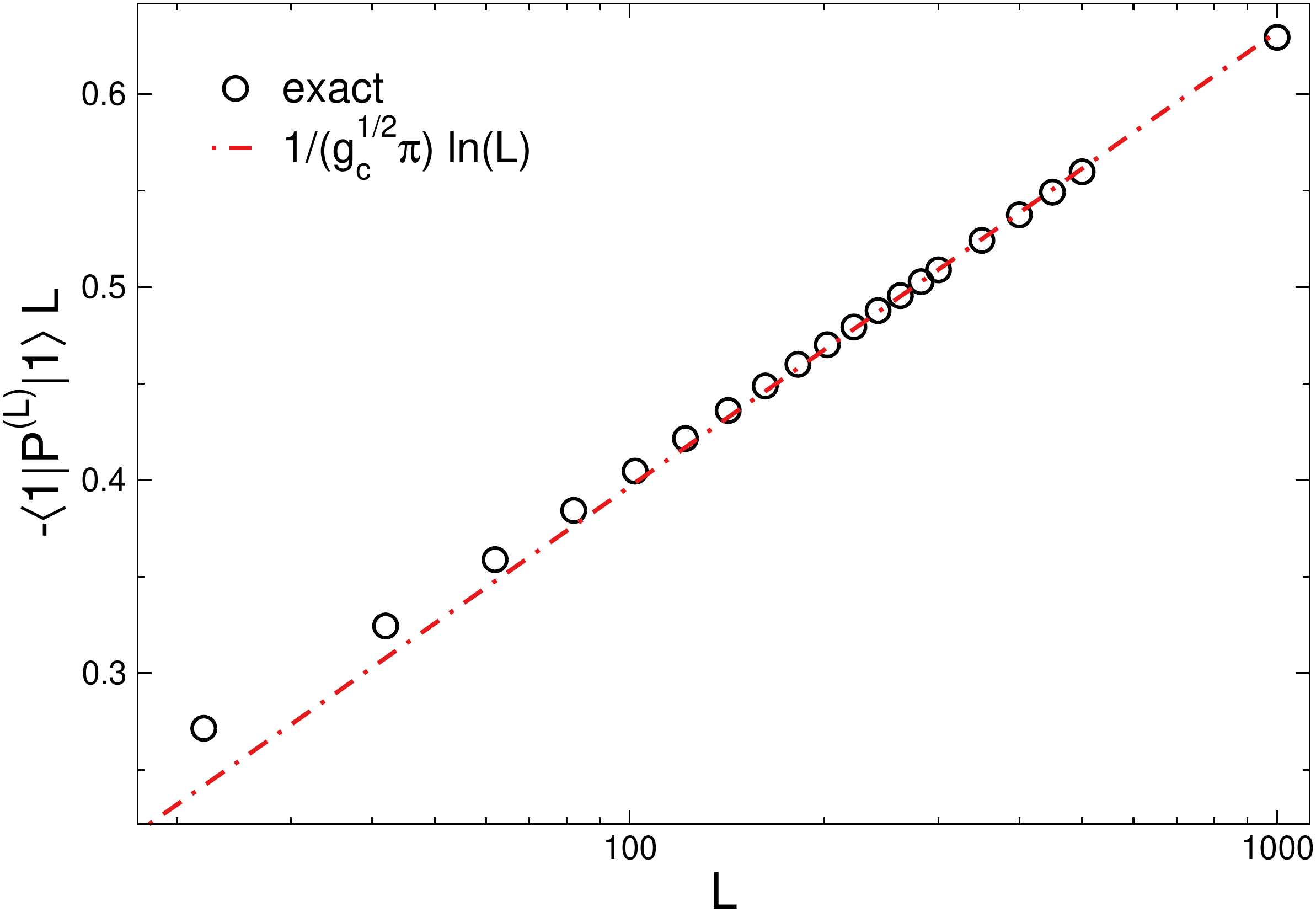}
\caption{ Large $L$ behaviour of $\langle\vec{1}|\mathbb{P}^{(L)}|\vec{1}\rangle$ 
 (cf.~\eqref{eq:pp} for its definition). 
 The symbols are numerical data obtained by using~\eqref{eq:q2-app}. The dashed-dotted line 
 is the analytical result~\eqref{eq:p-fs-fin}. All the results are for the 
 bipartition with $\omega_x=1/2$ and $\omega_y=1$. 
}
\label{fig:sum-1}
\end{figure}
%%%%%%%%%%%%%%%%%%%%%%%%%%%%%%%%%%
%
The validity of~\eqref{eq:p-fs-fin} is numerically confirmed in Fig.~\ref{fig:sum-1}. 

\bibliography{bibliography}

\begin{thebibliography}{114}%
\makeatletter
\providecommand \@ifxundefined [1]{%
 \@ifx{#1\undefined}
}%
\providecommand \@ifnum [1]{%
 \ifnum #1\expandafter \@firstoftwo
 \else \expandafter \@secondoftwo
 \fi
}%
\providecommand \@ifx [1]{%
 \ifx #1\expandafter \@firstoftwo
 \else \expandafter \@secondoftwo
 \fi
}%
\providecommand \natexlab [1]{#1}%
\providecommand \enquote  [1]{``#1''}%
\providecommand \bibnamefont  [1]{#1}%
\providecommand \bibfnamefont [1]{#1}%
\providecommand \citenamefont [1]{#1}%
\providecommand \href@noop [0]{\@secondoftwo}%
\providecommand \href [0]{\begingroup \@sanitize@url \@href}%
\providecommand \@href[1]{\@@startlink{#1}\@@href}%
\providecommand \@@href[1]{\endgroup#1\@@endlink}%
\providecommand \@sanitize@url [0]{\catcode `\\12\catcode `\$12\catcode
  `\&12\catcode `\#12\catcode `\^12\catcode `\_12\catcode `\%12\relax}%
\providecommand \@@startlink[1]{}%
\providecommand \@@endlink[0]{}%
\providecommand \url  [0]{\begingroup\@sanitize@url \@url }%
\providecommand \@url [1]{\endgroup\@href {#1}{\urlprefix }}%
\providecommand \urlprefix  [0]{URL }%
\providecommand \Eprint [0]{\href }%
\providecommand \doibase [0]{http://dx.doi.org/}%
\providecommand \selectlanguage [0]{\@gobble}%
\providecommand \bibinfo  [0]{\@secondoftwo}%
\providecommand \bibfield  [0]{\@secondoftwo}%
\providecommand \translation [1]{[#1]}%
\providecommand \BibitemOpen [0]{}%
\providecommand \bibitemStop [0]{}%
\providecommand \bibitemNoStop [0]{.\EOS\space}%
\providecommand \EOS [0]{\spacefactor3000\relax}%
\providecommand \BibitemShut  [1]{\csname bibitem#1\endcsname}%
\let\auto@bib@innerbib\@empty
%</preamble>
\bibitem [{\citenamefont {Eisert}\ \emph {et~al.}(2010)\citenamefont {Eisert},
  \citenamefont {Cramer},\ and\ \citenamefont {Plenio}}]{area}%
  \BibitemOpen
  \bibfield  {author} {\bibinfo {author} {\bibfnamefont {J.}~\bibnamefont
  {Eisert}}, \bibinfo {author} {\bibfnamefont {M.}~\bibnamefont {Cramer}}, \
  and\ \bibinfo {author} {\bibfnamefont {M.~B.}\ \bibnamefont {Plenio}},\
  }\href {\doibase 10.1103/RevModPhys.82.277} {\bibfield  {journal} {\bibinfo
  {journal} {Rev. Mod. Phys.}\ }\textbf {\bibinfo {volume} {82}},\ \bibinfo
  {pages} {277} (\bibinfo {year} {2010})}\BibitemShut {NoStop}%
\bibitem [{\citenamefont {Amico}\ \emph {et~al.}(2008)\citenamefont {Amico},
  \citenamefont {Fazio}, \citenamefont {Osterloh},\ and\ \citenamefont
  {Vedral}}]{amico-2008}%
  \BibitemOpen
  \bibfield  {author} {\bibinfo {author} {\bibfnamefont {L.}~\bibnamefont
  {Amico}}, \bibinfo {author} {\bibfnamefont {R.}~\bibnamefont {Fazio}},
  \bibinfo {author} {\bibfnamefont {A.}~\bibnamefont {Osterloh}}, \ and\
  \bibinfo {author} {\bibfnamefont {V.}~\bibnamefont {Vedral}},\ }\href
  {\doibase 10.1103/RevModPhys.80.517} {\bibfield  {journal} {\bibinfo
  {journal} {Rev. Mod. Phys.}\ }\textbf {\bibinfo {volume} {80}},\ \bibinfo
  {pages} {517} (\bibinfo {year} {2008})}\BibitemShut {NoStop}%
\bibitem [{\citenamefont {Calabrese}\ \emph {et~al.}(2009)\citenamefont
  {Calabrese}, \citenamefont {Cardy},\ and\ \citenamefont
  {Doyon}}]{calabrese-2009}%
  \BibitemOpen
  \bibfield  {author} {\bibinfo {author} {\bibfnamefont {P.}~\bibnamefont
  {Calabrese}}, \bibinfo {author} {\bibfnamefont {J.}~\bibnamefont {Cardy}}, \
  and\ \bibinfo {author} {\bibfnamefont {B.}~\bibnamefont {Doyon}},\ }\href
  {\doibase 10.1088/1751-8121/42/50/500301} {\bibfield  {journal} {\bibinfo
  {journal} {Journal of Physics A: Mathematical and Theoretical}\ }\textbf
  {\bibinfo {volume} {42}},\ \bibinfo {pages} {500301} (\bibinfo {year}
  {2009})}\BibitemShut {NoStop}%
\bibitem [{\citenamefont {Laflorencie}(2016)}]{laflorencie-2016}%
  \BibitemOpen
  \bibfield  {author} {\bibinfo {author} {\bibfnamefont {N.}~\bibnamefont
  {Laflorencie}},\ }\href {\doibase 10.1016/j.physrep.2016.06.008} {\bibfield
  {journal} {\bibinfo  {journal} {Physics Reports}\ }\textbf {\bibinfo {volume}
  {646}},\ \bibinfo {pages} {1} (\bibinfo {year} {2016})}\BibitemShut {NoStop}%
\bibitem [{\citenamefont {Peschel}\ \emph {et~al.}(1999)\citenamefont
  {Peschel}, \citenamefont {Kaulke},\ and\ \citenamefont
  {Legeza}}]{peschel-1999}%
  \BibitemOpen
  \bibfield  {author} {\bibinfo {author} {\bibfnamefont {I.}~\bibnamefont
  {Peschel}}, \bibinfo {author} {\bibfnamefont {M.}~\bibnamefont {Kaulke}}, \
  and\ \bibinfo {author} {\bibfnamefont {Ã.}~\bibnamefont {Legeza}},\ }\href
  {\doibase 10.1002/(SICI)1521-3889(199902)8:2<153::AID-ANDP153>3.0.CO;2-N}
  {\bibfield  {journal} {\bibinfo  {journal} {Annalen der Physik}\ }\textbf
  {\bibinfo {volume} {8}},\ \bibinfo {pages} {153} (\bibinfo {year}
  {1999})}\BibitemShut {NoStop}%
\bibitem [{\citenamefont {Chung}\ and\ \citenamefont
  {Peschel}(2000)}]{chung-2000}%
  \BibitemOpen
  \bibfield  {author} {\bibinfo {author} {\bibfnamefont {M.-C.}\ \bibnamefont
  {Chung}}\ and\ \bibinfo {author} {\bibfnamefont {I.}~\bibnamefont
  {Peschel}},\ }\href {\doibase 10.1103/PhysRevB.62.4191} {\bibfield  {journal}
  {\bibinfo  {journal} {Phys. Rev. B}\ }\textbf {\bibinfo {volume} {62}},\
  \bibinfo {pages} {4191} (\bibinfo {year} {2000})}\BibitemShut {NoStop}%
\bibitem [{\citenamefont {Peschel}(2004)}]{peschel-2004}%
  \BibitemOpen
  \bibfield  {author} {\bibinfo {author} {\bibfnamefont {I.}~\bibnamefont
  {Peschel}},\ }\href {\doibase 10.1088/1742-5468/2004/06/p06004} {\bibfield
  {journal} {\bibinfo  {journal} {Journal of Statistical Mechanics: Theory and
  Experiment}\ }\textbf {\bibinfo {volume} {2004}},\ \bibinfo {pages} {P06004}
  (\bibinfo {year} {2004})}\BibitemShut {NoStop}%
\bibitem [{\citenamefont {Peschel}\ and\ \citenamefont
  {Eisler}(2009)}]{viktor}%
  \BibitemOpen
  \bibfield  {author} {\bibinfo {author} {\bibfnamefont {I.}~\bibnamefont
  {Peschel}}\ and\ \bibinfo {author} {\bibfnamefont {V.}~\bibnamefont
  {Eisler}},\ }\href {\doibase 10.1088/1751-8113/42/50/504003} {\bibfield
  {journal} {\bibinfo  {journal} {Journal of Physics A: Mathematical and
  Theoretical}\ }\textbf {\bibinfo {volume} {42}},\ \bibinfo {pages} {504003}
  (\bibinfo {year} {2009})}\BibitemShut {NoStop}%
\bibitem [{\citenamefont {White}(1992)}]{white-1992}%
  \BibitemOpen
  \bibfield  {author} {\bibinfo {author} {\bibfnamefont {S.~R.}\ \bibnamefont
  {White}},\ }\href {\doibase 10.1103/PhysRevLett.69.2863} {\bibfield
  {journal} {\bibinfo  {journal} {Phys. Rev. Lett.}\ }\textbf {\bibinfo
  {volume} {69}},\ \bibinfo {pages} {2863} (\bibinfo {year}
  {1992})}\BibitemShut {NoStop}%
\bibitem [{\citenamefont {Schollw\"{o}ck}(2011)}]{uli-2011}%
  \BibitemOpen
  \bibfield  {author} {\bibinfo {author} {\bibfnamefont {U.}~\bibnamefont
  {Schollw\"{o}ck}},\ }\href {\doibase 10.1016/j.aop.2010.09.012} {\bibfield
  {journal} {\bibinfo  {journal} {Annals of Physics}\ }\textbf {\bibinfo
  {volume} {326}},\ \bibinfo {pages} {96} (\bibinfo {year} {2011})}\BibitemShut
  {NoStop}%
\bibitem [{\citenamefont {Li}\ and\ \citenamefont {Haldane}(2008)}]{li-2008}%
  \BibitemOpen
  \bibfield  {author} {\bibinfo {author} {\bibfnamefont {H.}~\bibnamefont
  {Li}}\ and\ \bibinfo {author} {\bibfnamefont {F.~D.~M.}\ \bibnamefont
  {Haldane}},\ }\href {\doibase 10.1103/PhysRevLett.101.010504} {\bibfield
  {journal} {\bibinfo  {journal} {Phys. Rev. Lett.}\ }\textbf {\bibinfo
  {volume} {101}},\ \bibinfo {pages} {010504} (\bibinfo {year}
  {2008})}\BibitemShut {NoStop}%
\bibitem [{\citenamefont {Thomale}\ \emph
  {et~al.}(2010{\natexlab{a}})\citenamefont {Thomale}, \citenamefont {Arovas},\
  and\ \citenamefont {Bernevig}}]{thomale-2010}%
  \BibitemOpen
  \bibfield  {author} {\bibinfo {author} {\bibfnamefont {R.}~\bibnamefont
  {Thomale}}, \bibinfo {author} {\bibfnamefont {D.~P.}\ \bibnamefont {Arovas}},
  \ and\ \bibinfo {author} {\bibfnamefont {B.~A.}\ \bibnamefont {Bernevig}},\
  }\href {\doibase 10.1103/PhysRevLett.105.116805} {\bibfield  {journal}
  {\bibinfo  {journal} {Phys. Rev. Lett.}\ }\textbf {\bibinfo {volume} {105}},\
  \bibinfo {pages} {116805} (\bibinfo {year} {2010}{\natexlab{a}})}\BibitemShut
  {NoStop}%
\bibitem [{\citenamefont {L\"auchli}\ \emph {et~al.}(2010)\citenamefont
  {L\"auchli}, \citenamefont {Bergholtz}, \citenamefont {Suorsa},\ and\
  \citenamefont {Haque}}]{andreas-2010}%
  \BibitemOpen
  \bibfield  {author} {\bibinfo {author} {\bibfnamefont {A.~M.}\ \bibnamefont
  {L\"auchli}}, \bibinfo {author} {\bibfnamefont {E.~J.}\ \bibnamefont
  {Bergholtz}}, \bibinfo {author} {\bibfnamefont {J.}~\bibnamefont {Suorsa}}, \
  and\ \bibinfo {author} {\bibfnamefont {M.}~\bibnamefont {Haque}},\ }\href
  {\doibase 10.1103/PhysRevLett.104.156404} {\bibfield  {journal} {\bibinfo
  {journal} {Phys. Rev. Lett.}\ }\textbf {\bibinfo {volume} {104}},\ \bibinfo
  {pages} {156404} (\bibinfo {year} {2010})}\BibitemShut {NoStop}%
\bibitem [{\citenamefont {Haque}\ \emph {et~al.}(2007)\citenamefont {Haque},
  \citenamefont {Zozulya},\ and\ \citenamefont {Schoutens}}]{haque-2007}%
  \BibitemOpen
  \bibfield  {author} {\bibinfo {author} {\bibfnamefont {M.}~\bibnamefont
  {Haque}}, \bibinfo {author} {\bibfnamefont {O.}~\bibnamefont {Zozulya}}, \
  and\ \bibinfo {author} {\bibfnamefont {K.}~\bibnamefont {Schoutens}},\ }\href
  {\doibase 10.1103/PhysRevLett.98.060401} {\bibfield  {journal} {\bibinfo
  {journal} {Phys. Rev. Lett.}\ }\textbf {\bibinfo {volume} {98}},\ \bibinfo
  {pages} {060401} (\bibinfo {year} {2007})}\BibitemShut {NoStop}%
\bibitem [{\citenamefont {Thomale}\ \emph
  {et~al.}(2010{\natexlab{b}})\citenamefont {Thomale}, \citenamefont
  {Sterdyniak}, \citenamefont {Regnault},\ and\ \citenamefont
  {Bernevig}}]{thomale-2010a}%
  \BibitemOpen
  \bibfield  {author} {\bibinfo {author} {\bibfnamefont {R.}~\bibnamefont
  {Thomale}}, \bibinfo {author} {\bibfnamefont {A.}~\bibnamefont {Sterdyniak}},
  \bibinfo {author} {\bibfnamefont {N.}~\bibnamefont {Regnault}}, \ and\
  \bibinfo {author} {\bibfnamefont {B.~A.}\ \bibnamefont {Bernevig}},\ }\href
  {\doibase 10.1103/PhysRevLett.104.180502} {\bibfield  {journal} {\bibinfo
  {journal} {Phys. Rev. Lett.}\ }\textbf {\bibinfo {volume} {104}},\ \bibinfo
  {pages} {180502} (\bibinfo {year} {2010}{\natexlab{b}})}\BibitemShut
  {NoStop}%
\bibitem [{\citenamefont {Hermanns}\ \emph {et~al.}(2011)\citenamefont
  {Hermanns}, \citenamefont {Chandran}, \citenamefont {Regnault},\ and\
  \citenamefont {Bernevig}}]{hermanns-2011}%
  \BibitemOpen
  \bibfield  {author} {\bibinfo {author} {\bibfnamefont {M.}~\bibnamefont
  {Hermanns}}, \bibinfo {author} {\bibfnamefont {A.}~\bibnamefont {Chandran}},
  \bibinfo {author} {\bibfnamefont {N.}~\bibnamefont {Regnault}}, \ and\
  \bibinfo {author} {\bibfnamefont {B.~A.}\ \bibnamefont {Bernevig}},\ }\href
  {\doibase 10.1103/PhysRevB.84.121309} {\bibfield  {journal} {\bibinfo
  {journal} {Phys. Rev. B}\ }\textbf {\bibinfo {volume} {84}},\ \bibinfo
  {pages} {121309(R)} (\bibinfo {year} {2011})}\BibitemShut {NoStop}%
\bibitem [{\citenamefont {Chandran}\ \emph {et~al.}(2011)\citenamefont
  {Chandran}, \citenamefont {Hermanns}, \citenamefont {Regnault},\ and\
  \citenamefont {Bernevig}}]{chandran-2011}%
  \BibitemOpen
  \bibfield  {author} {\bibinfo {author} {\bibfnamefont {A.}~\bibnamefont
  {Chandran}}, \bibinfo {author} {\bibfnamefont {M.}~\bibnamefont {Hermanns}},
  \bibinfo {author} {\bibfnamefont {N.}~\bibnamefont {Regnault}}, \ and\
  \bibinfo {author} {\bibfnamefont {B.~A.}\ \bibnamefont {Bernevig}},\ }\href
  {\doibase 10.1103/PhysRevB.84.205136} {\bibfield  {journal} {\bibinfo
  {journal} {Phys. Rev. B}\ }\textbf {\bibinfo {volume} {84}},\ \bibinfo
  {pages} {205136} (\bibinfo {year} {2011})}\BibitemShut {NoStop}%
\bibitem [{\citenamefont {Qi}\ \emph {et~al.}(2012)\citenamefont {Qi},
  \citenamefont {Katsura},\ and\ \citenamefont {Ludwig}}]{qi-2012}%
  \BibitemOpen
  \bibfield  {author} {\bibinfo {author} {\bibfnamefont {X.-L.}\ \bibnamefont
  {Qi}}, \bibinfo {author} {\bibfnamefont {H.}~\bibnamefont {Katsura}}, \ and\
  \bibinfo {author} {\bibfnamefont {A.~W.~W.}\ \bibnamefont {Ludwig}},\ }\href
  {\doibase 10.1103/PhysRevLett.108.196402} {\bibfield  {journal} {\bibinfo
  {journal} {Phys. Rev. Lett.}\ }\textbf {\bibinfo {volume} {108}},\ \bibinfo
  {pages} {196402} (\bibinfo {year} {2012})}\BibitemShut {NoStop}%
\bibitem [{\citenamefont {Liu}\ \emph {et~al.}(2012)\citenamefont {Liu},
  \citenamefont {Bergholtz}, \citenamefont {Fan},\ and\ \citenamefont
  {L\"auchli}}]{liu-2012}%
  \BibitemOpen
  \bibfield  {author} {\bibinfo {author} {\bibfnamefont {Z.}~\bibnamefont
  {Liu}}, \bibinfo {author} {\bibfnamefont {E.~J.}\ \bibnamefont {Bergholtz}},
  \bibinfo {author} {\bibfnamefont {H.}~\bibnamefont {Fan}}, \ and\ \bibinfo
  {author} {\bibfnamefont {A.~M.}\ \bibnamefont {L\"auchli}},\ }\href {\doibase
  10.1103/PhysRevB.85.045119} {\bibfield  {journal} {\bibinfo  {journal} {Phys.
  Rev. B}\ }\textbf {\bibinfo {volume} {85}},\ \bibinfo {pages} {045119}
  (\bibinfo {year} {2012})}\BibitemShut {NoStop}%
\bibitem [{\citenamefont {Sterdyniak}\ \emph {et~al.}(2012)\citenamefont
  {Sterdyniak}, \citenamefont {Chandran}, \citenamefont {Regnault},
  \citenamefont {Bernevig},\ and\ \citenamefont {Bonderson}}]{sterdyniak-2012}%
  \BibitemOpen
  \bibfield  {author} {\bibinfo {author} {\bibfnamefont {A.}~\bibnamefont
  {Sterdyniak}}, \bibinfo {author} {\bibfnamefont {A.}~\bibnamefont
  {Chandran}}, \bibinfo {author} {\bibfnamefont {N.}~\bibnamefont {Regnault}},
  \bibinfo {author} {\bibfnamefont {B.~A.}\ \bibnamefont {Bernevig}}, \ and\
  \bibinfo {author} {\bibfnamefont {P.}~\bibnamefont {Bonderson}},\ }\href
  {\doibase 10.1103/PhysRevB.85.125308} {\bibfield  {journal} {\bibinfo
  {journal} {Phys. Rev. B}\ }\textbf {\bibinfo {volume} {85}},\ \bibinfo
  {pages} {125308} (\bibinfo {year} {2012})}\BibitemShut {NoStop}%
\bibitem [{\citenamefont {Dubail}\ \emph
  {et~al.}(2012{\natexlab{a}})\citenamefont {Dubail}, \citenamefont {Read},\
  and\ \citenamefont {Rezayi}}]{dubail-2012}%
  \BibitemOpen
  \bibfield  {author} {\bibinfo {author} {\bibfnamefont {J.}~\bibnamefont
  {Dubail}}, \bibinfo {author} {\bibfnamefont {N.}~\bibnamefont {Read}}, \ and\
  \bibinfo {author} {\bibfnamefont {E.~H.}\ \bibnamefont {Rezayi}},\ }\href
  {\doibase 10.1103/PhysRevB.85.115321} {\bibfield  {journal} {\bibinfo
  {journal} {Phys. Rev. B}\ }\textbf {\bibinfo {volume} {85}},\ \bibinfo
  {pages} {115321} (\bibinfo {year} {2012}{\natexlab{a}})}\BibitemShut
  {NoStop}%
\bibitem [{\citenamefont {Dubail}\ \emph
  {et~al.}(2012{\natexlab{b}})\citenamefont {Dubail}, \citenamefont {Read},\
  and\ \citenamefont {Rezayi}}]{dubail-2012a}%
  \BibitemOpen
  \bibfield  {author} {\bibinfo {author} {\bibfnamefont {J.}~\bibnamefont
  {Dubail}}, \bibinfo {author} {\bibfnamefont {N.}~\bibnamefont {Read}}, \ and\
  \bibinfo {author} {\bibfnamefont {E.~H.}\ \bibnamefont {Rezayi}},\ }\href
  {\doibase 10.1103/PhysRevB.86.245310} {\bibfield  {journal} {\bibinfo
  {journal} {Phys. Rev. B}\ }\textbf {\bibinfo {volume} {86}},\ \bibinfo
  {pages} {245310} (\bibinfo {year} {2012}{\natexlab{b}})}\BibitemShut
  {NoStop}%
\bibitem [{\citenamefont {Chandran}\ \emph {et~al.}(2014)\citenamefont
  {Chandran}, \citenamefont {Khemani},\ and\ \citenamefont {Sondhi}}]{Chan14}%
  \BibitemOpen
  \bibfield  {author} {\bibinfo {author} {\bibfnamefont {A.}~\bibnamefont
  {Chandran}}, \bibinfo {author} {\bibfnamefont {V.}~\bibnamefont {Khemani}}, \
  and\ \bibinfo {author} {\bibfnamefont {S.~L.}\ \bibnamefont {Sondhi}},\
  }\href {\doibase 10.1103/PhysRevLett.113.060501} {\bibfield  {journal}
  {\bibinfo  {journal} {Phys. Rev. Lett.}\ }\textbf {\bibinfo {volume} {113}},\
  \bibinfo {pages} {060501} (\bibinfo {year} {2014})}\BibitemShut {NoStop}%
\bibitem [{\citenamefont {Pollmann}\ \emph {et~al.}(2010)\citenamefont
  {Pollmann}, \citenamefont {Turner}, \citenamefont {Berg},\ and\ \citenamefont
  {Oshikawa}}]{pollmann-2010}%
  \BibitemOpen
  \bibfield  {author} {\bibinfo {author} {\bibfnamefont {F.}~\bibnamefont
  {Pollmann}}, \bibinfo {author} {\bibfnamefont {A.~M.}\ \bibnamefont
  {Turner}}, \bibinfo {author} {\bibfnamefont {E.}~\bibnamefont {Berg}}, \ and\
  \bibinfo {author} {\bibfnamefont {M.}~\bibnamefont {Oshikawa}},\ }\href
  {\doibase 10.1103/PhysRevB.81.064439} {\bibfield  {journal} {\bibinfo
  {journal} {Phys. Rev. B}\ }\textbf {\bibinfo {volume} {81}},\ \bibinfo
  {pages} {064439} (\bibinfo {year} {2010})}\BibitemShut {NoStop}%
\bibitem [{\citenamefont {Turner}\ \emph {et~al.}(2011)\citenamefont {Turner},
  \citenamefont {Pollmann},\ and\ \citenamefont {Berg}}]{turner-2011}%
  \BibitemOpen
  \bibfield  {author} {\bibinfo {author} {\bibfnamefont {A.~M.}\ \bibnamefont
  {Turner}}, \bibinfo {author} {\bibfnamefont {F.}~\bibnamefont {Pollmann}}, \
  and\ \bibinfo {author} {\bibfnamefont {E.}~\bibnamefont {Berg}},\ }\href
  {\doibase 10.1103/PhysRevB.83.075102} {\bibfield  {journal} {\bibinfo
  {journal} {Phys. Rev. B}\ }\textbf {\bibinfo {volume} {83}},\ \bibinfo
  {pages} {075102} (\bibinfo {year} {2011})}\BibitemShut {NoStop}%
\bibitem [{\citenamefont {Bauer}\ \emph {et~al.}(2014)\citenamefont {Bauer},
  \citenamefont {Cincio}, \citenamefont {Keller}, \citenamefont {Dolfi},
  \citenamefont {Vidal}, \citenamefont {Trebst},\ and\ \citenamefont
  {Ludwig}}]{bauer-2014}%
  \BibitemOpen
  \bibfield  {author} {\bibinfo {author} {\bibfnamefont {B.}~\bibnamefont
  {Bauer}}, \bibinfo {author} {\bibfnamefont {L.}~\bibnamefont {Cincio}},
  \bibinfo {author} {\bibfnamefont {B.}~\bibnamefont {Keller}}, \bibinfo
  {author} {\bibfnamefont {M.}~\bibnamefont {Dolfi}}, \bibinfo {author}
  {\bibfnamefont {G.}~\bibnamefont {Vidal}}, \bibinfo {author} {\bibfnamefont
  {S.}~\bibnamefont {Trebst}}, \ and\ \bibinfo {author} {\bibfnamefont
  {A.}~\bibnamefont {Ludwig}},\ }\href {\doibase 10.1038/ncomms6137} {\bibfield
   {journal} {\bibinfo  {journal} {Nature Communications}\ }\textbf {\bibinfo
  {volume} {5}} (\bibinfo {year} {2014}),\ 10.1038/ncomms6137}\BibitemShut
  {NoStop}%
\bibitem [{\citenamefont {Poilblanc}(2010)}]{poilblanc-2010}%
  \BibitemOpen
  \bibfield  {author} {\bibinfo {author} {\bibfnamefont {D.}~\bibnamefont
  {Poilblanc}},\ }\href {\doibase 10.1103/PhysRevLett.105.077202} {\bibfield
  {journal} {\bibinfo  {journal} {Phys. Rev. Lett.}\ }\textbf {\bibinfo
  {volume} {105}},\ \bibinfo {pages} {077202} (\bibinfo {year}
  {2010})}\BibitemShut {NoStop}%
\bibitem [{\citenamefont {Cirac}\ \emph {et~al.}(2011)\citenamefont {Cirac},
  \citenamefont {Poilblanc}, \citenamefont {Schuch},\ and\ \citenamefont
  {Verstraete}}]{cirac-2011}%
  \BibitemOpen
  \bibfield  {author} {\bibinfo {author} {\bibfnamefont {J.~I.}\ \bibnamefont
  {Cirac}}, \bibinfo {author} {\bibfnamefont {D.}~\bibnamefont {Poilblanc}},
  \bibinfo {author} {\bibfnamefont {N.}~\bibnamefont {Schuch}}, \ and\ \bibinfo
  {author} {\bibfnamefont {F.}~\bibnamefont {Verstraete}},\ }\href {\doibase
  10.1103/PhysRevB.83.245134} {\bibfield  {journal} {\bibinfo  {journal} {Phys.
  Rev. B}\ }\textbf {\bibinfo {volume} {83}},\ \bibinfo {pages} {245134}
  (\bibinfo {year} {2011})}\BibitemShut {NoStop}%
\bibitem [{\citenamefont {De~Chiara}\ \emph {et~al.}(2012)\citenamefont
  {De~Chiara}, \citenamefont {Lepori}, \citenamefont {Lewenstein},\ and\
  \citenamefont {Sanpera}}]{de-chiara-2012}%
  \BibitemOpen
  \bibfield  {author} {\bibinfo {author} {\bibfnamefont {G.}~\bibnamefont
  {De~Chiara}}, \bibinfo {author} {\bibfnamefont {L.}~\bibnamefont {Lepori}},
  \bibinfo {author} {\bibfnamefont {M.}~\bibnamefont {Lewenstein}}, \ and\
  \bibinfo {author} {\bibfnamefont {A.}~\bibnamefont {Sanpera}},\ }\href
  {\doibase 10.1103/PhysRevLett.109.237208} {\bibfield  {journal} {\bibinfo
  {journal} {Phys. Rev. Lett.}\ }\textbf {\bibinfo {volume} {109}},\ \bibinfo
  {pages} {237208} (\bibinfo {year} {2012})}\BibitemShut {NoStop}%
\bibitem [{\citenamefont {Alba}\ \emph
  {et~al.}(2012{\natexlab{a}})\citenamefont {Alba}, \citenamefont {Haque},\
  and\ \citenamefont {L\"auchli}}]{alba-2011}%
  \BibitemOpen
  \bibfield  {author} {\bibinfo {author} {\bibfnamefont {V.}~\bibnamefont
  {Alba}}, \bibinfo {author} {\bibfnamefont {M.}~\bibnamefont {Haque}}, \ and\
  \bibinfo {author} {\bibfnamefont {A.~M.}\ \bibnamefont {L\"auchli}},\ }\href
  {\doibase 10.1103/PhysRevLett.108.227201} {\bibfield  {journal} {\bibinfo
  {journal} {Phys. Rev. Lett.}\ }\textbf {\bibinfo {volume} {108}},\ \bibinfo
  {pages} {227201} (\bibinfo {year} {2012}{\natexlab{a}})}\BibitemShut
  {NoStop}%
\bibitem [{\citenamefont {Metlitski}\ and\ \citenamefont
  {Grover}(2011)}]{metlitski-2011}%
  \BibitemOpen
  \bibfield  {author} {\bibinfo {author} {\bibfnamefont {M.~A.}\ \bibnamefont
  {Metlitski}}\ and\ \bibinfo {author} {\bibfnamefont {T.}~\bibnamefont
  {Grover}},\ }\href@noop {} {\enquote {\bibinfo {title} {Entanglement entropy
  of systems with spontaneously broken continuous symmetry},}\ } (\bibinfo
  {year} {2011}),\ \Eprint {http://arxiv.org/abs/1112.5166} {arXiv:1112.5166
  [cond-mat.str-el]} \BibitemShut {NoStop}%
\bibitem [{\citenamefont {Alba}\ \emph
  {et~al.}(2012{\natexlab{b}})\citenamefont {Alba}, \citenamefont {Haque},\
  and\ \citenamefont {L\"{a}uchli}}]{alba-2012}%
  \BibitemOpen
  \bibfield  {author} {\bibinfo {author} {\bibfnamefont {V.}~\bibnamefont
  {Alba}}, \bibinfo {author} {\bibfnamefont {M.}~\bibnamefont {Haque}}, \ and\
  \bibinfo {author} {\bibfnamefont {A.~M.}\ \bibnamefont {L\"{a}uchli}},\
  }\href {\doibase 10.1088/1742-5468/2012/08/p08011} {\bibfield  {journal}
  {\bibinfo  {journal} {Journal of Statistical Mechanics: Theory and
  Experiment}\ }\textbf {\bibinfo {volume} {2012}},\ \bibinfo {pages} {P08011}
  (\bibinfo {year} {2012}{\natexlab{b}})}\BibitemShut {NoStop}%
\bibitem [{\citenamefont {Alba}\ \emph {et~al.}(2013)\citenamefont {Alba},
  \citenamefont {Haque},\ and\ \citenamefont {L\"auchli}}]{Alba13}%
  \BibitemOpen
  \bibfield  {author} {\bibinfo {author} {\bibfnamefont {V.}~\bibnamefont
  {Alba}}, \bibinfo {author} {\bibfnamefont {M.}~\bibnamefont {Haque}}, \ and\
  \bibinfo {author} {\bibfnamefont {A.~M.}\ \bibnamefont {L\"auchli}},\ }\href
  {\doibase 10.1103/PhysRevLett.110.260403} {\bibfield  {journal} {\bibinfo
  {journal} {Phys. Rev. Lett.}\ }\textbf {\bibinfo {volume} {110}},\ \bibinfo
  {pages} {260403} (\bibinfo {year} {2013})}\BibitemShut {NoStop}%
\bibitem [{\citenamefont {Lepori}\ \emph {et~al.}(2013)\citenamefont {Lepori},
  \citenamefont {De~Chiara},\ and\ \citenamefont {Sanpera}}]{lepori-2013}%
  \BibitemOpen
  \bibfield  {author} {\bibinfo {author} {\bibfnamefont {L.}~\bibnamefont
  {Lepori}}, \bibinfo {author} {\bibfnamefont {G.}~\bibnamefont {De~Chiara}}, \
  and\ \bibinfo {author} {\bibfnamefont {A.}~\bibnamefont {Sanpera}},\ }\href
  {\doibase 10.1103/PhysRevB.87.235107} {\bibfield  {journal} {\bibinfo
  {journal} {Phys. Rev. B}\ }\textbf {\bibinfo {volume} {87}},\ \bibinfo
  {pages} {235107} (\bibinfo {year} {2013})}\BibitemShut {NoStop}%
\bibitem [{\citenamefont {James}\ and\ \citenamefont
  {Konik}(2013)}]{james-2013}%
  \BibitemOpen
  \bibfield  {author} {\bibinfo {author} {\bibfnamefont {A.~J.~A.}\
  \bibnamefont {James}}\ and\ \bibinfo {author} {\bibfnamefont {R.~M.}\
  \bibnamefont {Konik}},\ }\href {\doibase 10.1103/PhysRevB.87.241103}
  {\bibfield  {journal} {\bibinfo  {journal} {Phys. Rev. B}\ }\textbf {\bibinfo
  {volume} {87}},\ \bibinfo {pages} {241103} (\bibinfo {year}
  {2013})}\BibitemShut {NoStop}%
\bibitem [{\citenamefont {Kolley}\ \emph {et~al.}(2013)\citenamefont {Kolley},
  \citenamefont {Depenbrock}, \citenamefont {McCulloch}, \citenamefont
  {Schollw\"ock},\ and\ \citenamefont {Alba}}]{kolley-2013}%
  \BibitemOpen
  \bibfield  {author} {\bibinfo {author} {\bibfnamefont {F.}~\bibnamefont
  {Kolley}}, \bibinfo {author} {\bibfnamefont {S.}~\bibnamefont {Depenbrock}},
  \bibinfo {author} {\bibfnamefont {I.~P.}\ \bibnamefont {McCulloch}}, \bibinfo
  {author} {\bibfnamefont {U.}~\bibnamefont {Schollw\"ock}}, \ and\ \bibinfo
  {author} {\bibfnamefont {V.}~\bibnamefont {Alba}},\ }\href {\doibase
  10.1103/PhysRevB.88.144426} {\bibfield  {journal} {\bibinfo  {journal} {Phys.
  Rev. B}\ }\textbf {\bibinfo {volume} {88}},\ \bibinfo {pages} {144426}
  (\bibinfo {year} {2013})}\BibitemShut {NoStop}%
\bibitem [{\citenamefont {Rademaker}(2015)}]{rademaker-2015}%
  \BibitemOpen
  \bibfield  {author} {\bibinfo {author} {\bibfnamefont {L.}~\bibnamefont
  {Rademaker}},\ }\href {\doibase 10.1103/PhysRevB.92.144419} {\bibfield
  {journal} {\bibinfo  {journal} {Phys. Rev. B}\ }\textbf {\bibinfo {volume}
  {92}},\ \bibinfo {pages} {144419} (\bibinfo {year} {2015})}\BibitemShut
  {NoStop}%
\bibitem [{\citenamefont {Kolley}\ \emph {et~al.}(2015)\citenamefont {Kolley},
  \citenamefont {Depenbrock}, \citenamefont {McCulloch}, \citenamefont
  {Schollw\"ock},\ and\ \citenamefont {Alba}}]{kolley-2015}%
  \BibitemOpen
  \bibfield  {author} {\bibinfo {author} {\bibfnamefont {F.}~\bibnamefont
  {Kolley}}, \bibinfo {author} {\bibfnamefont {S.}~\bibnamefont {Depenbrock}},
  \bibinfo {author} {\bibfnamefont {I.~P.}\ \bibnamefont {McCulloch}}, \bibinfo
  {author} {\bibfnamefont {U.}~\bibnamefont {Schollw\"ock}}, \ and\ \bibinfo
  {author} {\bibfnamefont {V.}~\bibnamefont {Alba}},\ }\href {\doibase
  10.1103/PhysRevB.91.104418} {\bibfield  {journal} {\bibinfo  {journal} {Phys.
  Rev. B}\ }\textbf {\bibinfo {volume} {91}},\ \bibinfo {pages} {104418}
  (\bibinfo {year} {2015})}\BibitemShut {NoStop}%
\bibitem [{\citenamefont {Fr\'erot}\ and\ \citenamefont
  {Roscilde}(2016)}]{frerot-2016}%
  \BibitemOpen
  \bibfield  {author} {\bibinfo {author} {\bibfnamefont {I.}~\bibnamefont
  {Fr\'erot}}\ and\ \bibinfo {author} {\bibfnamefont {T.}~\bibnamefont
  {Roscilde}},\ }\href {\doibase 10.1103/PhysRevLett.116.190401} {\bibfield
  {journal} {\bibinfo  {journal} {Phys. Rev. Lett.}\ }\textbf {\bibinfo
  {volume} {116}},\ \bibinfo {pages} {190401} (\bibinfo {year}
  {2016})}\BibitemShut {NoStop}%
\bibitem [{\citenamefont {Calabrese}\ and\ \citenamefont
  {Lefevre}(2008)}]{calabrese-2008}%
  \BibitemOpen
  \bibfield  {author} {\bibinfo {author} {\bibfnamefont {P.}~\bibnamefont
  {Calabrese}}\ and\ \bibinfo {author} {\bibfnamefont {A.}~\bibnamefont
  {Lefevre}},\ }\href {\doibase 10.1103/PhysRevA.78.032329} {\bibfield
  {journal} {\bibinfo  {journal} {Phys. Rev. A}\ }\textbf {\bibinfo {volume}
  {78}},\ \bibinfo {pages} {032329} (\bibinfo {year} {2008})}\BibitemShut
  {NoStop}%
\bibitem [{\citenamefont {L\"auchli}(2013)}]{lauchli-2013}%
  \BibitemOpen
  \bibfield  {author} {\bibinfo {author} {\bibfnamefont {A.~M.}\ \bibnamefont
  {L\"auchli}},\ }\href@noop {} {} (\bibinfo {year} {2013}),\ \Eprint
  {http://arxiv.org/abs/1303.0741} {arXiv:1303.0741 [cond-mat.stat-mech]}
  \BibitemShut {NoStop}%
\bibitem [{\citenamefont {Alba}\ \emph {et~al.}(2017)\citenamefont {Alba},
  \citenamefont {Calabrese},\ and\ \citenamefont {Tonni}}]{Alba-2017}%
  \BibitemOpen
  \bibfield  {author} {\bibinfo {author} {\bibfnamefont {V.}~\bibnamefont
  {Alba}}, \bibinfo {author} {\bibfnamefont {P.}~\bibnamefont {Calabrese}}, \
  and\ \bibinfo {author} {\bibfnamefont {E.}~\bibnamefont {Tonni}},\ }\href
  {\doibase 10.1088/1751-8121/aa9365} {\bibfield  {journal} {\bibinfo
  {journal} {Journal of Physics A: Mathematical and Theoretical}\ }\textbf
  {\bibinfo {volume} {51}},\ \bibinfo {pages} {024001} (\bibinfo {year}
  {2017})}\BibitemShut {NoStop}%
\bibitem [{\citenamefont {Cardy}(2015)}]{cardy-talk}%
  \BibitemOpen
  \bibfield  {author} {\bibinfo {author} {\bibfnamefont {J.}~\bibnamefont
  {Cardy}},\ }\href {http://online.kitp.ucsb.edu/online/entangled-c15/cardy/}
  {\enquote {\bibinfo {title} {The entanglement gap in cfts, talk at the kitp
  conference "closing the entanglement gap: Quantum information, quantum
  matter, and quantum fields".}}\ } (\bibinfo {year} {2015})\BibitemShut
  {NoStop}%
\bibitem [{\citenamefont {Bayat}\ \emph {et~al.}(2014)\citenamefont {Bayat},
  \citenamefont {Johannesson}, \citenamefont {Bose},\ and\ \citenamefont
  {Sodano}}]{bayat-2014}%
  \BibitemOpen
  \bibfield  {author} {\bibinfo {author} {\bibfnamefont {A.}~\bibnamefont
  {Bayat}}, \bibinfo {author} {\bibfnamefont {H.}~\bibnamefont {Johannesson}},
  \bibinfo {author} {\bibfnamefont {S.}~\bibnamefont {Bose}}, \ and\ \bibinfo
  {author} {\bibfnamefont {P.}~\bibnamefont {Sodano}},\ }\href {\doibase
  10.1038/ncomms4784} {\bibfield  {journal} {\bibinfo  {journal} {Nature
  Communications}\ }\textbf {\bibinfo {volume} {5}} (\bibinfo {year} {2014}),\
  10.1038/ncomms4784}\BibitemShut {NoStop}%
\bibitem [{\citenamefont {Truong}\ and\ \citenamefont
  {Peschel}(1989)}]{truong-1987}%
  \BibitemOpen
  \bibfield  {author} {\bibinfo {author} {\bibfnamefont {T.~T.}\ \bibnamefont
  {Truong}}\ and\ \bibinfo {author} {\bibfnamefont {I.}~\bibnamefont
  {Peschel}},\ }\href {\doibase 10.1007/bf01313574} {\bibfield  {journal}
  {\bibinfo  {journal} {Zeitschrift f\"ur Physik B Condensed Matter}\ }\textbf
  {\bibinfo {volume} {75}},\ \bibinfo {pages} {119} (\bibinfo {year}
  {1989})}\BibitemShut {NoStop}%
\bibitem [{\citenamefont {Giulio}\ and\ \citenamefont
  {Tonni}(2020)}]{di-giulio-2020}%
  \BibitemOpen
  \bibfield  {author} {\bibinfo {author} {\bibfnamefont {G.~D.}\ \bibnamefont
  {Giulio}}\ and\ \bibinfo {author} {\bibfnamefont {E.}~\bibnamefont {Tonni}},\
  }\href {\doibase 10.1088/1742-5468/ab7129} {\bibfield  {journal} {\bibinfo
  {journal} {Journal of Statistical Mechanics: Theory and Experiment}\ }\textbf
  {\bibinfo {volume} {2020}},\ \bibinfo {pages} {033102} (\bibinfo {year}
  {2020})}\BibitemShut {NoStop}%
\bibitem [{\citenamefont {Lundgren}\ \emph {et~al.}(2014)\citenamefont
  {Lundgren}, \citenamefont {Blair}, \citenamefont {Greiter}, \citenamefont
  {L\"auchli}, \citenamefont {Fiete},\ and\ \citenamefont
  {Thomale}}]{lundgren-2014}%
  \BibitemOpen
  \bibfield  {author} {\bibinfo {author} {\bibfnamefont {R.}~\bibnamefont
  {Lundgren}}, \bibinfo {author} {\bibfnamefont {J.}~\bibnamefont {Blair}},
  \bibinfo {author} {\bibfnamefont {M.}~\bibnamefont {Greiter}}, \bibinfo
  {author} {\bibfnamefont {A.}~\bibnamefont {L\"auchli}}, \bibinfo {author}
  {\bibfnamefont {G.~A.}\ \bibnamefont {Fiete}}, \ and\ \bibinfo {author}
  {\bibfnamefont {R.}~\bibnamefont {Thomale}},\ }\href {\doibase
  10.1103/PhysRevLett.113.256404} {\bibfield  {journal} {\bibinfo  {journal}
  {Phys. Rev. Lett.}\ }\textbf {\bibinfo {volume} {113}},\ \bibinfo {pages}
  {256404} (\bibinfo {year} {2014})}\BibitemShut {NoStop}%
\bibitem [{\citenamefont {Lhuillier}\ and\ \citenamefont
  {Misguich}(2002)}]{lhullier}%
  \BibitemOpen
  \bibfield  {author} {\bibinfo {author} {\bibfnamefont {C.}~\bibnamefont
  {Lhuillier}}\ and\ \bibinfo {author} {\bibfnamefont {G.}~\bibnamefont
  {Misguich}},\ }in\ \href {\doibase 10.1007/3-540-45649-x_6} {\emph {\bibinfo
  {booktitle} {High Magnetic Fields}}}\ (\bibinfo  {publisher} {Springer Berlin
  Heidelberg},\ \bibinfo {year} {2002})\ pp.\ \bibinfo {pages}
  {161--190}\BibitemShut {NoStop}%
\bibitem [{\citenamefont {Beekman}\ \emph {et~al.}(2019)\citenamefont
  {Beekman}, \citenamefont {Rademaker},\ and\ \citenamefont {van
  Wezel}}]{beekman-2020}%
  \BibitemOpen
  \bibfield  {author} {\bibinfo {author} {\bibfnamefont {A.~J.}\ \bibnamefont
  {Beekman}}, \bibinfo {author} {\bibfnamefont {L.}~\bibnamefont {Rademaker}},
  \ and\ \bibinfo {author} {\bibfnamefont {J.}~\bibnamefont {van Wezel}},\
  }\href {\doibase 10.21468/SciPostPhysLectNotes.11} {\bibfield  {journal}
  {\bibinfo  {journal} {SciPost Phys. Lect. Notes}\ ,\ \bibinfo {pages} {11}}
  (\bibinfo {year} {2019})}\BibitemShut {NoStop}%
\bibitem [{\citenamefont {Wietek}\ \emph {et~al.}(2017)\citenamefont {Wietek},
  \citenamefont {Schuler},\ and\ \citenamefont {Läuchli}}]{Wietek-2017}%
  \BibitemOpen
  \bibfield  {author} {\bibinfo {author} {\bibfnamefont {A.}~\bibnamefont
  {Wietek}}, \bibinfo {author} {\bibfnamefont {M.}~\bibnamefont {Schuler}}, \
  and\ \bibinfo {author} {\bibfnamefont {A.~M.}\ \bibnamefont {Läuchli}},\
  }\href@noop {} {\enquote {\bibinfo {title} {Studying continuous symmetry
  breaking using energy level spectroscopy},}\ } (\bibinfo {year} {2017}),\
  \Eprint {http://arxiv.org/abs/1704.08622} {arXiv:1704.08622
  [cond-mat.str-el]} \BibitemShut {NoStop}%
\bibitem [{\citenamefont {Obermair}(1972)}]{Ober72}%
  \BibitemOpen
  \bibfield  {author} {\bibinfo {author} {\bibfnamefont {G.}~\bibnamefont
  {Obermair}},\ }in\ \href@noop {} {\emph {\bibinfo {booktitle} {Dynamical
  Aspects of critical phenomena}}}\ (\bibinfo  {publisher} {Gordon and
  Breach},\ \bibinfo {address} {New York},\ \bibinfo {year} {1972})\
  p.~\bibinfo {pages} {10}\BibitemShut {NoStop}%
\bibitem [{\citenamefont {Henkel}\ and\ \citenamefont {Hoeger}(1984)}]{Henk84}%
  \BibitemOpen
  \bibfield  {author} {\bibinfo {author} {\bibfnamefont {M.}~\bibnamefont
  {Henkel}}\ and\ \bibinfo {author} {\bibfnamefont {C.}~\bibnamefont
  {Hoeger}},\ }\href {\doibase 10.1007/BF01307503} {\bibfield  {journal}
  {\bibinfo  {journal} {Zeitschrift f{\"{u}}r Physik B Condensed Matter}\
  }\textbf {\bibinfo {volume} {55}},\ \bibinfo {pages} {67} (\bibinfo {year}
  {1984})}\BibitemShut {NoStop}%
\bibitem [{\citenamefont {Vojta}(1996)}]{Vojta96}%
  \BibitemOpen
  \bibfield  {author} {\bibinfo {author} {\bibfnamefont {T.}~\bibnamefont
  {Vojta}},\ }\href {\doibase 10.1103/PhysRevB.53.710} {\bibfield  {journal}
  {\bibinfo  {journal} {Physical Review B}\ }\textbf {\bibinfo {volume} {53}},\
  \bibinfo {pages} {710} (\bibinfo {year} {1996})}\BibitemShut {NoStop}%
\bibitem [{\citenamefont {Wald}\ and\ \citenamefont {Henkel}(2015)}]{Wald15}%
  \BibitemOpen
  \bibfield  {author} {\bibinfo {author} {\bibfnamefont {S.}~\bibnamefont
  {Wald}}\ and\ \bibinfo {author} {\bibfnamefont {M.}~\bibnamefont {Henkel}},\
  }\href {\doibase 10.1088/1742-5468/2015/07/P07006} {\bibfield  {journal}
  {\bibinfo  {journal} {Journal of Statistical Mechanics: Theory and
  Experiment}\ }\textbf {\bibinfo {volume} {07006}},\ \bibinfo {pages} {34}
  (\bibinfo {year} {2015})},\ \Eprint {http://arxiv.org/abs/1503.06713}
  {arXiv:1503.06713} \BibitemShut {NoStop}%
\bibitem [{\citenamefont {Bienzobaz}\ and\ \citenamefont
  {Salinas}(2012)}]{Bien12}%
  \BibitemOpen
  \bibfield  {author} {\bibinfo {author} {\bibfnamefont {P.}~\bibnamefont
  {Bienzobaz}}\ and\ \bibinfo {author} {\bibfnamefont {S.}~\bibnamefont
  {Salinas}},\ }\href {\doibase https://doi.org/10.1016/j.physa.2012.07.027}
  {\bibfield  {journal} {\bibinfo  {journal} {Physica A: Statistical Mechanics
  and its Applications}\ }\textbf {\bibinfo {volume} {391}},\ \bibinfo {pages}
  {6399 } (\bibinfo {year} {2012})}\BibitemShut {NoStop}%
\bibitem [{\citenamefont {Br{\'{e}}zin}(1982)}]{brezin-1982}%
  \BibitemOpen
  \bibfield  {author} {\bibinfo {author} {\bibfnamefont {E.}~\bibnamefont
  {Br{\'{e}}zin}},\ }\href {\doibase 10.1051/jphys:0198200430101500} {\bibfield
   {journal} {\bibinfo  {journal} {Journal de Physique}\ }\textbf {\bibinfo
  {volume} {43}},\ \bibinfo {pages} {15} (\bibinfo {year} {1982})}\BibitemShut
  {NoStop}%
\bibitem [{\citenamefont {Zinn-Justin}(1998)}]{zinn-justin-1998}%
  \BibitemOpen
  \bibfield  {author} {\bibinfo {author} {\bibfnamefont {J.}~\bibnamefont
  {Zinn-Justin}},\ }\href@noop {} {\enquote {\bibinfo {title} {Vector models in
  the large $n$ limit: a few applications},}\ } (\bibinfo {year} {1998}),\
  \Eprint {http://arxiv.org/abs/hep-th/9810198} {arXiv:hep-th/9810198 [hep-th]}
  \BibitemShut {NoStop}%
\bibitem [{\citenamefont {Stanley}(1968)}]{Stan68}%
  \BibitemOpen
  \bibfield  {author} {\bibinfo {author} {\bibfnamefont {H.~E.}\ \bibnamefont
  {Stanley}},\ }\href {\doibase 10.1103/PhysRev.176.718} {\bibfield  {journal}
  {\bibinfo  {journal} {Phys. Rev.}\ }\textbf {\bibinfo {volume} {176}},\
  \bibinfo {pages} {718} (\bibinfo {year} {1968})}\BibitemShut {NoStop}%
\bibitem [{\citenamefont {Lu}\ and\ \citenamefont
  {Grover}(2019{\natexlab{a}})}]{Lu19}%
  \BibitemOpen
  \bibfield  {author} {\bibinfo {author} {\bibfnamefont {T.-C.}\ \bibnamefont
  {Lu}}\ and\ \bibinfo {author} {\bibfnamefont {T.}~\bibnamefont {Grover}},\
  }\href {\doibase 10.1103/PhysRevB.99.075157} {\bibfield  {journal} {\bibinfo
  {journal} {Physical Review B}\ }\textbf {\bibinfo {volume} {99}},\ \bibinfo
  {pages} {075157} (\bibinfo {year} {2019}{\natexlab{a}})},\ \Eprint
  {http://arxiv.org/abs/1808.04381} {arXiv:1808.04381} \BibitemShut {NoStop}%
\bibitem [{\citenamefont {Lu}\ and\ \citenamefont
  {Grover}(2019{\natexlab{b}})}]{Lu20}%
  \BibitemOpen
  \bibfield  {author} {\bibinfo {author} {\bibfnamefont {T.-C.}\ \bibnamefont
  {Lu}}\ and\ \bibinfo {author} {\bibfnamefont {T.}~\bibnamefont {Grover}},\
  }\href@noop {} {\enquote {\bibinfo {title} {Structure of quantum entanglement
  at a finite temperature critical point},}\ } (\bibinfo {year}
  {2019}{\natexlab{b}}),\ \Eprint {http://arxiv.org/abs/1907.01569}
  {arXiv:1907.01569 [cond-mat.str-el]} \BibitemShut {NoStop}%
\bibitem [{\citenamefont {Wald}\ \emph {et~al.}(2020)\citenamefont {Wald},
  \citenamefont {Arias},\ and\ \citenamefont {Alba}}]{Wald20}%
  \BibitemOpen
  \bibfield  {author} {\bibinfo {author} {\bibfnamefont {S.}~\bibnamefont
  {Wald}}, \bibinfo {author} {\bibfnamefont {R.}~\bibnamefont {Arias}}, \ and\
  \bibinfo {author} {\bibfnamefont {V.}~\bibnamefont {Alba}},\ }\href {\doibase
  10.1088/1742-5468/ab6b19} {\bibfield  {journal} {\bibinfo  {journal} {Journal
  of Statistical Mechanics: Theory and Experiment}\ }\textbf {\bibinfo {volume}
  {2020}},\ \bibinfo {pages} {033105} (\bibinfo {year} {2020})}\BibitemShut
  {NoStop}%
\bibitem [{\citenamefont {Casini}\ \emph {et~al.}(2009)\citenamefont {Casini},
  \citenamefont {Huerta},\ and\ \citenamefont {Leitao}}]{Casini:2008as}%
  \BibitemOpen
  \bibfield  {author} {\bibinfo {author} {\bibfnamefont {H.}~\bibnamefont
  {Casini}}, \bibinfo {author} {\bibfnamefont {M.}~\bibnamefont {Huerta}}, \
  and\ \bibinfo {author} {\bibfnamefont {L.}~\bibnamefont {Leitao}},\ }\href
  {\doibase 10.1016/j.nuclphysb.2009.02.003} {\bibfield  {journal} {\bibinfo
  {journal} {Nuclear Physics B}\ }\textbf {\bibinfo {volume} {814}},\ \bibinfo
  {pages} {594} (\bibinfo {year} {2009})}\BibitemShut {NoStop}%
\bibitem [{\citenamefont {Casini}\ and\ \citenamefont
  {Huerta}(2007)}]{Casini:2006hu}%
  \BibitemOpen
  \bibfield  {author} {\bibinfo {author} {\bibfnamefont {H.}~\bibnamefont
  {Casini}}\ and\ \bibinfo {author} {\bibfnamefont {M.}~\bibnamefont
  {Huerta}},\ }\href {\doibase 10.1016/j.nuclphysb.2006.12.012} {\bibfield
  {journal} {\bibinfo  {journal} {Nuclear Physics B}\ }\textbf {\bibinfo
  {volume} {764}},\ \bibinfo {pages} {183} (\bibinfo {year}
  {2007})}\BibitemShut {NoStop}%
\bibitem [{\citenamefont {Kallin}\ \emph {et~al.}(2013)\citenamefont {Kallin},
  \citenamefont {Hyatt}, \citenamefont {Singh},\ and\ \citenamefont
  {Melko}}]{PhysRevLett.110.135702}%
  \BibitemOpen
  \bibfield  {author} {\bibinfo {author} {\bibfnamefont {A.~B.}\ \bibnamefont
  {Kallin}}, \bibinfo {author} {\bibfnamefont {K.}~\bibnamefont {Hyatt}},
  \bibinfo {author} {\bibfnamefont {R.~R.~P.}\ \bibnamefont {Singh}}, \ and\
  \bibinfo {author} {\bibfnamefont {R.~G.}\ \bibnamefont {Melko}},\ }\href
  {\doibase 10.1103/PhysRevLett.110.135702} {\bibfield  {journal} {\bibinfo
  {journal} {Phys. Rev. Lett.}\ }\textbf {\bibinfo {volume} {110}},\ \bibinfo
  {pages} {135702} (\bibinfo {year} {2013})}\BibitemShut {NoStop}%
\bibitem [{\citenamefont {Stoudenmire}\ \emph {et~al.}(2014)\citenamefont
  {Stoudenmire}, \citenamefont {Gustainis}, \citenamefont {Johal},
  \citenamefont {Wessel},\ and\ \citenamefont {Melko}}]{pitch}%
  \BibitemOpen
  \bibfield  {author} {\bibinfo {author} {\bibfnamefont {E.~M.}\ \bibnamefont
  {Stoudenmire}}, \bibinfo {author} {\bibfnamefont {P.}~\bibnamefont
  {Gustainis}}, \bibinfo {author} {\bibfnamefont {R.}~\bibnamefont {Johal}},
  \bibinfo {author} {\bibfnamefont {S.}~\bibnamefont {Wessel}}, \ and\ \bibinfo
  {author} {\bibfnamefont {R.~G.}\ \bibnamefont {Melko}},\ }\href {\doibase
  10.1103/PhysRevB.90.235106} {\bibfield  {journal} {\bibinfo  {journal} {Phys.
  Rev. B}\ }\textbf {\bibinfo {volume} {90}},\ \bibinfo {pages} {235106}
  (\bibinfo {year} {2014})}\BibitemShut {NoStop}%
\bibitem [{\citenamefont {Kallin}\ \emph {et~al.}(2014)\citenamefont {Kallin},
  \citenamefont {Stoudenmire}, \citenamefont {Fendley}, \citenamefont {Singh},\
  and\ \citenamefont {Melko}}]{2014arXiv1401.3504K}%
  \BibitemOpen
  \bibfield  {author} {\bibinfo {author} {\bibfnamefont {A.~B.}\ \bibnamefont
  {Kallin}}, \bibinfo {author} {\bibfnamefont {E.~M.}\ \bibnamefont
  {Stoudenmire}}, \bibinfo {author} {\bibfnamefont {P.}~\bibnamefont
  {Fendley}}, \bibinfo {author} {\bibfnamefont {R.~R.~P.}\ \bibnamefont
  {Singh}}, \ and\ \bibinfo {author} {\bibfnamefont {R.~G.}\ \bibnamefont
  {Melko}},\ }\href {\doibase 10.1088/1742-5468/2014/06/p06009} {\bibfield
  {journal} {\bibinfo  {journal} {Journal of Statistical Mechanics: Theory and
  Experiment}\ }\textbf {\bibinfo {volume} {2014}},\ \bibinfo {pages} {P06009}
  (\bibinfo {year} {2014})}\BibitemShut {NoStop}%
\bibitem [{\citenamefont {Singh}\ \emph {et~al.}(2012)\citenamefont {Singh},
  \citenamefont {Melko},\ and\ \citenamefont {Oitmaa}}]{Singh2012}%
  \BibitemOpen
  \bibfield  {author} {\bibinfo {author} {\bibfnamefont {R.~R.~P.}\
  \bibnamefont {Singh}}, \bibinfo {author} {\bibfnamefont {R.~G.}\ \bibnamefont
  {Melko}}, \ and\ \bibinfo {author} {\bibfnamefont {J.}~\bibnamefont
  {Oitmaa}},\ }\href {\doibase 10.1103/PhysRevB.86.075106} {\bibfield
  {journal} {\bibinfo  {journal} {Phys. Rev. B}\ }\textbf {\bibinfo {volume}
  {86}},\ \bibinfo {pages} {075106} (\bibinfo {year} {2012})}\BibitemShut
  {NoStop}%
\bibitem [{\citenamefont {Helmes}\ and\ \citenamefont
  {Wessel}(2015)}]{Helmes2014}%
  \BibitemOpen
  \bibfield  {author} {\bibinfo {author} {\bibfnamefont {J.}~\bibnamefont
  {Helmes}}\ and\ \bibinfo {author} {\bibfnamefont {S.}~\bibnamefont
  {Wessel}},\ }\href {\doibase 10.1103/PhysRevB.92.125120} {\bibfield
  {journal} {\bibinfo  {journal} {Phys. Rev. B}\ }\textbf {\bibinfo {volume}
  {92}},\ \bibinfo {pages} {125120} (\bibinfo {year} {2015})}\BibitemShut
  {NoStop}%
\bibitem [{\citenamefont {Seminara}\ \emph {et~al.}(2017)\citenamefont
  {Seminara}, \citenamefont {Sisti},\ and\ \citenamefont
  {Tonni}}]{Seminara-2017}%
  \BibitemOpen
  \bibfield  {author} {\bibinfo {author} {\bibfnamefont {D.}~\bibnamefont
  {Seminara}}, \bibinfo {author} {\bibfnamefont {J.}~\bibnamefont {Sisti}}, \
  and\ \bibinfo {author} {\bibfnamefont {E.}~\bibnamefont {Tonni}},\ }\href
  {\doibase 10.1007/JHEP11(2017)076} {\bibfield  {journal} {\bibinfo  {journal}
  {Journal of High Energy Physics}\ }\textbf {\bibinfo {volume} {2017}},\
  \bibinfo {pages} {76} (\bibinfo {year} {2017})}\BibitemShut {NoStop}%
\bibitem [{\citenamefont {Berlin}\ and\ \citenamefont {Kac}(1952)}]{Berl52}%
  \BibitemOpen
  \bibfield  {author} {\bibinfo {author} {\bibfnamefont {T.~H.}\ \bibnamefont
  {Berlin}}\ and\ \bibinfo {author} {\bibfnamefont {M.}~\bibnamefont {Kac}},\
  }\href {\doibase 10.1103/PhysRev.86.821} {\bibfield  {journal} {\bibinfo
  {journal} {Physical Review}\ }\textbf {\bibinfo {volume} {86}},\ \bibinfo
  {pages} {821} (\bibinfo {year} {1952})}\BibitemShut {NoStop}%
\bibitem [{\citenamefont {Lewis}\ and\ \citenamefont {Wannier}(1952)}]{Lew52}%
  \BibitemOpen
  \bibfield  {author} {\bibinfo {author} {\bibfnamefont {H.~W.}\ \bibnamefont
  {Lewis}}\ and\ \bibinfo {author} {\bibfnamefont {G.~H.}\ \bibnamefont
  {Wannier}},\ }\href {\doibase 10.1103/PhysRev.88.682.2} {\bibfield  {journal}
  {\bibinfo  {journal} {Physical Review}\ }\textbf {\bibinfo {volume} {88}},\
  \bibinfo {pages} {682} (\bibinfo {year} {1952})}\BibitemShut {NoStop}%
\bibitem [{\citenamefont {Amit}(1984)}]{Amit84}%
  \BibitemOpen
  \bibfield  {author} {\bibinfo {author} {\bibfnamefont {D.}~\bibnamefont
  {Amit}},\ }\href {https://books.google.de/books?id=M4yqQgAACAAJ} {\emph
  {\bibinfo {title} {Field Theory, the Renormalization Group, and Critical
  Phenomena}}},\ International series in pure and applied physics\ (\bibinfo
  {publisher} {World Scientific},\ \bibinfo {year} {1984})\BibitemShut
  {NoStop}%
\bibitem [{\citenamefont {Botero}\ and\ \citenamefont
  {Reznik}(2004)}]{Botero04}%
  \BibitemOpen
  \bibfield  {author} {\bibinfo {author} {\bibfnamefont {A.}~\bibnamefont
  {Botero}}\ and\ \bibinfo {author} {\bibfnamefont {B.}~\bibnamefont
  {Reznik}},\ }\href {\doibase 10.1103/PhysRevA.70.052329} {\bibfield
  {journal} {\bibinfo  {journal} {Phys. Rev. A}\ }\textbf {\bibinfo {volume}
  {70}},\ \bibinfo {pages} {052329} (\bibinfo {year} {2004})}\BibitemShut
  {NoStop}%
\bibitem [{\citenamefont {Abramowitz}\ and\ \citenamefont
  {Stegun}(1964)}]{Abra65}%
  \BibitemOpen
  \bibfield  {author} {\bibinfo {author} {\bibfnamefont {M.}~\bibnamefont
  {Abramowitz}}\ and\ \bibinfo {author} {\bibfnamefont {I.~A.}\ \bibnamefont
  {Stegun}},\ }\href@noop {} {\emph {\bibinfo {title} {Handbook of Mathematical
  Functions}}},\ \bibinfo {edition} {10th}\ ed.\ (\bibinfo  {publisher}
  {Dover},\ \bibinfo {address} {New York},\ \bibinfo {year} {1964})\BibitemShut
  {NoStop}%
\bibitem [{\citenamefont {Singh}\ and\ \citenamefont
  {Pathria}(1987)}]{singh-1987}%
  \BibitemOpen
  \bibfield  {author} {\bibinfo {author} {\bibfnamefont {S.}~\bibnamefont
  {Singh}}\ and\ \bibinfo {author} {\bibfnamefont {R.~K.}\ \bibnamefont
  {Pathria}},\ }\href {\doibase 10.1103/physrevb.36.3769} {\bibfield  {journal}
  {\bibinfo  {journal} {Physical Review B}\ }\textbf {\bibinfo {volume} {36}},\
  \bibinfo {pages} {3769} (\bibinfo {year} {1987})}\BibitemShut {NoStop}%
\bibitem [{\citenamefont {Chamati}\ \emph {et~al.}(1998)\citenamefont
  {Chamati}, \citenamefont {Pisanova},\ and\ \citenamefont
  {Tonchev}}]{chamati-1997}%
  \BibitemOpen
  \bibfield  {author} {\bibinfo {author} {\bibfnamefont {H.}~\bibnamefont
  {Chamati}}, \bibinfo {author} {\bibfnamefont {E.~S.}\ \bibnamefont
  {Pisanova}}, \ and\ \bibinfo {author} {\bibfnamefont {N.~S.}\ \bibnamefont
  {Tonchev}},\ }\href {\doibase 10.1103/PhysRevB.57.5798} {\bibfield  {journal}
  {\bibinfo  {journal} {Phys. Rev. B}\ }\textbf {\bibinfo {volume} {57}},\
  \bibinfo {pages} {5798} (\bibinfo {year} {1998})}\BibitemShut {NoStop}%
\bibitem [{\citenamefont {Caracciolo}\ and\ \citenamefont
  {Pelissetto}(1998)}]{caracciolo-1998}%
  \BibitemOpen
  \bibfield  {author} {\bibinfo {author} {\bibfnamefont {S.}~\bibnamefont
  {Caracciolo}}\ and\ \bibinfo {author} {\bibfnamefont {A.}~\bibnamefont
  {Pelissetto}},\ }\href {\doibase 10.1103/PhysRevD.58.105007} {\bibfield
  {journal} {\bibinfo  {journal} {Phys. Rev. D}\ }\textbf {\bibinfo {volume}
  {58}},\ \bibinfo {pages} {105007} (\bibinfo {year} {1998})}\BibitemShut
  {NoStop}%
\bibitem [{\citenamefont {Chamati}\ and\ \citenamefont
  {Tonchev}(2000)}]{chamati-2000}%
  \BibitemOpen
  \bibfield  {author} {\bibinfo {author} {\bibfnamefont {H.}~\bibnamefont
  {Chamati}}\ and\ \bibinfo {author} {\bibfnamefont {N.~S.}\ \bibnamefont
  {Tonchev}},\ }\href {\doibase 10.1088/0305-4470/33/5/305} {\bibfield
  {journal} {\bibinfo  {journal} {Journal of Physics A: Mathematical and
  General}\ }\textbf {\bibinfo {volume} {33}},\ \bibinfo {pages} {873}
  (\bibinfo {year} {2000})}\BibitemShut {NoStop}%
\bibitem [{\citenamefont {Caracciolo}\ \emph {et~al.}(2001)\citenamefont
  {Caracciolo}, \citenamefont {Gambassi}, \citenamefont {Gubinelli},\ and\
  \citenamefont {Pelissetto}}]{caracciolo-2001}%
  \BibitemOpen
  \bibfield  {author} {\bibinfo {author} {\bibfnamefont {S.}~\bibnamefont
  {Caracciolo}}, \bibinfo {author} {\bibfnamefont {A.}~\bibnamefont
  {Gambassi}}, \bibinfo {author} {\bibfnamefont {M.}~\bibnamefont {Gubinelli}},
  \ and\ \bibinfo {author} {\bibfnamefont {A.}~\bibnamefont {Pelissetto}},\
  }\href {\doibase 10.1007/BF01352587} {\bibfield  {journal} {\bibinfo
  {journal} {The European Physical Journal B - Condensed Matter and Complex
  Systems}\ }\textbf {\bibinfo {volume} {20}},\ \bibinfo {pages} {255}
  (\bibinfo {year} {2001})}\BibitemShut {NoStop}%
\bibitem [{\citenamefont {Casini}\ and\ \citenamefont
  {Huerta}(2009)}]{Casini:2009sr}%
  \BibitemOpen
  \bibfield  {author} {\bibinfo {author} {\bibfnamefont {H.}~\bibnamefont
  {Casini}}\ and\ \bibinfo {author} {\bibfnamefont {M.}~\bibnamefont
  {Huerta}},\ }\href {\doibase 10.1088/1751-8113/42/50/504007} {\bibfield
  {journal} {\bibinfo  {journal} {Journal of Physics A: Mathematical and
  Theoretical}\ }\textbf {\bibinfo {volume} {42}},\ \bibinfo {pages} {504007}
  (\bibinfo {year} {2009})}\BibitemShut {NoStop}%
\bibitem [{\citenamefont {Murciano}\ \emph
  {et~al.}(2020{\natexlab{a}})\citenamefont {Murciano}, \citenamefont
  {Ruggiero},\ and\ \citenamefont {Calabrese}}]{sara2D}%
  \BibitemOpen
  \bibfield  {author} {\bibinfo {author} {\bibfnamefont {S.}~\bibnamefont
  {Murciano}}, \bibinfo {author} {\bibfnamefont {P.}~\bibnamefont {Ruggiero}},
  \ and\ \bibinfo {author} {\bibfnamefont {P.}~\bibnamefont {Calabrese}},\
  }\href {\doibase 10.1088/1742-5468/aba1e5} {\bibfield  {journal} {\bibinfo
  {journal} {Journal of Statistical Mechanics: Theory and Experiment}\ }\textbf
  {\bibinfo {volume} {2020}},\ \bibinfo {pages} {083102} (\bibinfo {year}
  {2020}{\natexlab{a}})}\BibitemShut {NoStop}%
\bibitem [{\citenamefont {Giulio}\ \emph {et~al.}(2019)\citenamefont {Giulio},
  \citenamefont {Arias},\ and\ \citenamefont {Tonni}}]{DiGiulio19}%
  \BibitemOpen
  \bibfield  {author} {\bibinfo {author} {\bibfnamefont {G.~D.}\ \bibnamefont
  {Giulio}}, \bibinfo {author} {\bibfnamefont {R.}~\bibnamefont {Arias}}, \
  and\ \bibinfo {author} {\bibfnamefont {E.}~\bibnamefont {Tonni}},\ }\href
  {\doibase 10.1088/1742-5468/ab4e8f} {\bibfield  {journal} {\bibinfo
  {journal} {Journal of Statistical Mechanics: Theory and Experiment}\ }\textbf
  {\bibinfo {volume} {2019}},\ \bibinfo {pages} {123103} (\bibinfo {year}
  {2019})}\BibitemShut {NoStop}%
\bibitem [{\citenamefont {Singh}\ and\ \citenamefont
  {Pathria}(1989)}]{singh-1989}%
  \BibitemOpen
  \bibfield  {author} {\bibinfo {author} {\bibfnamefont {S.}~\bibnamefont
  {Singh}}\ and\ \bibinfo {author} {\bibfnamefont {R.~K.}\ \bibnamefont
  {Pathria}},\ }\href {\doibase 10.1088/0305-4470/22/11/026} {\bibfield
  {journal} {\bibinfo  {journal} {Journal of Physics A: Mathematical and
  General}\ }\textbf {\bibinfo {volume} {22}},\ \bibinfo {pages} {1883}
  (\bibinfo {year} {1989})}\BibitemShut {NoStop}%
\bibitem [{\citenamefont {Contino}\ and\ \citenamefont
  {Gambassi}(2003)}]{contino-2002}%
  \BibitemOpen
  \bibfield  {author} {\bibinfo {author} {\bibfnamefont {R.}~\bibnamefont
  {Contino}}\ and\ \bibinfo {author} {\bibfnamefont {A.}~\bibnamefont
  {Gambassi}},\ }\href {\doibase 10.1063/1.1531215} {\bibfield  {journal}
  {\bibinfo  {journal} {Journal of Mathematical Physics}\ }\textbf {\bibinfo
  {volume} {44}},\ \bibinfo {pages} {570} (\bibinfo {year} {2003})}\BibitemShut
  {NoStop}%
\bibitem [{\citenamefont {Pelissetto}\ and\ \citenamefont
  {Vicari}(2002)}]{vicari}%
  \BibitemOpen
  \bibfield  {author} {\bibinfo {author} {\bibfnamefont {A.}~\bibnamefont
  {Pelissetto}}\ and\ \bibinfo {author} {\bibfnamefont {E.}~\bibnamefont
  {Vicari}},\ }\href {\doibase 10.1016/s0370-1573(02)00219-3} {\bibfield
  {journal} {\bibinfo  {journal} {Physics Reports}\ }\textbf {\bibinfo {volume}
  {368}},\ \bibinfo {pages} {549} (\bibinfo {year} {2002})}\BibitemShut
  {NoStop}%
\bibitem [{\citenamefont {Ruggiero}\ \emph {et~al.}(2016)\citenamefont
  {Ruggiero}, \citenamefont {Alba},\ and\ \citenamefont
  {Calabrese}}]{ruggiero-2016}%
  \BibitemOpen
  \bibfield  {author} {\bibinfo {author} {\bibfnamefont {P.}~\bibnamefont
  {Ruggiero}}, \bibinfo {author} {\bibfnamefont {V.}~\bibnamefont {Alba}}, \
  and\ \bibinfo {author} {\bibfnamefont {P.}~\bibnamefont {Calabrese}},\ }\href
  {\doibase 10.1103/PhysRevB.94.195121} {\bibfield  {journal} {\bibinfo
  {journal} {Phys. Rev. B}\ }\textbf {\bibinfo {volume} {94}},\ \bibinfo
  {pages} {195121} (\bibinfo {year} {2016})}\BibitemShut {NoStop}%
\bibitem [{\citenamefont {Mbeng}\ \emph {et~al.}(2017)\citenamefont {Mbeng},
  \citenamefont {Alba},\ and\ \citenamefont {Calabrese}}]{mbeng-2017}%
  \BibitemOpen
  \bibfield  {author} {\bibinfo {author} {\bibfnamefont {G.~B.}\ \bibnamefont
  {Mbeng}}, \bibinfo {author} {\bibfnamefont {V.}~\bibnamefont {Alba}}, \ and\
  \bibinfo {author} {\bibfnamefont {P.}~\bibnamefont {Calabrese}},\ }\href
  {\doibase 10.1088/1751-8121/aa6734} {\bibfield  {journal} {\bibinfo
  {journal} {Journal of Physics A: Mathematical and Theoretical}\ }\textbf
  {\bibinfo {volume} {50}},\ \bibinfo {pages} {194001} (\bibinfo {year}
  {2017})}\BibitemShut {NoStop}%
\bibitem [{\citenamefont {Shapourian}\ \emph {et~al.}(2019)\citenamefont
  {Shapourian}, \citenamefont {Ruggiero}, \citenamefont {Ryu},\ and\
  \citenamefont {Calabrese}}]{shapourian-2019}%
  \BibitemOpen
  \bibfield  {author} {\bibinfo {author} {\bibfnamefont {H.}~\bibnamefont
  {Shapourian}}, \bibinfo {author} {\bibfnamefont {P.}~\bibnamefont
  {Ruggiero}}, \bibinfo {author} {\bibfnamefont {S.}~\bibnamefont {Ryu}}, \
  and\ \bibinfo {author} {\bibfnamefont {P.}~\bibnamefont {Calabrese}},\ }\href
  {\doibase 10.21468/SciPostPhys.7.3.037} {\bibfield  {journal} {\bibinfo
  {journal} {SciPost Phys.}\ }\textbf {\bibinfo {volume} {7}},\ \bibinfo
  {pages} {37} (\bibinfo {year} {2019})}\BibitemShut {NoStop}%
\bibitem [{\citenamefont {Turkeshi}\ \emph
  {et~al.}(2020{\natexlab{a}})\citenamefont {Turkeshi}, \citenamefont
  {Ruggiero},\ and\ \citenamefont {Calabrese}}]{xhek-2020_A}%
  \BibitemOpen
  \bibfield  {author} {\bibinfo {author} {\bibfnamefont {X.}~\bibnamefont
  {Turkeshi}}, \bibinfo {author} {\bibfnamefont {P.}~\bibnamefont {Ruggiero}},
  \ and\ \bibinfo {author} {\bibfnamefont {P.}~\bibnamefont {Calabrese}},\
  }\href {\doibase 10.1103/PhysRevB.101.064207} {\bibfield  {journal} {\bibinfo
   {journal} {Phys. Rev. B}\ }\textbf {\bibinfo {volume} {101}},\ \bibinfo
  {pages} {064207} (\bibinfo {year} {2020}{\natexlab{a}})}\BibitemShut
  {NoStop}%
\bibitem [{\citenamefont {Laflorencie}\ and\ \citenamefont
  {Rachel}(2014)}]{Laflorencie2014}%
  \BibitemOpen
  \bibfield  {author} {\bibinfo {author} {\bibfnamefont {N.}~\bibnamefont
  {Laflorencie}}\ and\ \bibinfo {author} {\bibfnamefont {S.}~\bibnamefont
  {Rachel}},\ }\href {\doibase 10.1088/1742-5468/2014/11/p11013} {\bibfield
  {journal} {\bibinfo  {journal} {Journal of Statistical Mechanics: Theory and
  Experiment}\ }\textbf {\bibinfo {volume} {2014}},\ \bibinfo {pages} {P11013}
  (\bibinfo {year} {2014})}\BibitemShut {NoStop}%
\bibitem [{\citenamefont {Xavier}\ \emph {et~al.}(2018)\citenamefont {Xavier},
  \citenamefont {Alcaraz},\ and\ \citenamefont {Sierra}}]{Xavier2018}%
  \BibitemOpen
  \bibfield  {author} {\bibinfo {author} {\bibfnamefont {J.~C.}\ \bibnamefont
  {Xavier}}, \bibinfo {author} {\bibfnamefont {F.~C.}\ \bibnamefont {Alcaraz}},
  \ and\ \bibinfo {author} {\bibfnamefont {G.}~\bibnamefont {Sierra}},\ }\href
  {\doibase 10.1103/PhysRevB.98.041106} {\bibfield  {journal} {\bibinfo
  {journal} {Phys. Rev. B}\ }\textbf {\bibinfo {volume} {98}},\ \bibinfo
  {pages} {041106} (\bibinfo {year} {2018})}\BibitemShut {NoStop}%
\bibitem [{\citenamefont {Murciano}\ \emph
  {et~al.}(2020{\natexlab{b}})\citenamefont {Murciano}, \citenamefont
  {Giulio},\ and\ \citenamefont {Calabrese}}]{Murciano2019}%
  \BibitemOpen
  \bibfield  {author} {\bibinfo {author} {\bibfnamefont {S.}~\bibnamefont
  {Murciano}}, \bibinfo {author} {\bibfnamefont {G.~D.}\ \bibnamefont
  {Giulio}}, \ and\ \bibinfo {author} {\bibfnamefont {P.}~\bibnamefont
  {Calabrese}},\ }\href {\doibase 10.21468/SciPostPhys.8.3.046} {\bibfield
  {journal} {\bibinfo  {journal} {SciPost Phys.}\ }\textbf {\bibinfo {volume}
  {8}},\ \bibinfo {pages} {46} (\bibinfo {year}
  {2020}{\natexlab{b}})}\BibitemShut {NoStop}%
\bibitem [{\citenamefont {Goldstein}\ and\ \citenamefont
  {Sela}(2018)}]{Goldstein2018}%
  \BibitemOpen
  \bibfield  {author} {\bibinfo {author} {\bibfnamefont {M.}~\bibnamefont
  {Goldstein}}\ and\ \bibinfo {author} {\bibfnamefont {E.}~\bibnamefont
  {Sela}},\ }\href {\doibase 10.1103/PhysRevLett.120.200602} {\bibfield
  {journal} {\bibinfo  {journal} {Phys. Rev. Lett.}\ }\textbf {\bibinfo
  {volume} {120}},\ \bibinfo {pages} {200602} (\bibinfo {year}
  {2018})}\BibitemShut {NoStop}%
\bibitem [{\citenamefont {Cornfeld}\ \emph {et~al.}(2018)\citenamefont
  {Cornfeld}, \citenamefont {Goldstein},\ and\ \citenamefont
  {Sela}}]{Goldstein2018B}%
  \BibitemOpen
  \bibfield  {author} {\bibinfo {author} {\bibfnamefont {E.}~\bibnamefont
  {Cornfeld}}, \bibinfo {author} {\bibfnamefont {M.}~\bibnamefont {Goldstein}},
  \ and\ \bibinfo {author} {\bibfnamefont {E.}~\bibnamefont {Sela}},\ }\href
  {\doibase 10.1103/PhysRevA.98.032302} {\bibfield  {journal} {\bibinfo
  {journal} {Phys. Rev. A}\ }\textbf {\bibinfo {volume} {98}},\ \bibinfo
  {pages} {032302} (\bibinfo {year} {2018})}\BibitemShut {NoStop}%
\bibitem [{\citenamefont {Feldman}\ and\ \citenamefont
  {Goldstein}(2019)}]{Feld2019}%
  \BibitemOpen
  \bibfield  {author} {\bibinfo {author} {\bibfnamefont {N.}~\bibnamefont
  {Feldman}}\ and\ \bibinfo {author} {\bibfnamefont {M.}~\bibnamefont
  {Goldstein}},\ }\href {\doibase 10.1103/PhysRevB.100.235146} {\bibfield
  {journal} {\bibinfo  {journal} {Phys. Rev. B}\ }\textbf {\bibinfo {volume}
  {100}},\ \bibinfo {pages} {235146} (\bibinfo {year} {2019})}\BibitemShut
  {NoStop}%
\bibitem [{\citenamefont {Calabrese}\ \emph {et~al.}(2020)\citenamefont
  {Calabrese}, \citenamefont {Collura}, \citenamefont {Giulio},\ and\
  \citenamefont {Murciano}}]{Calabrese2020}%
  \BibitemOpen
  \bibfield  {author} {\bibinfo {author} {\bibfnamefont {P.}~\bibnamefont
  {Calabrese}}, \bibinfo {author} {\bibfnamefont {M.}~\bibnamefont {Collura}},
  \bibinfo {author} {\bibfnamefont {G.~D.}\ \bibnamefont {Giulio}}, \ and\
  \bibinfo {author} {\bibfnamefont {S.}~\bibnamefont {Murciano}},\ }\href
  {\doibase 10.1209/0295-5075/129/60007} {\bibfield  {journal} {\bibinfo
  {journal} {{EPL} (Europhysics Letters)}\ }\textbf {\bibinfo {volume} {129}},\
  \bibinfo {pages} {60007} (\bibinfo {year} {2020})}\BibitemShut {NoStop}%
\bibitem [{\citenamefont {Bonsignori}\ \emph {et~al.}(2019)\citenamefont
  {Bonsignori}, \citenamefont {Ruggiero},\ and\ \citenamefont
  {Calabrese}}]{Bonsignori2019}%
  \BibitemOpen
  \bibfield  {author} {\bibinfo {author} {\bibfnamefont {R.}~\bibnamefont
  {Bonsignori}}, \bibinfo {author} {\bibfnamefont {P.}~\bibnamefont
  {Ruggiero}}, \ and\ \bibinfo {author} {\bibfnamefont {P.}~\bibnamefont
  {Calabrese}},\ }\href {\doibase 10.1088/1751-8121/ab4b77} {\bibfield
  {journal} {\bibinfo  {journal} {Journal of Physics A: Mathematical and
  Theoretical}\ }\textbf {\bibinfo {volume} {52}},\ \bibinfo {pages} {475302}
  (\bibinfo {year} {2019})}\BibitemShut {NoStop}%
\bibitem [{\citenamefont {Fraenkel}\ and\ \citenamefont
  {Goldstein}(2020)}]{Fraenkel2020}%
  \BibitemOpen
  \bibfield  {author} {\bibinfo {author} {\bibfnamefont {S.}~\bibnamefont
  {Fraenkel}}\ and\ \bibinfo {author} {\bibfnamefont {M.}~\bibnamefont
  {Goldstein}},\ }\href {\doibase 10.1088/1742-5468/ab7753} {\bibfield
  {journal} {\bibinfo  {journal} {Journal of Statistical Mechanics: Theory and
  Experiment}\ }\textbf {\bibinfo {volume} {2020}},\ \bibinfo {pages} {033106}
  (\bibinfo {year} {2020})}\BibitemShut {NoStop}%
\bibitem [{\citenamefont {Capizzi}\ \emph {et~al.}(2020)\citenamefont
  {Capizzi}, \citenamefont {Ruggiero},\ and\ \citenamefont
  {Calabrese}}]{crc-20}%
  \BibitemOpen
  \bibfield  {author} {\bibinfo {author} {\bibfnamefont {L.}~\bibnamefont
  {Capizzi}}, \bibinfo {author} {\bibfnamefont {P.}~\bibnamefont {Ruggiero}}, \
  and\ \bibinfo {author} {\bibfnamefont {P.}~\bibnamefont {Calabrese}},\ }\href
  {\doibase 10.1088/1742-5468/ab96b6} {\bibfield  {journal} {\bibinfo
  {journal} {Journal of Statistical Mechanics: Theory and Experiment}\ }\textbf
  {\bibinfo {volume} {2020}},\ \bibinfo {pages} {073101} (\bibinfo {year}
  {2020})}\BibitemShut {NoStop}%
\bibitem [{\citenamefont {Cornfeld}\ \emph {et~al.}(2019)\citenamefont
  {Cornfeld}, \citenamefont {Landau}, \citenamefont {Shtengel},\ and\
  \citenamefont {Sela}}]{clss-19}%
  \BibitemOpen
  \bibfield  {author} {\bibinfo {author} {\bibfnamefont {E.}~\bibnamefont
  {Cornfeld}}, \bibinfo {author} {\bibfnamefont {L.~A.}\ \bibnamefont
  {Landau}}, \bibinfo {author} {\bibfnamefont {K.}~\bibnamefont {Shtengel}}, \
  and\ \bibinfo {author} {\bibfnamefont {E.}~\bibnamefont {Sela}},\ }\href
  {\doibase 10.1103/PhysRevB.99.115429} {\bibfield  {journal} {\bibinfo
  {journal} {Phys. Rev. B}\ }\textbf {\bibinfo {volume} {99}},\ \bibinfo
  {pages} {115429} (\bibinfo {year} {2019})}\BibitemShut {NoStop}%
\bibitem [{\citenamefont {Caputa}\ \emph {et~al.}(2013)\citenamefont {Caputa},
  \citenamefont {Mandal},\ and\ \citenamefont {Sinha}}]{cms-13}%
  \BibitemOpen
  \bibfield  {author} {\bibinfo {author} {\bibfnamefont {P.}~\bibnamefont
  {Caputa}}, \bibinfo {author} {\bibfnamefont {G.}~\bibnamefont {Mandal}}, \
  and\ \bibinfo {author} {\bibfnamefont {R.}~\bibnamefont {Sinha}},\ }\href
  {\doibase 10.1007/jhep11(2013)052} {\bibfield  {journal} {\bibinfo  {journal}
  {Journal of High Energy Physics}\ }\textbf {\bibinfo {volume} {2013}}
  (\bibinfo {year} {2013}),\ 10.1007/jhep11(2013)052}\BibitemShut {NoStop}%
\bibitem [{\citenamefont {Dowker}(2016)}]{d-16}%
  \BibitemOpen
  \bibfield  {author} {\bibinfo {author} {\bibfnamefont {J.~S.}\ \bibnamefont
  {Dowker}},\ }\href {\doibase 10.1088/1751-8113/49/14/145401} {\bibfield
  {journal} {\bibinfo  {journal} {Journal of Physics A: Mathematical and
  Theoretical}\ }\textbf {\bibinfo {volume} {49}},\ \bibinfo {pages} {145401}
  (\bibinfo {year} {2016})}\BibitemShut {NoStop}%
\bibitem [{\citenamefont {Dowker}(2017)}]{d-16-1}%
  \BibitemOpen
  \bibfield  {author} {\bibinfo {author} {\bibfnamefont {J.~S.}\ \bibnamefont
  {Dowker}},\ }\href {\doibase 10.1088/1751-8121/aa6178} {\bibfield  {journal}
  {\bibinfo  {journal} {Journal of Physics A: Mathematical and Theoretical}\
  }\textbf {\bibinfo {volume} {50}},\ \bibinfo {pages} {165401} (\bibinfo
  {year} {2017})}\BibitemShut {NoStop}%
\bibitem [{\citenamefont {Belin}\ \emph {et~al.}(2013)\citenamefont {Belin},
  \citenamefont {Hung}, \citenamefont {Maloney}, \citenamefont {Matsuura},
  \citenamefont {Myers},\ and\ \citenamefont {Sierens}}]{matsuura}%
  \BibitemOpen
  \bibfield  {author} {\bibinfo {author} {\bibfnamefont {A.}~\bibnamefont
  {Belin}}, \bibinfo {author} {\bibfnamefont {L.-Y.}\ \bibnamefont {Hung}},
  \bibinfo {author} {\bibfnamefont {A.}~\bibnamefont {Maloney}}, \bibinfo
  {author} {\bibfnamefont {S.}~\bibnamefont {Matsuura}}, \bibinfo {author}
  {\bibfnamefont {R.~C.}\ \bibnamefont {Myers}}, \ and\ \bibinfo {author}
  {\bibfnamefont {T.}~\bibnamefont {Sierens}},\ }\href {\doibase
  10.1007/jhep12(2013)059} {\bibfield  {journal} {\bibinfo  {journal} {Journal
  of High Energy Physics}\ }\textbf {\bibinfo {volume} {2013}} (\bibinfo {year}
  {2013}),\ 10.1007/jhep12(2013)059}\BibitemShut {NoStop}%
\bibitem [{\citenamefont {Caputa}\ \emph {et~al.}(2016)\citenamefont {Caputa},
  \citenamefont {Nozaki},\ and\ \citenamefont {Numasawa}}]{SREE}%
  \BibitemOpen
  \bibfield  {author} {\bibinfo {author} {\bibfnamefont {P.}~\bibnamefont
  {Caputa}}, \bibinfo {author} {\bibfnamefont {M.}~\bibnamefont {Nozaki}}, \
  and\ \bibinfo {author} {\bibfnamefont {T.}~\bibnamefont {Numasawa}},\ }\href
  {\doibase 10.1103/PhysRevD.93.105032} {\bibfield  {journal} {\bibinfo
  {journal} {Phys. Rev. D}\ }\textbf {\bibinfo {volume} {93}},\ \bibinfo
  {pages} {105032} (\bibinfo {year} {2016})}\BibitemShut {NoStop}%
\bibitem [{\citenamefont {Turkeshi}\ \emph
  {et~al.}(2020{\natexlab{b}})\citenamefont {Turkeshi}, \citenamefont
  {Ruggiero}, \citenamefont {Alba},\ and\ \citenamefont
  {Calabrese}}]{xhek-2020_B}%
  \BibitemOpen
  \bibfield  {author} {\bibinfo {author} {\bibfnamefont {X.}~\bibnamefont
  {Turkeshi}}, \bibinfo {author} {\bibfnamefont {P.}~\bibnamefont {Ruggiero}},
  \bibinfo {author} {\bibfnamefont {V.}~\bibnamefont {Alba}}, \ and\ \bibinfo
  {author} {\bibfnamefont {P.}~\bibnamefont {Calabrese}},\ }\href {\doibase
  10.1103/PhysRevB.102.014455} {\bibfield  {journal} {\bibinfo  {journal}
  {Phys. Rev. B}\ }\textbf {\bibinfo {volume} {102}},\ \bibinfo {pages}
  {014455} (\bibinfo {year} {2020}{\natexlab{b}})}\BibitemShut {NoStop}%
\bibitem [{\citenamefont {Wolf}(2006)}]{Wolf-2004}%
  \BibitemOpen
  \bibfield  {author} {\bibinfo {author} {\bibfnamefont {M.~M.}\ \bibnamefont
  {Wolf}},\ }\href {\doibase 10.1103/PhysRevLett.96.010404} {\bibfield
  {journal} {\bibinfo  {journal} {Phys. Rev. Lett.}\ }\textbf {\bibinfo
  {volume} {96}},\ \bibinfo {pages} {010404} (\bibinfo {year}
  {2006})}\BibitemShut {NoStop}%
\bibitem [{\citenamefont {Gioev}\ and\ \citenamefont
  {Klich}(2006)}]{Gioev-2006}%
  \BibitemOpen
  \bibfield  {author} {\bibinfo {author} {\bibfnamefont {D.}~\bibnamefont
  {Gioev}}\ and\ \bibinfo {author} {\bibfnamefont {I.}~\bibnamefont {Klich}},\
  }\href {\doibase 10.1103/PhysRevLett.96.100503} {\bibfield  {journal}
  {\bibinfo  {journal} {Phys. Rev. Lett.}\ }\textbf {\bibinfo {volume} {96}},\
  \bibinfo {pages} {100503} (\bibinfo {year} {2006})}\BibitemShut {NoStop}%
\bibitem [{\citenamefont {Farkas}\ and\ \citenamefont
  {Zimbor\'as}(2007)}]{Farkas-2007}%
  \BibitemOpen
  \bibfield  {author} {\bibinfo {author} {\bibfnamefont {S.}~\bibnamefont
  {Farkas}}\ and\ \bibinfo {author} {\bibfnamefont {Z.}~\bibnamefont
  {Zimbor\'as}},\ }\href {\doibase 10.1063/1.2800167} {\bibfield  {journal}
  {\bibinfo  {journal} {Journal of Mathematical Physics}\ }\textbf {\bibinfo
  {volume} {48}},\ \bibinfo {pages} {102110} (\bibinfo {year} {2007})},\
  \Eprint {http://arxiv.org/abs/https://doi.org/10.1063/1.2800167}
  {https://doi.org/10.1063/1.2800167} \BibitemShut {NoStop}%
\bibitem [{\citenamefont {Li}\ \emph {et~al.}(2006)\citenamefont {Li},
  \citenamefont {Ding}, \citenamefont {Yu}, \citenamefont {Roscilde},\ and\
  \citenamefont {Haas}}]{Li-2006}%
  \BibitemOpen
  \bibfield  {author} {\bibinfo {author} {\bibfnamefont {W.}~\bibnamefont
  {Li}}, \bibinfo {author} {\bibfnamefont {L.}~\bibnamefont {Ding}}, \bibinfo
  {author} {\bibfnamefont {R.}~\bibnamefont {Yu}}, \bibinfo {author}
  {\bibfnamefont {T.}~\bibnamefont {Roscilde}}, \ and\ \bibinfo {author}
  {\bibfnamefont {S.}~\bibnamefont {Haas}},\ }\href {\doibase
  10.1103/PhysRevB.74.073103} {\bibfield  {journal} {\bibinfo  {journal} {Phys.
  Rev. B}\ }\textbf {\bibinfo {volume} {74}},\ \bibinfo {pages} {073103}
  (\bibinfo {year} {2006})}\BibitemShut {NoStop}%
\bibitem [{\citenamefont {Swingle}(2010)}]{Swingle-2010}%
  \BibitemOpen
  \bibfield  {author} {\bibinfo {author} {\bibfnamefont {B.}~\bibnamefont
  {Swingle}},\ }\href {\doibase 10.1103/PhysRevLett.105.050502} {\bibfield
  {journal} {\bibinfo  {journal} {Phys. Rev. Lett.}\ }\textbf {\bibinfo
  {volume} {105}},\ \bibinfo {pages} {050502} (\bibinfo {year}
  {2010})}\BibitemShut {NoStop}%
\bibitem [{\citenamefont {Calabrese}\ \emph {et~al.}(2012)\citenamefont
  {Calabrese}, \citenamefont {Mintchev},\ and\ \citenamefont
  {Vicari}}]{Calabrese-2012}%
  \BibitemOpen
  \bibfield  {author} {\bibinfo {author} {\bibfnamefont {P.}~\bibnamefont
  {Calabrese}}, \bibinfo {author} {\bibfnamefont {M.}~\bibnamefont {Mintchev}},
  \ and\ \bibinfo {author} {\bibfnamefont {E.}~\bibnamefont {Vicari}},\ }\href
  {\doibase 10.1209/0295-5075/97/20009} {\bibfield  {journal} {\bibinfo
  {journal} {{EPL} (Europhysics Letters)}\ }\textbf {\bibinfo {volume} {97}},\
  \bibinfo {pages} {20009} (\bibinfo {year} {2012})}\BibitemShut {NoStop}%
\bibitem [{\citenamefont {Ding}\ \emph {et~al.}(2012)\citenamefont {Ding},
  \citenamefont {Seidel},\ and\ \citenamefont {Yang}}]{Ding-2012}%
  \BibitemOpen
  \bibfield  {author} {\bibinfo {author} {\bibfnamefont {W.}~\bibnamefont
  {Ding}}, \bibinfo {author} {\bibfnamefont {A.}~\bibnamefont {Seidel}}, \ and\
  \bibinfo {author} {\bibfnamefont {K.}~\bibnamefont {Yang}},\ }\href {\doibase
  10.1103/PhysRevX.2.011012} {\bibfield  {journal} {\bibinfo  {journal} {Phys.
  Rev. X}\ }\textbf {\bibinfo {volume} {2}},\ \bibinfo {pages} {011012}
  (\bibinfo {year} {2012})}\BibitemShut {NoStop}%
\bibitem [{\citenamefont {Copson}(1965)}]{Cop65}%
  \BibitemOpen
  \bibfield  {author} {\bibinfo {author} {\bibfnamefont {E.~T.}\ \bibnamefont
  {Copson}},\ }\href {\doibase 10.1017/CBO9780511526121} {\emph {\bibinfo
  {title} {Asymptotic Expansions}}},\ Cambridge Tracts in Mathematics\
  (\bibinfo  {publisher} {Cambridge University Press},\ \bibinfo {year}
  {1965})\BibitemShut {NoStop}%
\end{thebibliography}%

\end{document}